\documentclass[11pt,a4paper]{article}

\usepackage[american]{babel}              
\usepackage[utf8]{inputenc}     
\usepackage[T1]{fontenc}
\usepackage{jheppub}
\usepackage{amsmath,amstext,amsfonts,amssymb}
\usepackage{bm,dsfont}
\usepackage{array,xcolor,graphicx}
\usepackage{floatrow,dcolumn,rotating}
\usepackage{braket,tensor,units}
\usepackage{booktabs,multirow}
\usepackage[above,below,section]{placeins}
\usepackage[version=3]{mhchem}

\definecolor{darkred}{rgb}{.7,0,0}
\definecolor{darkgreen}{rgb}{0,.5,0}       
\definecolor{darkblue}{rgb}{0,0,.55}
\definecolor{darkyellow}{rgb}{.9,.9,0}
\definecolor{yellowgreen}{rgb}{.8,1,0}
\definecolor{lightgray}{gray}{.8}

\definecolor{mdarkgreen}{HTML}{00aa00}
\definecolor{mpurple}{HTML}{800080}
\definecolor{scgreen}{rgb}{0,.882353,.227451}

\definecolor{DarkRainbow0}{rgb}{.237736,.340215,.575113}
\definecolor{DarkRainbow01}{rgb}{.253651,.344893,.558151}
\definecolor{DarkRainbow02}{rgb}{.264425,.423024,.3849}
\definecolor{DarkRainbow03}{rgb}{.291469,.47717,.271411}
\definecolor{DarkRainbow04}{rgb}{.416394,.555345,.24182}
\definecolor{DarkRainbow05}{rgb}{.624866,.673302,.264296}
\definecolor{DarkRainbow06}{rgb}{.813033,.766292,.303458}
\definecolor{DarkRainbow07}{rgb}{.877875,.731045,.326896}
\definecolor{DarkRainbow08}{rgb}{.812807,.518694,.303459}
\definecolor{DarkRainbow09}{rgb}{.72987,.239399,.230961}

\definecolor{DarkRainbow033}{rgb}{.3289465,.5006225,.2625337}
\definecolor{DarkRainbow066}{rgb}{.8519382,.7451438,.3175208}
\definecolor{DarkRainbow099}{rgb}{.72987,.239399,.230961}

\definecolor{DarkRainbow0p045}{rgb}{.244898,.34232,.56748}
\definecolor{DarkRainbow0p09}{rgb}{.25206,.344425,.559847}
\definecolor{DarkRainbow0p135}{rgb}{.257422,.372239,.497513}
\definecolor{DarkRainbow0p18}{rgb}{.26227,.407398,.41955}
\definecolor{DarkRainbow0p225}{rgb}{.271186,.43656,.356528}
\definecolor{DarkRainbow0p27}{rgb}{.283356,.460926,.305458}
\definecolor{DarkRainbow0p315}{rgb}{.310208,.488896,.266972}
\definecolor{DarkRainbow0p36}{rgb}{.366424,.524075,.253656}
\definecolor{DarkRainbow0p405}{rgb}{.426818,.561243,.242944}
\definecolor{DarkRainbow0p45}{rgb}{.52063,.614323,.253058}
\definecolor{DarkRainbow0p495}{rgb}{.614442,.667404,.263172}
\definecolor{DarkRainbow0p54}{rgb}{.700133,.710498,.279961}
\definecolor{DarkRainbow0p585}{rgb}{.784808,.752343,.297584}
\definecolor{DarkRainbow0p63}{rgb}{.832486,.755718,.310489}
\definecolor{DarkRainbow0p675}{rgb}{.861664,.739857,.321037}
\definecolor{DarkRainbow0p72}{rgb}{.864861,.688575,.322209}
\definecolor{DarkRainbow0p765}{rgb}{.835581,.593017,.311662}
\definecolor{DarkRainbow0p81}{rgb}{.804513,.490765,.296209}
\definecolor{DarkRainbow0p855}{rgb}{.767192,.365082,.263585}
\definecolor{DarkRainbow0p9}{rgb}{.72987,.239399,.230961}

\definecolor{SunsetColors0p6}{rgb}{.991751,.5902,.0983158}
\definecolor{SunsetColors0p55}{rgb}{.985564,.520833,.0747243}
\definecolor{SunsetColors0p54}{rgb}{.984876,.513126,.0721031}
\definecolor{SunsetColors0p5}{rgb}{.979377,.451467,.0511329}
\definecolor{SunsetColors0p49}{rgb}{.966638,.43869,.0657759}
\definecolor{SunsetColors0p45}{rgb}{.92205,.393972,.117026}
\definecolor{SunsetColors0p43}{rgb}{.902941,.374807,.138991}
\definecolor{SunsetColors0p4}{rgb}{.864723,.336476,.18292}
\definecolor{SunsetColors0p38}{rgb}{.839244,.310923,.212206}
\definecolor{SunsetColors0p35}{rgb}{.807396,.278981,.248813}
\definecolor{SunsetColors0p32}{rgb}{.760587,.251548,.286493}
\definecolor{SunsetColors0p3}{rgb}{.705188,.235013,.317923}
\definecolor{SunsetColors0p27}{rgb}{.622089,.21021,.365068}
\definecolor{SunsetColors0p25}{rgb}{.58054,.197808,.388641}
\definecolor{SunsetColors0p21}{rgb}{.483591,.168871,.443643}
\definecolor{SunsetColors0p2}{rgb}{.455892,.160603,.459358}
\definecolor{SunsetColors0p16}{rgb}{.34794,.126747,.472736}
\definecolor{SunsetColors0p15}{rgb}{.335514,.12222,.455853}
\definecolor{SunsetColors0p1}{rgb}{.223676,.08148,.303902}

\newcolumntype{C}{>{$}c<{$}}
\newcolumntype{L}{>{$}l<{$}}
\newcolumntype{R}{>{$}r<{$}}             

\newcolumntype{S}[1]{>{$}c<{$}@{\hspace{#1}}}
\newcolumntype{T}[1]{>{$}l<{$}@{\hspace{#1}}}
\newcolumntype{U}[1]{>{$}r<{$}@{\hspace{#1}}}

\newcommand\figref[1]{\figurename~\ref{#1}}
\newcommand\tabref[1]{\tablename~\ref{#1}}

\newfloatcommand{figcapside}{figure}
    [{\capbeside\captionsetup[capbesidefigure]{format=plain}%
      \thisfloatsetup{facing=yes,capbesideposition={inside,center},%
        capbesidesep=qquad}}][\FBwidth]

\newfloatcommand{tabcapside}{table}
    [{\capbeside\captionsetup[capbesidetable]{format=plain}%
      \thisfloatsetup{facing=yes,capbesideposition={inside,center},%
        capbesidesep=qquad}}][\FBwidth]

\makeatletter
\newcommand\etc{\textit{etc}\@ifnextchar.{}{.\@\space}}
\newcommand\cff{\textit{c.f}\@ifnextchar.{}{.\@\space}}
\newcommand\ie{\textit{i.e}\@ifnextchar.{}{.\@\space}}
\newcommand\eg{\textit{e.g}\@ifnextchar.{}{.\@\space}}
\newcommand\ia{\textit{i.a}\@ifnextchar.{}{.\@\space}}
\makeatother

\DeclareMathAlphabet{\mathbfit}{T1}{cmbr}{b}{n}

\newcommand\ci{\mathrm i}
\newcommand\dd[2][]{\operatorname{d^{#1}\mathit{#2}}}
\newcommand\e[1][]{\,\mathrm{e}^{#1}\,}

\newcommand\vvec[1]{\mathbfit{#1}}
\newcommand\abs[1]{\left\lvert#1\right\rvert}

\DeclareMathOperator\im{Im}
\DeclareMathOperator\re{Re}
\DeclareMathOperator\sgn{sgn}

\newcommand\expect[1]{\left\langle#1\right\rangle}

\newcommand\eqcomma{\text{,}}

\allowdisplaybreaks

\title{S-Wave Superconductivity in Anisotropic Holographic Insulators}

\author[a]{Johanna Erdmenger,}
\author[a,b]{Benedikt Herwerth,}
\author[c]{Steffen Klug,}
\author[d]{René Meyer,}
\author[c]{and Koenraad Schalm}

\affiliation[a]{\href{http://www.mppmu.mpg.de}{Max Planck Institute for Physics}
  (Werner-Heisenberg-Institut), F\"ohringer Ring 6, 80805 Munich, Germany}

\affiliation[b]{\href{http://www.mpq.mpg.de}{Max Planck Institute of Quantum Optics},
  Hans-Kopfermann-Str. 1, 85748 Garching, Germany}

\affiliation[c]{\href{http://www.lorentz.leidenuniv.nl}
  {Institute Lorentz for Theoretical Physics}, Leiden University, P.O. Box 9506, Leiden
  2300RA, The Netherlands}

\affiliation[d]{\href{http://www.ipmu.jp}
  {Kavli Institute for the Physics and Mathematics of the Universe}
  (WPI), Todai Institutes for Advanced Study, The University of Tokyo,
  Kashiwa, Chiba 277-8568, Japan}

\emailAdd{jke@mpp.mpg.de}
\emailAdd{benedikt.herwerth@mpq.mpg.de}
\emailAdd{klug@lorentz.leidenuniv.nl}
\emailAdd{rene.meyer@ipmu.jp}
\emailAdd{kschalm@lorentz.leidenuniv.nl}

\abstract{
  Within gauge/gravity duality, we consider finite density systems in a helical lattice
  dual to asymptotically anti-de Sitter space-times with Bianchi VII symmetry. These
  systems can become an anisotropic insulator in one direction while retaining metallic
  behavior in others. To this model, we add a $U(1)$ charged scalar and show that below
  a critical temperature, it forms a spatially homogeneous condensate that restores
  isotropy in a new superconducting ground state. We determine the phase diagram in
  terms of the helix parameters and perform a stability analysis on its IR fixed point
  corresponding to a finite density condensed phase at zero temperature. Moreover, by
  analyzing fluctuations about the gravity background, we study the optical
  conductivity. Due to the lattice, this model provides an example for a holographic
  insulator-superfluid transition in which there is no unrealistic delta-function peak
  in the normal phase DC conductivity. Our results suggest that in the zero temperature
  limit, all degrees of freedom present in the normal phase condense. This, together
  with the breaking of translation invariance, has implications for Homes' and Uemuras's
  relations. This is of relevance for applications to real world condensed
  matter systems. We find a range of parameters in this system where
  Homes' relation holds.
}

\preprint{\parbox[c]{3cm}{\flushright IPMU14-0284\\MPP-2014-581}}


\arxivnumber{1501.07615}

\hypersetup{pdfauthor={Johanna Erdmenger, Benedikt Herwerth, Steffen Klug, René Meyer,
    Koenraad Schalm},pdftitle={S-Wave Superconductivity in Anisotropic Holographic
    Insulators},pdfsubject={Holography and Condensed Matter Physics (AdS/CMT),
    Gauge-Gravity Correspondence}
}

\begin{document}

\maketitle

\section{Introduction\label{sec:introduction}}

Significant progress has recently been achieved in applying gauge/gravity duality to
strongly coupled systems of relevance to condensed matter physics. In particular,
different approaches were proposed to include a lattice into the dual gravity
background, in order to holographically study the conductivity in systems with broken
translation invariance. Systems with manifest translation invariance
display an unrealistic $\delta$-function at zero-frequency in the conductivity; breaking
the symmetry weakly broadens this into a realistic Drude peak known from condensed
matter physics as a consequence of momentum dissipation.
Holography allows moreover the exploration of the consequences of translational symmetry
breaking for strongly correlated systems beyond the Drude peak, both in the weak Drude regime
and for stronger lattice potentials. We will follow this avenue in the
present paper.
\\[\baselineskip]
Within holography, translation breaking approaches include explicit breaking by a
modulated Ansatz for the chemical potential \cite{Horowitz2012,Liu:2012tr,Ling:2013aya,
  Horowitz2013}, the use of massive gravity \cite{Vegh2013,Blake:2013owa} or linear
axions \cite{Andrade:2013gsa,Gouteraux:2014hca,Taylor:2014tka}, or other lattice
Ans\"atze such as the Q-lattices \cite{Donos:2013eha} or  the method used in this work,
Bianchi symmetric solutions \cite{Iizuka:2012iv,Iizuka:2012pn}. In some 
cases, translation invariance is also spontaneously broken, for instance when
Chern-Simons or $F \wedge F$ terms are present in the gravity action
\cite{Domokos:2007kt,Nakamura:2009tf,Ooguri:2010kt,Donos:2011bh,Donos:2013wia,
  Withers:2013loa,Withers:2014sja,Ling:2014saa}, or in an external $SU(2)$ magnetic
field \cite{Ammon:2011je,Bu:2012mq}. Explicit breaking with an interesting
phenomenological consequence is realized in the \emph{helical lattice} approach
\cite{Iizuka:2012iv,Iizuka:2012pn,Donos2011b,Donos2012b,Donos2012a,Donos2013c,
  Donos:2014gya}. The original motivation to study this model was that the helical
symmetry allows for momentum relaxation along one spatial direction without the need to
solve PDEs. The helix in one of the field theory directions along the boundary is
encoded in a non-trivial background $U(1)$ gauge field on the gravity side of the
holographic duality, and its shape is protected by a so-called Bianchi $\text{VII}_0$
symmetry. In addition to that `helix $U(1)$', the five-dimensional gravity action
\eqref{eq:helix-action-old} that we study involves a separate `charge $U(1)$' dual to a
globally conserved charge current in the boundary theory. This is needed to encode a
field theory at finite density. As discussed in \cite{Donos2013c}, the natural finite
density state of this model is a conducting metal, but it can display a transition to an
insulating phase as a function of the helix momentum. The remarkable aspect is that this
new phase is uni-directional and anisotropic: It is an insulator only along the
direction of broken translation invariance along the helical axis. In the orthogonal
directions the system remains a metal. From a condensed matter point of view, this
system resembles a so-called quantum smectic.
\\[\baselineskip]
The specific objective we shall be interested in this paper is the consequences of
translational symmetry breaking for the transition to superconductivity. This was also
recently studied in a holographic Q-lattice in \cite{Donos:2013eha,Ling:2014laa} and in
axion and related holographic superconductor models in \cite{Andrade:2014xca,
  Kim:2015dna}. In both cases only isotropic  models were considered, though both models
can support anisotropic lattices \cite{Donos:2014uba,Koga:2014hwa}. In our intrinsically
anisotropic helical lattice, the dual gravitational dynamics imply that the favored
ground state will nevertheless be an isotropic s-wave superconductor. In fact the only
Bianchi $\text{VII}_0$ symmetric and time-independent Ansatz for the scalar field dual
to the order parameter is just a constant in boundary direction. This is the system we
shall study. We show that a scalar field added to the helical lattice action and charged
under the second $U(1)$ gauge field condenses below a critical temperature, both in the
insulating and in the conducting phase. We explore the phase diagram which is determined
by the amplitude and the momentum of the translationally symmetry breaking helix, both
at finite and at vanishing temperature. Moreover, we analyse the thermodynamical as well
as transport properties of the different phases --- metallic, insulating, condensed ---
of the helical lattice model, and obtain the finite temperature phase diagram of the
system.\\[\baselineskip]
Our findings can be summarized as follows:

\bigskip

\emph{The superconducting phase transition:} In Section 2 we investigate the finite
temperature phase diagram of our model, which displays a second order mean field
transition from both the insulating as well as metallic phase to a superfluid phase. We
in particular show that the critical temperature $T_c $ depends strongly on the
amplitude of the helix, but to a first approximation rather weakly on its momentum. This
indicates that the strength of the translational symmetry breaking (the depth of the
potential wells) is more important than their spatial distribution. For large amplitude,
$T_c$ can in principle be suppressed all the way to zero, and a quantum phase transition
to the uncondensed phase may be expected.
\\[\baselineskip]
The critical temperature $T_c$ does depend mildly on the helix momentum  $p$,  in a curious way.
Starting in the phase at small helix momentum which is originally a
zero-temperature insulator in the normal phase, $T_c$ decreases with increasing helix
momentum. However, $T_c$ grows again for even larger helix momentum. This might be
understood from the observation that initially with increasing momentum the underlying
original insulating system changes to a zero temperature conductor in the normal 
phase, but then for even larger momentum turns back into an insulator.

\bigskip

\emph{The optical conductivity:}
In Section 3 we calculate the optical conductivity in the direction of translational symmetry
breaking in the insulating, conducting as well as condensed phases. In the insulating
and conducting phases we reproduce the results of \cite{Donos2013c}. In the condensed
phase we observe the appearance of a gap at low frequencies as expected for spontaneous
symmetry breaking. The spectral weight is transferred to a $\delta$-function
contribution at zero frequency: this is confirmed with the Ferrell-Glover-Tinkham sum 
rule.
\\[\baselineskip]
The virtue of the helix model is that this $\delta$-function is now cleanly interpreted
as the consequence of spontaneous symmetry breaking. There is no artificial contribution
due to translational symmetry. The strength of this $\delta$-peak therefore defines the
superfluid density in the condensed phase. For weak momentum relaxation,
$\nicefrac\lambda\mu\ll 1$ we find that in the limit $T\rightarrow 0$, the
superfluid density coincides with the total charge density in the system, as measured by
the second $U(1)$ gauge field. This can be understood by the fact that the
zero-temperature normal state of our system is already in a cohesive phase,
in which no uncondensed charged degrees of freedom are present in the
deep IR. 
This however does not mean that we are dealing with a plain vanilla superconductor. At
any finite temperature, the horizon does carry charge. This reflects itself in the
temperature dependence of the superconducting gap. We find that the low $T$ behavior of
the superconducting gap is algebraic, \ie $\sigma(\omega^\ast)\sim T^\#$, rather than
exponential. Nevertheless as stated earlier, computing the thermodynamical charge
density $n_s$ and the superfluid density $\rho_s$ independently, we find that they
coincide in the limit of zero temperature, in the regime of weak momentum relaxation.   
\\[\baselineskip]
The helical system considered has therefore two important properties:
Translation symmetry is broken and all charged degrees of freedom
condense at very low temperatures. The combination of both these facts
enables us to take a further look at {\it Homes' relation} in the context
of holography. This empirical law, found experimentally \cite{Homes2004,Homes2005}, 
states that there is a universal behaviour for classes of
superconductors that relates the superconducting density $\rho_s$ at zero
temperature to the DC conductivity $\sigma_{\mathrm{DC}}$ at $T_c$
\begin{equation}
  \rho_s(T=0)=C\sigma_{\mathrm{DC}}(T_c)\cdot T_c.
\end{equation}
The constant $C$, which is dimensionless in suitable units, is experimentally found
to be around $C=4.4$ for in-plane high-$T_c$ superconductors as well as clean BCS
superconductors and around $C=8.1$ for c-axis high-$T_c$ materials and BCS superconductors
in the dirty limit \cite{Homes2004,Homes2005}. Generally, a relation of this type is
expected for systems which are {\it Planckian dissipators} \cite{Zaanen2004}. Homes'
relation was first considered in the context of holography in \cite{Erdmenger2012a},
where it was found that for a holographic realization, both translation symmetry
breaking and the condensation of all charged degrees of freedom are necessary
conditions. Both of these conditions are realized in the helical lattice system in the
present paper. It was found in \cite{Horowitz2013} that a simple breaking of translation
invariance by a modulated chemical potential is not sufficient for a holographic
realization of Homes' relation, essentially since in the limit of vanishing chemical
potential, the DC conductivity diverges while the superconducting density remains
finite. Indeed, Homes' relation cannot hold for weak momentum relaxation. However,
motivated by the arguments given above, we considered Homes' relation in the context of
the helical lattice model for strong momentum relaxation.
In a parameter region around the minimum of $T_c$ found in Section 2, $C$ appears to be
roughly constant for a significant region in parameter space. We find a value of about 
\begin{equation}
  C=6.2\pm 0.3,
\end{equation}
which lies between the experimental results for high $T_c$ and dirty limit BCS
superconductors. These encouraging results call for further detailed analysis, which we
leave for future work.

\bigskip
\begin{figure}[t]
  \includegraphics[width=0.9\textwidth]{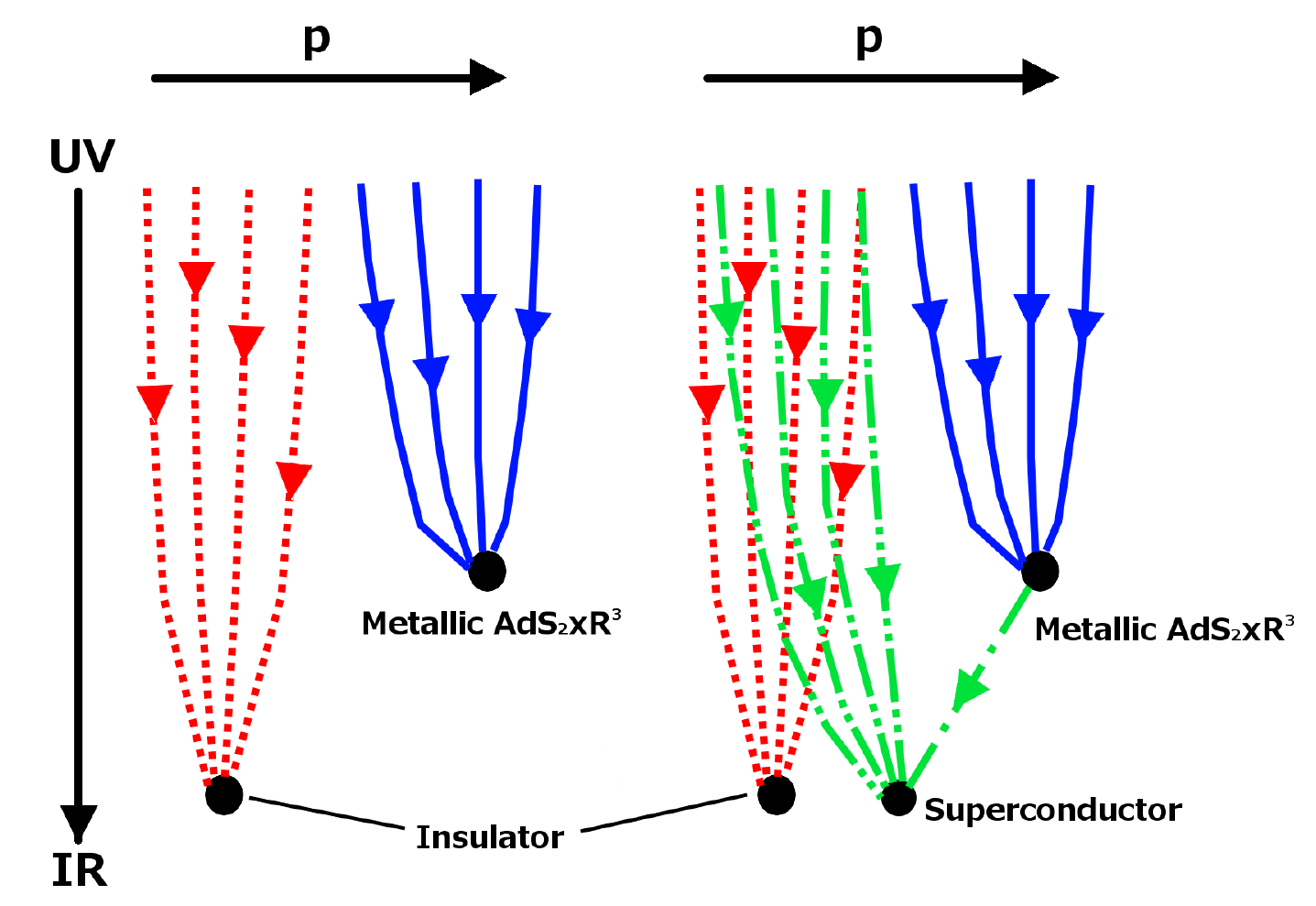}
  \caption{Sketch of the holographic renormalisation group flows without (left panel)
    and with superconducting order parameter (right panel), for the case of large
    $\kappa \geq 0.57$ in which no additional unstable bifurcation fixed points appear
    in the model of \cite{Donos2013c}. Without the superconducting order parameter, the
    transition from the insulating ({\color{red}red} dotted) to the metallic
    $\text{AdS}_2\times\mathds{R}^3$ ({\color{blue}blue} solid) ground states occurs as
    the helix pitch $\nicefrac p\mu$ is increased \cite{Donos2013c}. With the
    superconducting order parameter a new superconducting ({\color{scgreen}green}
    dash-dot-dotted) ground state, \eqref{eq:app-t-zer-power}, appears. While the
    metallic ground states becomes dynamically unstable and presumably flows to the
    superconductor, the insulating states stay dynamically stable but are presumably
    thermodynamically disfavoured compared to flows to the superconducting fixed point.
    \label{fig:rgflowsuco}}
\end{figure}
\emph{The zero-temperature ground state:} In Section 4 we provide a preliminary analysis
of the zero-temperature ground states of the condensed system. The starting point is the
IR geometry of the insulating ground state geometry of the helical model in the absence
of a condensate \cite{Donos2013c}. We show that this Ansatz for the IR geometry can
naturally be extended to the superconducting solution. Besides this, the insulating as
well as metallic ground states of \cite{Donos2013c} continue to exist.
We perform the usual IR fluctuation analysis and delineate the various IR relevant
directions, if present, and the IR irrelevant directions. This allows us to understand
the RG flow of the model in principle (\figref{fig:rgflowsuco}). We in particular find
a difference in the instability mechanisms of the insulating and metallic fixed points
of \cite{Donos2013c}: While the metallic $\text{AdS}_2\times\mathds{R}^3$ becomes
dynamically unstable at low temperatures towards condensation of the superconducting
order parameter, the insulating fixed point stays dynamically stable, but most
presumably becomes thermodynamically disfavored compared with flows to the
superconducting fixed point. Since our finite temperature phase diagram
(\figref{fig:helix-finite-t-phase-trans-phase-diag}) indicates the possibility of a
zero temperature phase transition between the condensed and insulating solution, these
results calls for a more detailed study of the phase diagram at finite and zero
temperature in a future work \cite{WIP}.

\bigskip

We conclude in Section 5 where we discuss in particular the implications of our results
for the holographic realization of Homes' relation. We end by giving an outlook to
further investigations.

\section{Holographic S-Wave Superconductors on a Helical
  Lattice\label{sec:helical-s-wave}}


In this section we first explain our setup, which is based on the model of \cite{Donos2013c}.
We then discuss and present our results for the finite temperature phase diagram. 


\subsection{Holographic Setup\label{ssec:holographic-setup}}

\begin{figure}[t]
  \centering
  \includegraphics[width=.8\textwidth]{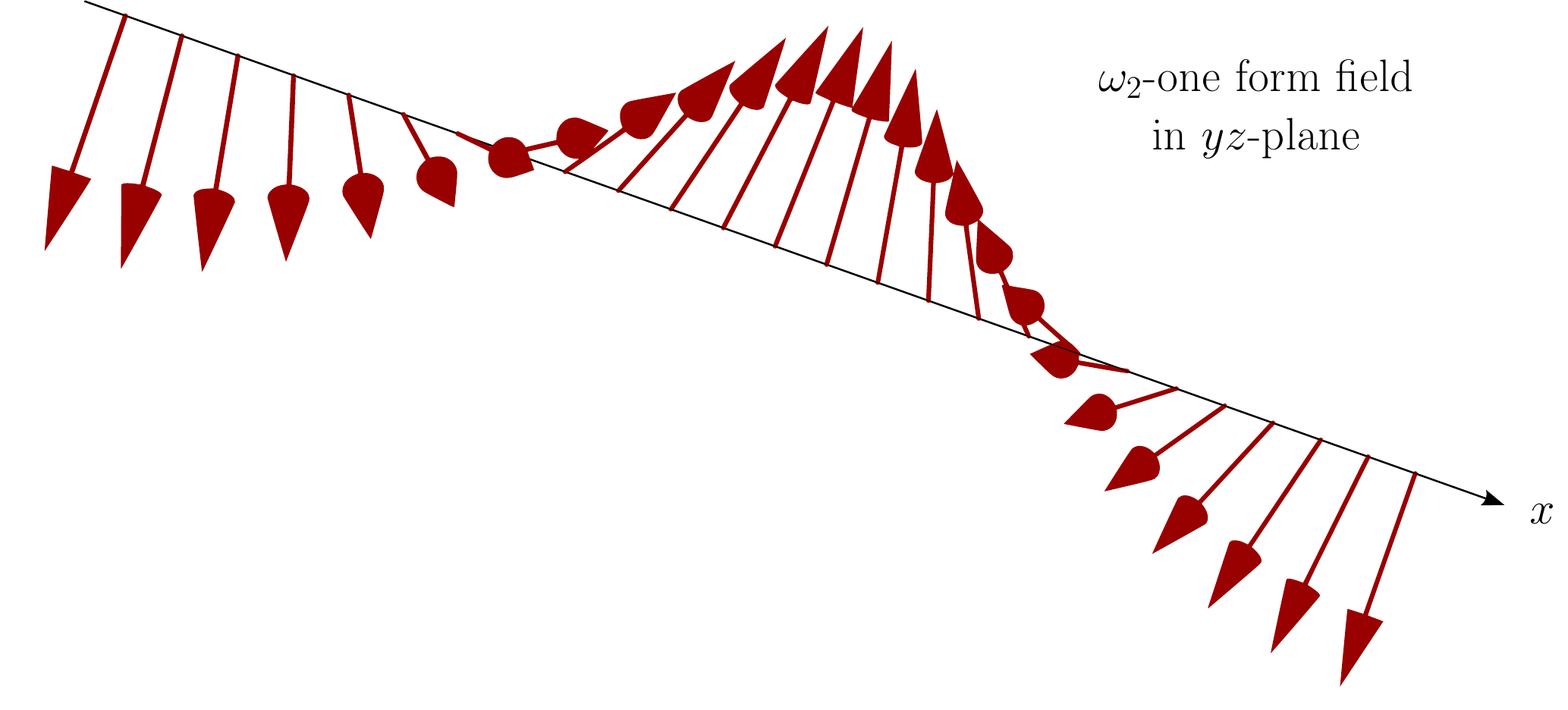}
  \caption{Plot of the one-form $\omega_2$ along the $x$-axis for one
    period. Being periodic with period $2 \pi/p$, $\omega_2$ is not
    translationally invariant for $p \neq 0$. The vector field $B = w(r)
    \omega_2$ acts as a source for the helix and imprints the helical,
    translational symmetry breaking structure on the system.
    \label{fig:plot-helix}}
\end{figure}

The holographic model that dualizes to a field theory in the presence
of a helical lattice has the action \cite{Donos2013c}
\begin{align}
  S_{\text{helix}}={}&\int\dd[4+1]x\sqrt{-g}\bigg[R+12-\frac14F^{\mu\nu}F_{\mu\nu}
  -\frac14W^{\mu\nu}W_{\mu\nu}-m^2B_{\mu}B^{\mu}\notag
  \bigg]
  \notag\\
  &-\frac\kappa2\int B\wedge F\wedge W.
  \label{eq:helix-action-old}
\end{align}
Here $g_{\mu\nu}$ is the metric of a 5-dimensional asymptotically anti-de-Sitter
spacetime including the $3+1$ field theory dimensions and the additional radial
coordinate $r$. $R$ is the Ricci scalar of this metric. There are two field strengths: 
$F_{\mu\nu}=\partial_\mu A_{\nu}-\partial_{\nu}A_{\mu}$ is the Maxwell field which
accounts for the $U(1)$ charge dynamics. The additional massive Proca
field $B_{\mu}$ generates the `helix U(1)' with field strength $W_{\mu\nu}=\partial_\mu B_{\nu}-\partial_{\nu}
B_{\mu}$, and supports the helical structure. In addition, there is a
Chern-Simons term which couples the fields $A_{\mu}$ and $B_{\mu}$ with
coupling constant $\kappa$. In the above action, the AdS radius $L$ has been set to
one. Furthermore, Newton's constant has been fixed to $\kappa_5^2 = 1/2$. This can be
achieved by redefining the remaining couplings such that $\nicefrac1{(2\kappa_5^2)}$
becomes a total factor multiplying the action.\\[\baselineskip]
To encode the $U(1)$ order parameter, we add to this action a scalar field with charge
$q$ and mass $m_{\rho}$ minimally coupled to $A_{\mu}$,
\begin{align}
  S_{\text{total}}=S_{\text{helix}}
  +&\int\dd[4+1]x\sqrt{-g}\bigg[-|\partial\rho-\ci qA\rho|^2-m_{\rho}^2|\rho|^2\bigg].
  \label{eq:helix-action}
\end{align}
The equations of motion following from the action \eqref{eq:helix-action} are
\begin{align}
  R_{\mu\nu}-\frac12Rg_{\mu\nu}-6g_{\mu\nu}=T^{(A)}_{\mu\nu}+T^{(B)}_{\mu \nu}+T^{(\rho)}_{\mu\nu},
  \label{eq:helix-Einstein-EOMS}
\end{align}
where
\begin{align}
  T^{(A)}_{\mu\nu}&=\frac12F_{\mu\alpha}F\indices{_\nu^\alpha}-\frac18g_{\mu\nu}F^2, \notag\\
  T^{(B)}_{\mu\nu}&=\frac12W_{\mu\alpha}W\indices{_\nu^\alpha}-\frac18g_{\mu\nu}W^2
  -\frac{m^2}2B_\mu B_\nu, \notag\\ 
  T^{(\rho)}_{\mu\nu}&=\re\big[(\nabla_{\mu}\rho^\ast+\ci qA_\mu\rho^\ast)
  (\nabla_{\nu}\rho-\ci qA_\nu \rho)\big]-\frac12g_{\mu \nu}
  \big(|\partial\rho-\ci qA\rho|^2+m_\rho^2|\rho|^2\big), 
  \label{eq:helix-energy-mom-matter}
\end{align}
are the energy-momentum tensors of the two vector fields $A$ and $B$, and of the complex
scalar $\rho$. Furthermore, we have the scalar equation
\begin{align}
  0&=\left[(\nabla^\mu -\ci qA^\mu)(\nabla_\mu-\ci qA_\mu)-m_\rho^2\right]\rho,
  \label{eq:helix-scalar-EOMS}
  \intertext{and the Maxwell equations}
  \nabla_{\mu}F^{\mu \nu}&=\ci q\left[\rho^\ast(\partial^\nu-\ci qA^\nu)\rho
    -\rho(\partial^\nu+\ci qA^\nu)\rho^\ast\right] 
  +\frac\kappa{4\sqrt{-g}}\tilde{\epsilon}^{\mu\nu\alpha\beta\gamma}\partial_{\alpha}
  (B_\mu W_{\beta\gamma}), \label{eq:Maxwell-EOM}\\ 
  \nabla_{\mu}W^{\mu\nu}&=m^2B^\nu+\frac{\kappa}{8\sqrt{-g}}
  \tilde{\epsilon}^{\mu\nu\alpha\beta\gamma}\left[2\partial_\gamma(B_\mu F_{\alpha\beta})
    -F_{\mu\alpha}W_{\beta\gamma}\right].
  \label{eq:helix-Maxwell-EOM}
\end{align}
Here $\tilde{\epsilon}^{\mu\nu\alpha\beta\gamma}$ is the totally antisymmetric
Levi-Civita symbol in 5 dimensions with $\tilde{\epsilon}^{01234}=1$. As in
\cite{Donos2013c}, the wedge product in the action \eqref{eq:helix-action} is normalized
such that the Chern-Simons term evaluated on the chosen Ansatz equals
$S_{\text{CS}}=\int dr\, p\kappa w^2a'/2$.\\[\baselineskip]
We now construct solutions to the equations that have the following properties. First we
aim to study the system with the helix structure in order to break translational
symmetry. For this purpose, the one-forms
\begin{align}
  \omega_1&=\dd x\eqcomma\notag\\
  \omega_2&=\cos(px)\dd y-\sin(px)\dd z,\notag\\
  \omega_3&=\sin(p x)\dd y+\cos(px)\dd z,
  \label{eq:helix-omega-forms}
\end{align}
are introduced. They provide a basis for the spatial  $(x,y,z)$ part of the
metric and the two vector fields $A_\mu$ and $B_\mu$. In
\figref{fig:plot-helix}, one period of $\omega_2$ is plotted along the
$x$-coordinate. The forms $\omega_2$ and $\omega_3$ have the structure of a
helix with periodicity $\nicefrac{2\pi}p$.
%
In the following, we focus on the case $m=0$, \ie we are considering a massless helix
field $B$.  In our setup, the role of $B$ is to introduce a lattice in a
phenomenological way and thus break translational symmetry. Since this can be achieved
with a massless helix field, $m=0$ is chosen for simplicity. This choice follows
\cite{Donos2013c}. 
Using these one-forms we make the Ansatz for the helix field $B=B_{\mu}dx^{\mu}$
to be
\begin{alignat}{2}
  B&=w(r)\omega_2, &\qquad\qquad w(\infty)&=\lambda,
  \label{eq:helix-ansatz-B}
\end{alignat}
where $r=\infty$ denotes the boundary of the asymptotically anti-de-Sitter space. Since
this Ansatz shows that $B_{y}$ and $B_z$ do not vanish at the boundary, the field theory
interpretation is that we explicitly introduce a source $\lambda$ for the operator dual
to $B$, \ie we are deforming the homogeneous theory by a lattice operator. $\lambda$
can be interpreted as the lattice strength. The field $B$ extends along $\omega_2$ and
therefore breaks translational symmetry in the $x$-direction for $p \neq 0$. Via
backreaction on the metric, this helical structure is imprinted on the whole
gravitational system. This is manifested in a metric Ansatz 
\cite{Iizuka:2012iv}
\begin{align}
  \dd s^2=-U(r)\dd t^2+\frac{\dd r^2}{U(r)}+\e[2 v_1(r)]\omega_1^2
  +\e[2 v_2(r)]\omega_2^2+\e[2 v_3(r)]\omega_3^2.
  \label{eq:helix-metric-ansatz}
\end{align}
From a technical point of view, the usefulness of this Ansatz is that it is compatible
with the so-called Bianchi $\text{VII}_0$ symmetry and is therefore guaranteed to be
self-consistent --- naïvely the spatial dependence in the one-forms should imply that
all the components of the metric become spatially dependent. Thanks to the symmetry this
is not so. Instead, all $x$-dependence is carried by the one-forms of
Eq.\@ \eqref{eq:helix-omega-forms} such that the resulting equations of motion are
ordinary differential equations in the radial coordinate $r$. It is therefore consistent
to assume that all fields are functions of $r$ only, even though translational symmetry
is broken.
\\[\baselineskip]
In the dual field theory the blackening factor $U$ encodes the energy density
and the $v_i$ are related to the pressures in the system. In the finite temperature phase, the metric
function $U$ has a zero at a finite value of $r$, which defines the thermal horizon
radius $r_h$,
\begin{align}
  U(r_h)=0.
 \label{eq:helix-U-zero-horizon}
\end{align}
We will consider solutions which, for large values of $r$, satisfy
\begin{alignat}{3}
  U(r) &= r^2+\ldots, &\qquad\qquad v_i(r)&=\ln(r)+\ldots, &\qquad\qquad\text{for } i&=1,2,3.
  \label{eq:helix-asymptotically-AdS-bc}
\end{alignat}
This guarantees that at the boundary (for $r\to\infty$) the metric is of anti-de
Sitter form, $\dd s^2=\dd r^2/r^2+r^2(-\dd t^2+\dd{\vvec x}^2)$. Field theoretically,
this means that the theory has an ultraviolet fixed point with
conformal symmetry. The introduction of the helix through the source for the operator
dual to $B_{\mu}$ deforms away from this UV fixed point, as we will see below, \cff
Eq.\@ \eqref{eq:helix-conformal-anomaly}.\\[\baselineskip]
To model a field theory at finite density, we make the additional Ansatz
\begin{alignat}{2}
  A&=a(r)\dd t, &\qquad\qquad a(\infty)&=\mu.
  \label{eq:helix-ansatz-A}
\end{alignat}
Field theoretically the operator dual to $A_{\mu}$ is a conserved current $j^{\mu}$. As
this Ansatz again does not vanish at the AdS boundary it means that we introduce a
source for the zero component of $j^\mu$. We work in the grand canonical ensemble with
chemical potential $\mu$ rather then a system at fixed density.
\\[\baselineskip]
We furthermore choose the charged scalar, which is the gravitational encoding of our
superconducting order parameter, to respect the Bianchi VII${}_0$ symmetry, \ie to be
invariant under the vector fields dual to the Bianchi VII${}_0$ one-forms in Eq.\@
\eqref{eq:helix-omega-forms},
\begin{alignat}{2}
  \omega_i^\mu\partial_\mu\rho(t,x,y,z,r)&=0, &\qquad i&=1,2,3.
  \label{eq:scalaransatz}
\end{alignat}
Since the Bianchi one-forms \eqref{eq:helix-omega-forms} are linearly
independent, this restricts the charged scalar to be at most of form
$\rho(t,r)$. For the background we choose a time-independent Ansatz
$\rho(r)$. Note that choosing a superconducting order parameter
compatible with the helix symmetries corresponds, in condensed matter
terminology, to the statement that the superconducting order parameter
must respect the crystal symmetry of the underlying lattice. The resulting coupled
ordinary differential equations for the functions $a$, $w$, $U$, $v_1$, $v_2$, $v_3$ and
$\rho$ can be found in Appendix\@ \ref{app:eq-helix-background}. Making use of the
$U(1)$ symmetry associated with $A_\mu$, $\rho$ is chosen to be real. The total
differential order of the equations of motion is 13, corresponding to the three second
order equations for $a,w$ and $\rho$, three second order Einstein equations and one
first order Einstein equation, the constraint equation.\\[\baselineskip]
It will be shown that this system exhibits a second order phase transition at finite
temperature: above some critical temperature $T_c$, the system is in a state with a
vanishing scalar field; below $T_c$ a phase with a non-vanishing scalar field is
thermodynamically preferred. The order parameter of the phase transition is the vacuum
expectation value $\Braket{\mathcal{O}_\rho}$ of the operator dual to $\rho$, and the
low temperature phase is characterized by breaking the global $U(1)$ under which
$\mathcal{O}_\rho$ is charged. Since we will only consider the case of
$\mathcal{O}_\rho$ not being sourced explicitly, the $U(1)$ symmetry is broken
spontaneously. The superconducting phase transition can be understood in terms of an
effective scalar mass becoming sufficiently negative: from the scalar field equation
\begin{align}
  0=\rho''+\rho'\left(\frac{U'}{U}+v_1'+v_2'+v_3'\right)
  +\rho\left(\frac{a^2q^2}{U^2}-\frac{m_{\rho }^2}{U}\right), 
\label{eq:helix-scalar-eq-effective-mass}
\end{align}
we can deduce the effective $r$-dependent scalar mass
$m_{\text{eff}}^2={m_\rho^2-q^2a^2}/U$. The second term shifts the effective
mass squared towards smaller values and causes an instability as $m_{\text{eff}}^2$
becomes sufficiently negative. Both larger values of $q$ and more negative values of
$m_\rho^2$ favor the appearance of the instability. Most studies on holographic
superconductors with a scalar condensate focus on tachyonic scalar masses corresponding
to relevant operators.\footnote{Tachyonic scalar masses are allowed in asymptotically
  anti-de Sitter spaces as long as they are above the Breitenlohner-Freedman
  bound \cite{Breitenlohner1982a} $m_\rho^2\geq-d^2/4$. For such fields the tendency
  towards an instability driven by the negative mass squared is stabilized by the
  curvature of the AdS space.} From studies on translationally invariant s-wave
superconductors it is however known that condensation to a superconducting state also
appears for a massless (and even irrelevant) scalar \cite{Horowitz2008}. As in the case
of translationally invariant s-wave superconductors studied in \cite{Horowitz2009}, it
is expected that the zero temperature infrared geometry depends on the mass of the
scalar field and that the massless scalar represents one of several distinct cases. In
the following, we focus on the case $m_\rho=0$ and leave the case of general scalar
masses to further studies. One reason for this choice will become clear in
Section \ref{sec:zero-temperature}: We have only been able to construct extremal (zero
temperature) geometries for this value of the scalar mass. The mass of the scalar field
$m_\rho$ is related to the scaling dimension $\Delta$ of the operator dual to $\rho$ via
$\Delta=2+\sqrt{4+m_\rho^2}$. Therefore, the case $m_\rho=0$ corresponds to a marginal
operator with $\Delta=4$.\\[\baselineskip]
The Chern-Simons term couples the Maxwell field $A_\mu$ and the helix field $B_\mu$. The
authors of \cite{Donos2013c} point out that in their case the Chern-Simons coupling
allows for the existence of a {cohesive phase}, \ie a phase without electric flux
through the horizon. This requires some field or coupling that sources the electric field
outside of the horizon. The {insulating geometry} of \cite{Donos2013c} is of this
type and it is stabilized by the Chern-Simons interaction, inducing the necessary charge
density via the coupling to the helix field. In particular, for larger values of the
Chern-Simons coupling $\kappa>0.57$ \cite{Donos2013c} the phase diagram simplifies due
to the absence of an under RG flow unstable bifurcation critical point. Due to this
added simplicity, and in order to be compatible with the results of \cite{Donos2013c},
we focus in this work on the case of relatively large Chern-Simons coupling and choose
$\kappa=\nicefrac1{\sqrt2}$. Furthermore, in order to be able to compare the physics of
our model better to the usual holographic superconductor
\cite{Hartnoll2008,Hartnoll2008a}, we also investigate the case of vanishing
Chern-Simons coupling $\kappa=0$.


\subsubsection{Asymptotic Expansions\label{sssec:helix-asymp-exp}}

As an important step towards solving the equations of motion, asymptotic expansions are
calculated at the thermal horizon $r_h$ and at the boundary, \ie for large values of
$r$. These are the two points where the boundary conditions on the fields are
imposed. In addition to its defining property as the zero of the metric function $U$,
the horizon $r_h$ is also a zero of the Maxwell potential $a$ due to regularity (\cff for
example \cite{Horowitz2011})
\begin{alignat}{4}
  && U(r_h)&= 0, & a(r_h)&=0. &&
  \label{eq:helix-bc-horizon}
  \intertext{The remaining fields are finite at the horizon. At the boundary $r \to
    \infty$, the conditions}
  a&=\mu, &\qquad\qquad w&=\lambda, &\qquad\qquad \rho&=0, &\qquad\qquad v_i&=\ln(r),
  \label{eq:helix-bc-bdy}
\end{alignat}
are imposed. The first three conditions determine the sources of the operators dual to
$a, w$ and $\rho$. For general scalar field masses $m_\rho$, the leading power of $\rho$
at the boundary is $r^{-(4-\Delta)}$, where $\Delta$ is the scaling dimension of
$\rho$. The massless scalar considered here has $\Delta=4$, such that it is constant to
leading order in $\nicefrac1r$. Therefore, the source of the operator dual to $\rho$ is chosen
to vanish by imposing $\rho = 0$ at the boundary. A solution with a non-vanishing scalar
field in the bulk then breaks the $U(1)$ symmetry associated with the Maxwell field
{spontaneously}. As discussed before, the conditions on $v_i$ ensure that the
metric is asymptotically anti-de Sitter. For an asymptotically AdS space it is also
required that $U(r) = r^2$ at the boundary. This condition, however, follows from the
equations of motion (in particular the constraint, the sixth line in
\eqref{eq:appendix-helix-background-eoms}) and does not need to be imposed explicitly. 
In order to determine the asymptotic horizon expansion respecting the conditions stated
above, we make the Ansatz 
\begin{equation}
  \begin{alignedat}{2}
    a&=a^h_1(r-r_h)+a^h_2(r-r_h)^2+\dotsb, &\qquad
    w&=w^h_0+w^h_1(r-r_h)+\dotsb, \\ 
    U&=U^h_1(r-r_h)+U^h_2(r-r_h)^2+\dotsb, &
    v_i&=v^h_{(i,0)}+v^h_{(i,1)}(r-r_h)+\dotsb, \\ 
    \rho&=\rho^h_0+\rho^h_1(r-r_h)+\dotsb,
  \end{alignedat}
  \label{eq:helix-asymp-exp-hor-ansatz}
\end{equation}
and solve the equations of motion order by order in $r-r_h$. The horizon expansion has
seven free parameters which are chosen to be 
\begin{align}
  (a^h_1,w^h_0,\rho^h_0,U^h_1,v^h_{(i,0)}).
  \label{eq:helix-indep-horizon-params}
\end{align}
All higher order expansion coefficients can be expressed in terms of these
parameters. $U^h_1$ is related to the Hawking temperature
by 
\begin{equation}
  T=\frac{U'(r_h)}{4\pi}=\frac{U^h_1}{4\pi}.
  \label{eq:TU1h}
\end{equation}
The asymptotic horizon expansion will be used to define initial conditions for the
numerical integration of the equations of motion.\\[\baselineskip]
At the boundary, a double expansion in $\nicefrac1r$ and in $\nicefrac{\ln(r)}r$ is
carried out. The $\ln(r)$-terms introduce a scale and indicate the presence of a scaling
anomaly, which is related to the lattice structure. This will be discussed in more
detail in Section \ref{sssec:helix-thermodyn}, where the stress-energy tensor of the
system is calculated. The following Ansatz is used as a building block for the
asymptotic expansion at the boundary
\begin{align}
  f_{N}(r)={}&\sum^{N}_{k = 0}\sum^{N-k}_{j=0} a_{k,j}
  \left(\frac1r\right)^k\left(\frac{\ln r}r\right)^j \notag\\ 
  ={}&a_{00}+\left(a_{10}\frac1r+a_{01}\frac{\ln r}r+\dotsb\right) \notag\\ 
  &+\left(a_{20}\frac1{r^2}+a_{11}\frac{\ln r}{r^2}+a_{02}\frac{(\ln r)^2}{r^2}
    +\dotsb\right)+\dotsb.
  \label{eq:helix-ansatz-bdy}
\end{align}
The constant $N$ defines the order of the expansion. Since both $\nicefrac1r$ and
$\nicefrac{\ln(r)}r$ are small for large values of $r$, higher powers in the expansion
are truly subleading. $f_N(r)$ is used to define an Ansatz for the matter fields $a,w$
and $\rho$. The Ansatz for $U$ is given by $r^2f_N(r)$, and the Ansatz for the metric
functions $v_i$ by $\ln(r)+f_N(r)$. This accounts for the fact that $r^2$ is the
leading behavior of $U$ and of $\e[2 v_i]$ for large values of $r$. Solving the
equations of motion order by order in $\nicefrac1r$, we obtain an asymptotic expansion
with the leading terms
\begin{equation}
  \begin{alignedat}{2}
    a&=\mu+\frac\nu{r^2}+\dotsb, &\qquad 
    w&=\lambda+\frac{\beta-p^2\lambda\,{\ln(r)}/2}{r^2}+\dotsb, \\ 
    U&=r^2-\frac{\epsilon/3+p^2\lambda^2\,{\ln(r)}/6}{r^2}+\dotsb, & 
    v_1&=\ln(r)+\frac{g_1+p^2\lambda^2\,{\ln(r)}/{24}}{r^4}+\dotsb, \\
    v_2&=\ln(r)+\frac{g_2-p^2\lambda^2\,{\ln(r)}/{12}}{r^4}+\dotsb, & 
    v_3&=\ln(r)+\frac{g_3+p^2\lambda^2\,{\ln(r)}/{24}}{r^4}+\dotsb, \\
    \rho&=\frac{\rho _b}{r^4}-\frac{q^2\mu^2\rho_b}{12r^6}+\dotsb. 
  \end{alignedat}
  \label{eq:helix-bdy-expansion}
\end{equation}
%
%
Here we find that $g_3$ satisfies 
\begin{equation}
  g_3=-g_1-g_2.
  \label{eq:g3g1g2}
\end{equation}
As we will see later, the parameters $g_i$ are related to the pressure of the system and
$\epsilon$ is the energy density. The subleading mode of $a$ denoted by $\nu$, is
related to the charge density and $\beta$, the subleading mode of $w$, to the vacuum
expectation value of the operator dual to $w$. The leading mode of $\rho$ has been set
to zero and the subleading mode, $\rho_b$, is proportional to the vacuum expectation
value $\Braket{\mathcal{O}_\rho}$. In total there are 8 physical parameters at the
boundary, namely
\begin{align}
  (\epsilon, g_1, g_2, \mu, \nu, \lambda, \beta, \rho_b).
  \label{eq:helix-bdry-params}
\end{align}
There is another, non-physical parameter present in the asymptotic boundary expansion,
which has been set to zero above to keep the expressions clear. This parameter is
related to a shift in the $r$ coordinate. It can be reinstated into the boundary
expansion by the transformation
\begin{align}
  r&\to r+\frac\alpha2, & g_1&\to g_1+\frac{\alpha^4}{64}, & g_2&\to g_2+\frac{\alpha^4}{64}. 
  \label{eq:helix-reinstate-alpha}
\end{align}
$\alpha$ gives rise to odd powers of $\nicefrac1r$ in the boundary expansion. These odd 
powers are in general present in the (numerical) solutions to the equations of
motion. However, in contrast to the remaining parameters of the asymptotic expansion,
$\alpha$ has no physical meaning. It can be thought of as an artifact of the coordinate
choice. Note that the Ansatz of Eq.\@ \eqref{eq:helix-metric-ansatz}, and therefore also
the equations of motion, do not explicitly depend on the radial coordinate $r$. As a
consequence, the system exhibits a shift-symmetry $r \to r + \text{const.} $ in the radial
direction. The presence of $\alpha$ in the boundary expansion reflects precisely
this shift-symmetry.


\subsubsection{Thermodynamics and the Conformal Anomaly\label{sssec:helix-thermodyn}}

The temperature of our strongly coupled field theory is given by the Hawking temperature
of the bulk black hole, Eq.\@ \eqref{eq:TU1h}.
%
%
%
%
The Bekenstein-Hawking entropy is calculated according to the area law as
\begin{align}
  S&=4\pi A_h=4\pi\int\dd\tau\dd[3]{\vvec x}\sqrt\gamma\bigg\vert_{r \to r_h}
  =4\pi\e[v_1(r_h)+v_2(r_h)+v_3(r_h)]V.
  \label{eq:helix-entropy}
\end{align}
Here $A_h$ denotes the area of the black hole horizon, $\gamma_{ab}$ is the induced
metric and $V=\int\dd[3]{\vvec x}$ denotes the $3$-dimensional volume of the field theory.
In order to determine the grand canonical potential, we calculate the Euclidean
(imaginary time) on-shell action. Applying appropriate integrations by part with respect
to $r$, the Euclidean action reduces to a boundary term upon use of the equations of
motion. We obtain the two versions
\begin{align}
  I_{\text{bulk}}&=\frac VT\left[-a\e[v_1+v_2+v_3]a'-\frac12a\kappa pw^2
    +\e[v_1+v_2+v_3]U'\right]^{r=r_b}_{r=r_h},\notag \\ 
  I_{\text{bulk}}&=\frac VT\left[\frac12U\e[v_1-v_2+v_3]ww'+U\e[v_1+v_2+v_3]v_2'
    +U\e[v_1+v_2+v_3]v_3'\right]^{r=r_b}_{r=r_h},
  \label{eq:bulk-on-shell}
\end{align}
which will give rise to two expressions for the grand canonical potential $\Omega$. A
regularizing ultraviolet cutoff $r_b\gg r_h$ has been introduced. Eventually, after
determining appropriate counterterms, the limit $r_b\to\infty$ will be taken. The
renormalized Euclidean on-shell action is given by 
$I_{\text{os}}=I_{\text{bulk}}+I_{\text{GH}}+I_{\text{ct}}$, where $I_{\text{GH}}$ is
the Gibbons-Hawking boundary term and $I_{\text{ct}}$ stands for the counterterms
necessary to make $I_{\text{os}}$ finite.\footnote{For a review of holographic
  renormalization, see \cite{Skenderis2002}.} The factor $\nicefrac1T=\beta$ originates
from the $\tau$-integration and $V=\int\dd[3]{\vvec{x}}$ is the three-dimensional
volume. Note that the second expression for $I_{\text{bulk}}$ receives no contribution
from the horizon since $U(r_h)=0$. The Gibbons-Hawking term is evaluated as
\begin{align}
  I_{\text{GH}}&=-2\int\dd\tau\dd[3]{\vvec x}\sqrt\gamma\;\nabla_{\mu}n^{\mu} \notag\\ 
  &=\frac VT\left[-\e[v_1+v_2+v_3]U'-2U\e[v_1+v_2+v_3]v_1'-2U\e[v_1+v_2+v_3]v_2'
    -2U\e[v_1+v_2+v_3]v_3'\right].
   \label{eq:helix-gibbons-hawking}
\end{align}
Here $\gamma_{ab}$ is the induced metric at $r=r_b$, and $n=\sqrt U\partial_r$ is the
outward pointing normal vector to the surface at $r=r_b$. The sum
$I_{\text{bulk}}+I_{\text{GH}}$ is divergent in the limit $r_b\to\infty$. Using the
asymptotic expansion of Section \ref{sssec:helix-asymp-exp}, the diverging terms can be
written as\footnote{For the sake of clarity, the shift parameter $\alpha$ is set to zero
  in this expression. It can be reinstated using the transformation of
  Eq. \eqref{eq:helix-reinstate-alpha}.}
\begin{equation}
  I_{\text{bulk}}+I_{\text{GH}}
  =\frac VT\left[-6r_b^4-p^2\lambda^2\ln\left(r_b\right)-\frac12p\kappa\mu\lambda^2
    +2\epsilon+2\mu\nu+O\left(\frac1{r_b}\right)\right].
 \label{eq:I-bulk-plus-I-GH}
\end{equation}
Here the first expression for $I_{\text{bulk}}$ of Eq.\@ \eqref{eq:bulk-on-shell} was
used. The divergent terms are, however, the same for both expressions. They only differ
in terms that are finite at the boundary. There are two types of divergences: the $-6
r_b^4$-term is due to the infinite volume of the asymptotically anti-de Sitter space and
can be canceled by a counterterm proportional to the volume
$\int\dd\tau\dd[3]{\vvec{x}}\sqrt\gamma$ of the surface at $r=r_b$. The second type of
divergence is logarithmic in $r_b$ and, in order to cancel it, a counterterm
proportional to $\ln(r_b)$ needs to be introduced. As shown later, this causes a
scaling anomaly. Roughly speaking, the logarithm present in the counterterm introduces a
scale and therefore breaks conformal symmetry. The total on-shell action
$I_{\text{os}}=I_{\text{bulk}}+I_{\text{GH}}+I_{\text{ct}}$ can be made finite by means
of the counterterm
\begin{equation}
  I_{\text{ct}}=\int\dd\tau\dd[3]{\vvec x}\sqrt\gamma
  \left[6-\frac14\ln(r_b)W^{ab}W_{ab}\right].
  \label{eq:helix-counterterm-action}
\end{equation}
In this expression, indices are contracted using the induced metric $\gamma_{ab}$. Since
the scalar field $\rho$ does not cause any new divergences, these counterterms are
identical to the one of \cite{Donos2013c}.\footnote{In the corresponding expression of
  \cite{Donos2013c}, an additional term proportional to
  $\sqrt\gamma\ln(r_b)F^{ab}F_{ab}$ is included. This term is present for general gauge
  fields $A_{\mu}$ but vanishes if $A_{\mu}$ has only a time component.}
\\[\baselineskip]
According to the AdS/CFT correspondence, the field theory partition function is related
to the on-shell action by $Z = \exp (-I_{\text{os}})$. Therefore, the grand canonical
potential can be expressed as $\Omega = T I_{\text{os}}$. Using the asymptotic boundary
expansion of Eq.\@ \eqref{eq:helix-bdy-expansion}, $\Omega$ can be written as
\begin{align}
  \frac\Omega V&=\epsilon+2\mu\nu-\frac12\kappa\lambda^2\mu p
  -\e[v_1\left(r_h\right)+v_2\left(r_h\right)+v_3\left(r_h\right)]U'\left(r_h\right),
  \notag\\
  \frac\Omega V&=4g_1+\frac{\alpha^4}{16}-\frac\epsilon3-\beta\lambda
  -\frac{\lambda^2p^2}8,
 \label{eq:helix-grand-canonical-pot}
\end{align}
corresponding to the two expressions for $I_{\text{bulk}}$ of
Eq.\@ \eqref{eq:bulk-on-shell}. The non-physical shift-parameter $\alpha$, which is
explicitly indicated here, can be removed by the redefinition
\eqref{eq:helix-reinstate-alpha}.
%
%
The first expression for $\Omega$ can be brought into the more
familiar form 
\begin{align}
  \frac\Omega V&=\epsilon-\mu n-Ts,
  \label{eq:helix-omega-familiar}
  \intertext{by rewriting the horizon contribution in terms of the temperature
    $T$ and the entropy $S$ and by defining the particle density as}
  n&=-2\nu+\frac{\kappa\lambda^2p}2,
  \label{eq:ndef}
\end{align}
where $s=\nicefrac SV$ denotes the entropy density, 
and $\epsilon$ is the energy density. This will be confirmed by an explicit computation
of the field theory stress-energy tensor. Since the scalar $\rho$ has a vanishing
source, there is no explicit scalar contribution to the grand canonical free
energy.\footnote{
  The grand canonical free energy does, however,
  implicitly depend on the scalar. When existent, a solution with a non-vanishing scalar
  has lower free energy than the normal phase solution, as the superfluid transition is
  second order in our model.} Therefore the above expression agrees with the one given in
\cite{Donos2013c}.\\[\baselineskip] 
The expectation value of the field theory stress-energy
tensor \cite{Balasubramanian1999,Myers1999} is calculated from the extrinsic curvature
$\theta_{ab}$ at $r=r_b$, its trace $\theta=\gamma^{ab}\theta_{ab}$, and the real-time
counterterm action $S_{\text{ct}}$ as
\begin{equation}
  \Braket{T_{ab}}=\lim_{r_b\to\infty}2r_b^2\left(-\theta_{ab}-\gamma_{ab}\theta
    -\frac2{\sqrt{-\gamma}}\frac{\delta S_{\text{ct}}}{\delta\gamma^{ab}}\right).
  \label{eq:helix-stress-energy}
\end{equation}
Using the asymptotic boundary expansion, the resulting expression for $\Braket{T_{ab}}$
can be expressed in terms of boundary parameters as
\begin{equation}
  \Braket T= 
  \begin{pmatrix}
    \epsilon & 0 & 0 & 0 \\
    0 &  8g_1-\frac14\lambda^2p^2+\frac\epsilon3 & 0 & 0 \\
    0 & 0 & T_{yy} & T_{yz} \\
    0 & 0 & T_{yz} & T_{zz} \\
  \end{pmatrix},
  \label{eq:helix-stress-energy-in-modes}
\end{equation}
with
\begin{align}
  T_{yy}&=\frac{\epsilon}3-4g_1-\frac18\lambda^2p^2
  +\left(4g_1+8g_2+\frac18\lambda^2p^2\right)\cos(2px), \notag\\
  T_{yz}&=-\left(4g_1+8g_2+\frac18\lambda^2p^2\right)\sin(2px), \notag\\
  T_{zz}&=\frac{\epsilon}3-4g_1-\frac18\lambda^2p^2
  -\left(4g_1+8g_2+\frac18\lambda^2p^2\right)\cos(2px).
   \label{eq:helix-stress-energy-in-modes-rest}
\end{align}
For $p\neq0$, the stress-energy tensor is spatially modulated and anisotropic. Indeed,
$\epsilon = T_{00}$ is the energy density, and the parameters $g_i$ are related to the
pressure terms $T_{11}$, $T_{22}$, and $T_{33}$. The trace of the stress-energy tensor
$\Braket{T\indices{^a_a}}=\eta^{ab}\Braket{T_{ab}}$ determines the conformal anomaly. It
is given by
\begin{equation}
  \Braket{T\indices{^a_a}}=-\frac12\lambda^2p^2.
  \label{eq:helix-conformal-anomaly}
\end{equation}
In the presence of the lattice, \ie for $p\neq0$ and $\lambda\neq0$, the conformal
symmetry of the ultraviolet fixed point is broken by an anomaly. The fact
that $\Braket{T\indices{^a_a}}$ is proportional to $p^2$ has the intuitive
interpretation that the lattice momentum $p$ introduces a scale and therefore breaks
conformal symmetry.\footnote{Note that the $T\indices{^\mu_\mu}$ transforms under the
  scaling transformation \eqref{eq:helix-scaling-symmetries} with weight 4, as it
  should.} In addition to the temperature $T$ and the sources $\mu$ and $\lambda$, the
system is characterized by the dimensionful scale $p$. Out of these four dimensionful
parameters, we can form the dimensionless ratios $\nicefrac T\mu$, $\nicefrac\lambda\mu$
and $\nicefrac p\mu$. These ratios will be used to parameterize the system (in addition to
the scalar charge $q$, which is dimensionless). Note however that $p$ itself is a
parameter in the state of the boundary field theory, but not a source switched on and
hence, as explained in detail in Appendix \ref{app:helix-numerical-method}, should not
be counted as an independent UV integration constant.


\subsection{Phase Transition with a Scalar Order  Parameter
  \label{ssec:helix-finite-t-phase-trans}}

\begin{figure}[t]
  \centering
  \includegraphics[width=\textwidth]{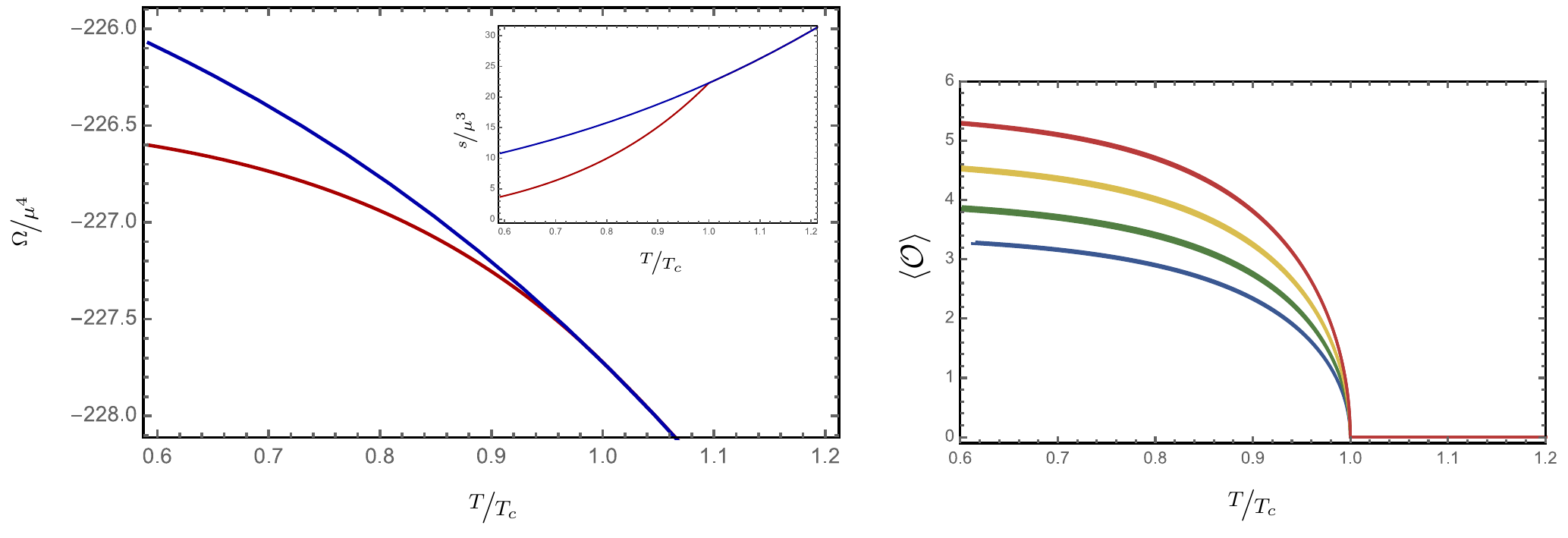}
  \caption{In the left panel, the free energy for solutions with and without a scalar
    condensate are shown. The {\color{darkblue}blue} curve corresponds
    to solutions without and the {\color{darkred}red} curve with a scalar condensate,
    respectively. Having a lower value of the grand canonical potential $\Omega$, the
    solution with a scalar condensate is thermodynamically preferred. The phase
    transition is second order since the derivative of the entropy is discontinuous at
    $T=T_c$, see inset. In both graphs $\kappa=\nicefrac1{\sqrt2}$, $q=5$, 
    $\nicefrac p\mu=5.2$, and $\nicefrac\lambda\mu=3.3$. In the right panel, the scalar
    condensate order parameter $\expect{\mathcal O}$ is shown for $\nicefrac p\mu=4.8$,
    $5.0$, $5.2$ and $\nicefrac\lambda\mu=\color{DarkRainbow0}3.0$,
    $\color{DarkRainbow033}3.3$, $\color{DarkRainbow066}3.6$,
    $\color{DarkRainbow099}3.9$. The graph shows that the strength of the order
    parameter is affected much stronger by changes in $\nicefrac\lambda\mu$ than by
    changes in $\nicefrac p\mu$: different $\lambda$ leads to each of the four curves
    shown, while the different $\nicefrac p\mu$ lead to nearly the same curves for each
    value of $\nicefrac\lambda\mu$.\label{fig:helix-finite-t-phase-trans-thermodyn}
  }
\end{figure}

\begin{figure}[t]
  \includegraphics[width=\textwidth]{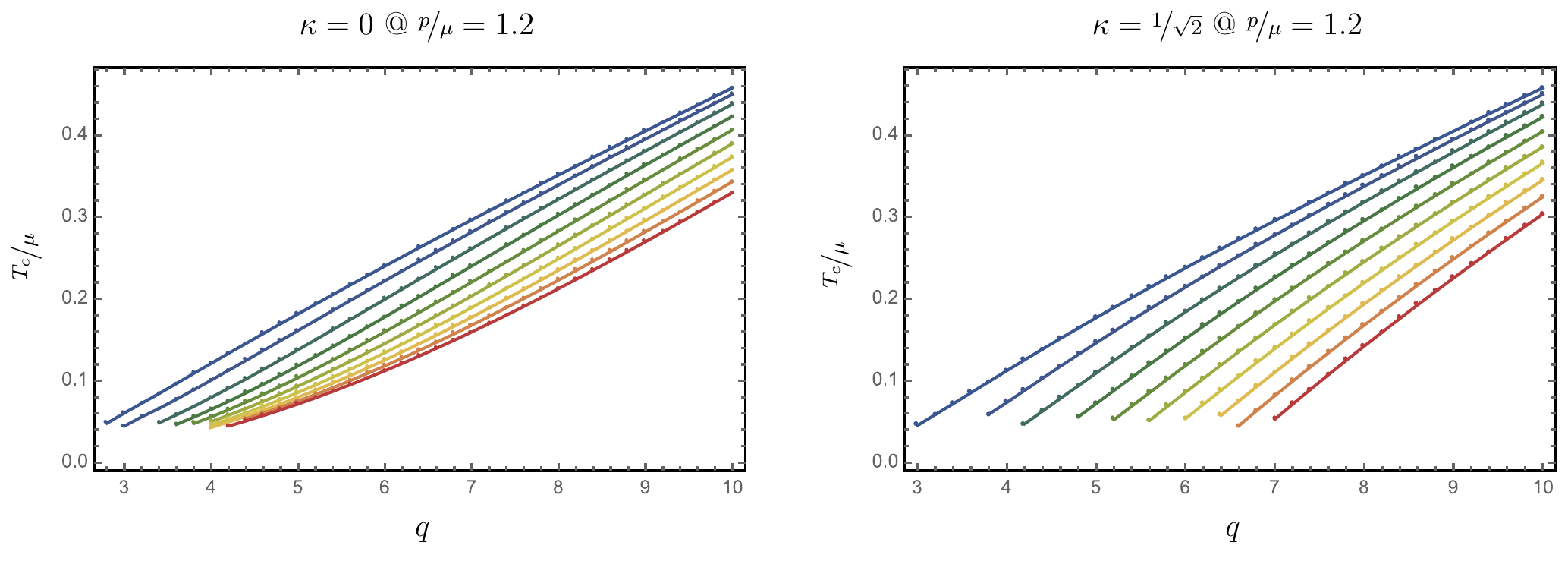}
  \caption{The superfluid phase diagram as a function of the scalar charge $q$ close to
    the minimal $T_c$-value in the phase diagram, \textit{i.e.}\@ $\nicefrac p\mu=1.2$ for
    $\kappa=0$ (left figure) and $\kappa=\nicefrac1{\sqrt2}$ (right figure). As one can
    see the critical temperature is decreasing with decreasing scalar charge $q$ or
    increasing backreaction $\sim\nicefrac1q$, respectively. This result agrees nicely
    with the plain s-wave superconductor behavior. Moreover for
    $\kappa=\nicefrac1{\sqrt2}$ and for sufficiently high values
    of $\nicefrac\lambda\mu$ (color coding of the lines represent the following values
    of $\nicefrac\lambda\mu=\color{DarkRainbow0}1$, $\color{DarkRainbow01}2$,
    $\color{DarkRainbow02}3$, $\color{DarkRainbow03}4$, $\color{DarkRainbow04}5$,
    $\color{DarkRainbow05}6$, $\color{DarkRainbow06}7$, $\color{DarkRainbow07}8$,
    $\color{DarkRainbow08}9$, $\color{DarkRainbow09}10$) we expect a quantum
    critical point, where the superconducting phase transition happens at zero
    temperature. On the other hand, for $\kappa=0$ it seems that the curves are
    asymptotically approaching zero charge as $T_c=0$, and thus there may be no quantum
    critical point in the $q$ direction.
    \label{fig:helix-finite-t-phase-trans-tc-as-function-of-q}}
\end{figure}

\begin{figure}
  \includegraphics[width=\textwidth]{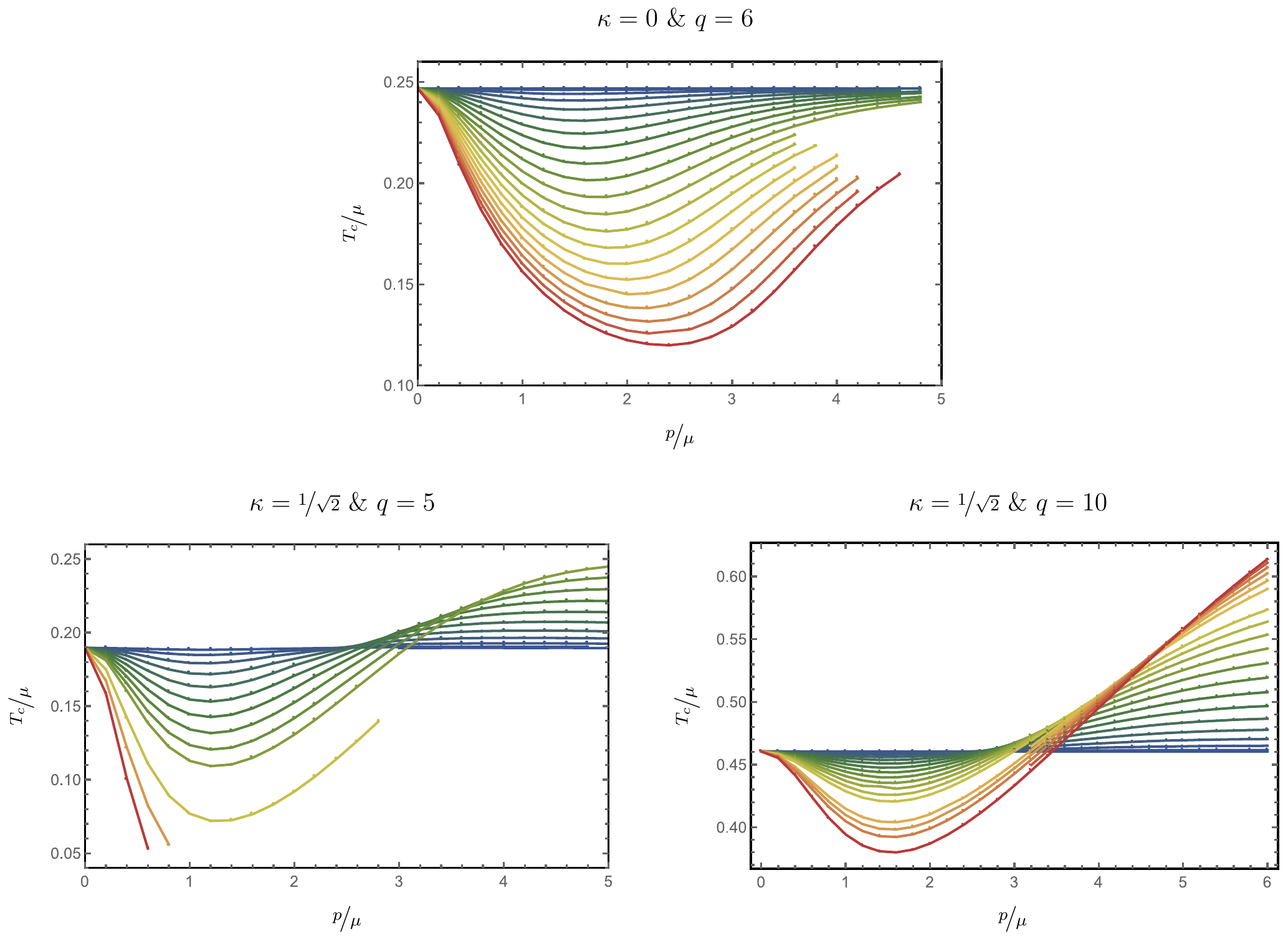}
  \caption{The superfluid phase diagram computed for $\kappa=0$ and scalar charge (or
    inverse backreaction) $q=6$ in the top panel and for $\kappa=\nicefrac1{\sqrt2}$ and
    $q=5,10$ in the lower panel. Below the respective $T_c$-curve, the condensate is
    non-zero indicating a superfluid phase. Different colored curves correspond to
    different values of $\nicefrac\lambda\mu$ with the following color coding:
    $\color{DarkRainbow0}0$, $\color{DarkRainbow0p045}0.3$,
    $\color{DarkRainbow0p09}0.6$, $\color{DarkRainbow0p135}0.9$,
    $\color{DarkRainbow0p18}1.2$, $\color{DarkRainbow0p225}1.5$,
    $\color{DarkRainbow0p27}1.8$, $\color{DarkRainbow0p315}2.1$,
    $\color{DarkRainbow0p36}2.4$, $\color{DarkRainbow0p405}2.7$,
    $\color{DarkRainbow0p45}3$, $\color{DarkRainbow0p495}3.3$,
    $\color{DarkRainbow0p54}3.6$, $\color{DarkRainbow0p585}3.9$,
    $\color{DarkRainbow0p63}4.2$, $\color{DarkRainbow0p675}4.5$,
    $\color{DarkRainbow0p72}4.8$, $\color{DarkRainbow0p765}5.1$,
    $\color{DarkRainbow0p81}5.4$, $\color{DarkRainbow0p855}5.7$,
    $\color{DarkRainbow0p9}6$. Note that the case of $\kappa=0$ the minimum of the
    critical temperature is moving from $\nicefrac p\mu=1.8$ at
    $\nicefrac\lambda\mu=1.2$ over $\nicefrac p\mu=2$ at $\nicefrac\lambda\mu=3$ to
    $\nicefrac p\mu=2.4$ at $\nicefrac\lambda\mu=6$, whereas in the case of
    $\kappa=\nicefrac1{\sqrt2}$ the minimum is fixed respect to $\lambda$. However
    for different values of the scalar charge $q=4,\dotsc,10$ we find a slight increase
    starting from $\nicefrac p\mu=1.2$ and ending at $\nicefrac p\mu=1.6$.
    \label{fig:helix-finite-t-phase-trans-phase-diag}}
\end{figure}

Numerically, we find that above some critical temperature $T_c$, there is only a
solution to the equations of motion with a vanishing scalar field.\footnote{The
  numerical method used is a simple shooting method adjusting the free parameters of
  the horizon expansion in such a way that we arrive for certain fixed values on the
  boundary. For a detailed exposition, see Appendix \ref{app:helix-numerical-method}.}
For $T<T_c$, a second branch of black hole solutions arises which has a non-vanishing
scalar field and therefore a non-vanishing vacuum expectation value
$\Braket{\mathcal{O}_\rho}$. Since the source of $\mathcal{O}_\rho$ is set to zero, the
condensate $\Braket{\mathcal{O}_\rho}$ breaks the $U(1)$ symmetry associated with the
Maxwell field $A_\mu$ spontaneously. By comparing the free energy in the grand canonical
ensemble, we find that the solution with a scalar condensate is thermodynamically
preferred, \cff \figref{fig:helix-finite-t-phase-trans-thermodyn}.
%
The phase transition is second order, since the derivative of the entropy is
discontinuous at the transition temperature. The order parameter of the phase transition
is the vacuum expectation value $\Braket{\mathcal{O}_\rho}$, which is proportional to
the boundary mode $\rho_b$ of $\rho$. It is plotted as a function of the temperature in
\figref{fig:helix-finite-t-phase-trans-thermodyn}. The numerical data are consistent
with a mean field behavior of the order parameter near the transition temperature,
$\Braket{\mathcal{O}_\rho}\sim\sqrt{1-T/T_c}$. This is expected in the large-$N$ limit,
which is intrinsic to our holographic model. For each set of parameters
($\nicefrac p\mu$,$\nicefrac\lambda\mu$,$q$), the transition temperature
$\nicefrac{T_c}\mu$ can be determined as the temperature where the order parameter
$\Braket{\mathcal{O}_\rho}$ vanishes. In this way, the phase diagram of the strongly
coupled field theory can be studied numerically. The transition temperature increases as
a function of $q$ as is shown in \figref{fig:helix-finite-t-phase-trans-tc-as-function-of-q}.
%
This can be understood in terms of the effective scalar mass
\begin{equation}
  m_{\text{eff}}^2=-\frac{q^2a^2}U. 
  \label{eq:effmass}
\end{equation}
As the scalar charge increases, the effective mass squared becomes more negative which
favors the instability. Therefore, the phase transition already appears at higher
temperatures. The same behavior was found for translationally invariant holographic
s-wave superconductors in \cite{Hartnoll2008a}.\\[\baselineskip]
Analyzing the transition temperature as a function of $\nicefrac p\mu$ and
$\nicefrac\lambda\mu$ for different $q$ reveals a more interesting structure as shown
for $q=5$ and $q=10$ in \figref{fig:helix-finite-t-phase-trans-phase-diag}.
%
In the case of $\kappa=0$ the critical temperature is observed to decrease monotonously
as a function of $\nicefrac\lambda\mu$. As a function of $\nicefrac p\mu$, it first
decreases for small $\nicefrac p\mu$, then assumes a minimum at $\nicefrac p\mu\approx1.7$, which
is slightly shifted towards larger values \ie $\nicefrac p\mu\approx2.3$ for increasing
$\nicefrac\lambda\mu\leq10$, and then returns again to the homogeneous value
$T_c(\nicefrac p\mu=0)$ for large values of $\nicefrac p\mu$. This minimum is more
pronounced for larger values of $\nicefrac\lambda\mu$ which shows that larger values for
the source of the helix field indeed increases the effect of the lattice on the system,
while the helix momentum dependence has a smaller effect on the transition
temperature. This is consistent with the expectation that generally, it is the depth of
a lattice of potential valleys which influences the physical behaviour more than the
lattice constant or spacing between the individual potential depths. However, for large
values of $\nicefrac p\mu$ the critical temperature seems to, at least for $\kappa=0$,
asymptotically approach the $p=0$ value which might imply that
$T_c(p\to\infty)\sim T_c(p=0)$.\\[\baselineskip]
In the case of $\kappa=\nicefrac1{\sqrt2}$ there is not only a minimal value of $T_c$
for $\nicefrac p\mu\approx1.2$ that is robust under changes in $\nicefrac\lambda\mu$,
but allows for higher values of $T_c$ above $\nicefrac p\mu\approx2.6,\dotsc,3$, as
displayed in the lower row of
\figref{fig:helix-finite-t-phase-trans-phase-diag}. Curiously, there the behavior of 
$\nicefrac\lambda\mu$ is inverted, \ie with increasing $\nicefrac\lambda\mu$ the
transition temperature, $T_c$ is increasing. However, this regime might very well lie
far outside of the range of applicability of condensed matter physics since $p>\mu$,
suggesting that the ``energy stored'' in the lattice exceeds the chemical potential. It
appears that the critical temperature is unbounded from above unlike in the $\kappa=0$
case where it seems to be bounded by its initial value at $p=0$. 
\\[\baselineskip]
Finally, note that in both cases $\kappa=0$ and $\kappa=\nicefrac1{\sqrt2}$ the data
suggests the existence of a quantum critical point for high values of
$\nicefrac\lambda\mu\geq6$ and strong backreaction, \ie $q\leq5$, similar to the
observations made in \cite{Kim:2015dna}. One may hence speculate that for a finite range
of $q$ there exists a critical value of $\nicefrac\lambda\mu$ at zero temperature where
the superfluid phase breaks down above a certain $\nicefrac p\mu$. We will analyse this
possibility further in future work \cite{WIP}.  

\section{Optical Conductivity\label{sec:helix-optical-conductivity}}

A main focus of our work is to study the optical conductivity in the $x$-direction, in
which translation symmetry is broken for $p \neq 0$. The conductivity is given
by the Kubo formula in terms of the retarded Green function of the current
operator,
\begin{equation}
  \sigma_x(\omega)=\lim_{\vvec k\to0}\frac{G^R_{xx}(\omega,\vvec k)}{\ci\omega}.
  \label{eq:helix-kubo-form}
\end{equation}
%


\subsection{Numerical computation of the optical conductivity
  \label{ssec:numerical-conductivity}}

We calculate the retarded Green function using the well-established extension of the
gauge/\-gravity correspondence to real-time problems pioneered in
\cite{Son2002}. Accordingly, the conductivity is determined by linearized perturbations
around the background solutions. In a physical picture, the system is slightly perturbed
around thermal equilibrium by an external force, the external electric field. The
conductivity describes the induced response of the system to these small
perturbations. Since the gauge field $A_{\mu}$ corresponds to a conserved current on the
field theory side, the electric conductivity is related to the perturbations of
$A_\mu$. In particular, the fluctuation $\delta A=\mathcal A\dd x$ governs the
conductivity in the $x$-direction. There are certain fields coupling to $\delta
A=\mathcal A\dd x$. To determine them, we write all fields including the metric as a sum
of the background solution and a perturbation. Then, the action of
\eqref{eq:helix-action} is expanded to second order in the perturbations. In this 
way, a quadratic action $S_{q}$ for the perturbations is obtained, which determines the
linearized equations of motion for the perturbations. By analyzing $S_q$, all fields
coupling to $\delta A=\mathcal A\dd x$ are determined.\footnote{The full set of the most
  general couplings at vanishing momentum is shown in the Appendix
  \ref{ssec:appendix-helix-fluct}, \tabref{tab:appendix-helix-matrix-of-couplings}.} The
block of coupled perturbations containing $\delta A=\mathcal A\dd x$ is
\begin{equation}
  \begin{aligned}
    \delta\left(\dd s^2\right)
    &=h_{t1}(t,r)\dd t\otimes\omega_1+h_{23}(t,r)\omega_2\otimes\omega_3 
    +h_{r1}(t,r)\dd r\otimes\omega_1, \\ 
    \delta A&=\mathcal A(t,r)\omega_1, \\
    \delta B&=\mathcal B(t,r)\omega_3.
  \end{aligned}
  \label{eq:helix-block-of-perturbations}
\end{equation}
The remaining perturbations can be set to zero, consistently. The fluctuation fields are
chosen to depend on $r$ and $t$ only since the conductivity is evaluated in the limit of
vanishing spatial momentum for the  Kubo formula \eqref{eq:helix-kubo-form}. The
equations of motion for the above perturbations are obtained by varying the quadratic
action.
After variation, we impose a \emph{radial gauge} in which $h_{r1} \equiv 0$. In this
gauge, the equation for $h_{r1}$ becomes a constraint for the remaining
fields. Furthermore, the equations become ordinary differential equations in the radial
coordinate after Fourier transforming the time coordinate. In total, we obtain one first
order equation (the constraint originating from the $h_{r1}$-equation after choosing
radial gauge), and four second order equations for $\mathcal A, \mathcal B, h_{t1}$ and
$h_{23}$. One of the second order equations can be replaced by the constraint, hence the
total differential order of the system is $7=1+2\times3$. The equation of motion are
given in Appendix \ref{ssec:appendix-helix-fluct}. The asymptotic expansions of the
fluctuation fields, which are necessary for obtaining numerical solutions, are discussed
in Appendix \ref{app:appendix-helix-fluct-asymp-exp}. As usual we implement
\emph{infalling} wave boundary conditions at the thermal horizon $r_h$. For
$\mathcal{A}$, we find that 
\begin{equation}
  \mathcal A=(r-r_h)^{\pm\nicefrac{\ci\omega}{(4\pi T)}}\left(A^h_0+A^h_1(r-r_h)+\dotsb\right).
  \label{eq:helix-fluct-near-horizon-example-a}
\end{equation}
The exponent in the prefactor can assume the values $\pm\nicefrac{\ci\omega}{(4\pi T)}$,
which correspond to outgoing and infalling waves, respectively. This can be seen
by taking into account the phase factor $\e[-\ci\omega t]$ of the Fourier transform,
 \begin{alignat}{3}
  \e[-\ci\omega t](r-r_h)^{\pm\nicefrac{\ci\omega}{(4\pi T)}}
  &=\e[\ci\omega(\pm\bar r-t)] &\quad &\text{with} &\quad 
  \bar r&=\frac{\ln(r-r_h)}{(4\pi T)}.
  \label{eq:helix-wave-near-hor}
\end{alignat}
The \emph{infalling} solution, \ie the one with the minus sign, should be used in order
to obtain the \emph{retarded} Green function. The leading behavior of $\mathcal A$ near
the boundary is found to be
\begin{equation}
  \mathcal A=A^b_0+\frac{A^b_2+A^b_0\omega^2\ln(r)/2}{r^2}+\dotsb.
  \label{eq:helix-asymp-exp-fluct-only-a}
\end{equation}
For calculating the conductivity, we have to identify the degrees of freedom coupling to
$\mathcal A$ by imposing gauge invariance. Even after imposing radial gauge, in which
all radial fluctuations vanish, there are still residual gauge transformations
left. These consist of the diffeomorphisms and $U(1)$ transformations that do not change
the radial gauge $h_{r\mu}\equiv A_r\equiv B_r\equiv0$. The residual gauge
transformations are worked out in detail in
Appendix \ref{ssec:appendix-helix-residual-gauge-tr}, following a calculation carried
out in \cite{Erdmenger2012} in the framework of the holographic $p$-wave system. We find
that the relevant physical fields are (i) $\mathcal A$, which is already gauge
invariant, (ii) $h_{23}$, which is gauge invariant at the boundary $r \to \infty$, and
(iii) the linear combination
\begin{equation}
  \mathcal G =-\ci\omega\mathcal B+wp\e[-2v_1]h_{t1}.
 \label{eq:helix-fluct-gauge-inv-linear-comb}
\end{equation}
The field $h_{t1}$ is not gauge invariant and does therefore not carry physical
degrees of freedom. In order to calculate the Green function corresponding to 
$\mathcal A$, we impose the condition that the remaining physical fields, $h_{23}$ and
$\mathcal G$, have no source term, \ie that their leading modes for $r\to\infty$
vanish.\footnote{Alternatively, we can make use of a method devised
  for treating holographic operator mixing, 
as explained in Appendix \ref{app:helix-renorm-and-op-mixing}.} In
this case the renormalized on-shell action for the fluctuations
(\cff Appendix \ref{app:helix-renorm-and-op-mixing}), expressed in terms of the
asymptotic modes of $\mathcal{A}$, is\footnote{Only the boundary contribution is
  indicated. According to the prescription of \cite{Son2002}, the horizon contribution
  is to be discarded.}
\begin{equation}
  S_{\text{on-shell}}=\int\frac{\dd\omega}{2\pi}\dd[3]{\vvec x}A^b_0(-\omega)
  \left(\frac{A^b_2(\omega)}{A^b_0(\omega)}-\frac{\omega^2}4\right)A^b_0(\omega).
  \label{eq:helix-fluct-on-shell-only-a}
\end{equation}
In this expression, the frequency dependence of the modes is indicated
explicitly. The Green function does not follow directly from
\eqref{eq:helix-fluct-on-shell-only-a}. Following the prescription of \cite{Son2002},
one needs to analytically continue the kernel in
\eqref{eq:helix-fluct-on-shell-only-a}. It follows that
\begin{align}
  \label{eq:helix-res-retarded-gf}
  G^R_{xx}(\omega,0)&=2\left(\frac{A^b_2(\omega)}{A^b_0(\omega)}-\frac{\omega^2}4\right),
  \intertext{and, using the Kubo formula \eqref{eq:helix-kubo-form},}
  \sigma(\omega)&=\frac{2A^b_2(\omega)}{\ci\omega A^b_0(\omega)}
  +\frac{\ci\omega}2.
  \label{eq:helix-res-cond-in-modes}
\end{align}
The numerical steps in calculating the conductivity for a given solution to the
background equations of motion are described in Appendix \ref{app:helix-numerical-method}. 


\subsection{Comparison to the Drude-model and the Two-Fluid-model \label{ssec:drude-two-fluid}}

For holographic metallic systems in homogeneous translation invariant 
backgrounds, one finds an ideal metallic behavior related to the conservation of
momentum. Thus, strictly speaking the Drude model is not applicable. In the presence of
a lattice, momentum is not a conserved quantity and therefore charge carriers can
dissipate their momentum within a typical timescale $\tau$ by interactions with the
lattice. According to the Drude model (\cff for example~\cite{Erdmenger2012a} for a
review and more details),
\begin{alignat}{2}
  \sigma(\omega)&=\frac{\sigma_{\text{DC}}}{1-\ci\omega\tau}, &\qquad
  \re\sigma(\omega)&=\frac{\sigma_{\text{DC}}}{1+\left(\omega\tau\right)^2},
  \label{eq:drude-model}
\end{alignat}
the dissipation time scale $\tau$ is inversely proportional to the width of
$\re(\sigma)$ near $\omega=0$, and the Drude peak can be seen as a direct consequence of
the translation symmetry breaking lattice. The same reasoning carries over to
holographic superconductors. However, in the absence of a momentum dissipating mechanism
such as a lattice, the holographic system describes an ideal metal in the normal phase
and a mixture of a holographic superconductor and a remaining ideal metal in the
condensed phase. In the limit of restored translational symmetry, the Drude peak
degenerates into a delta peak at $\omega=0$
%
\begin{figure}[t]
  \includegraphics[width=\textwidth]{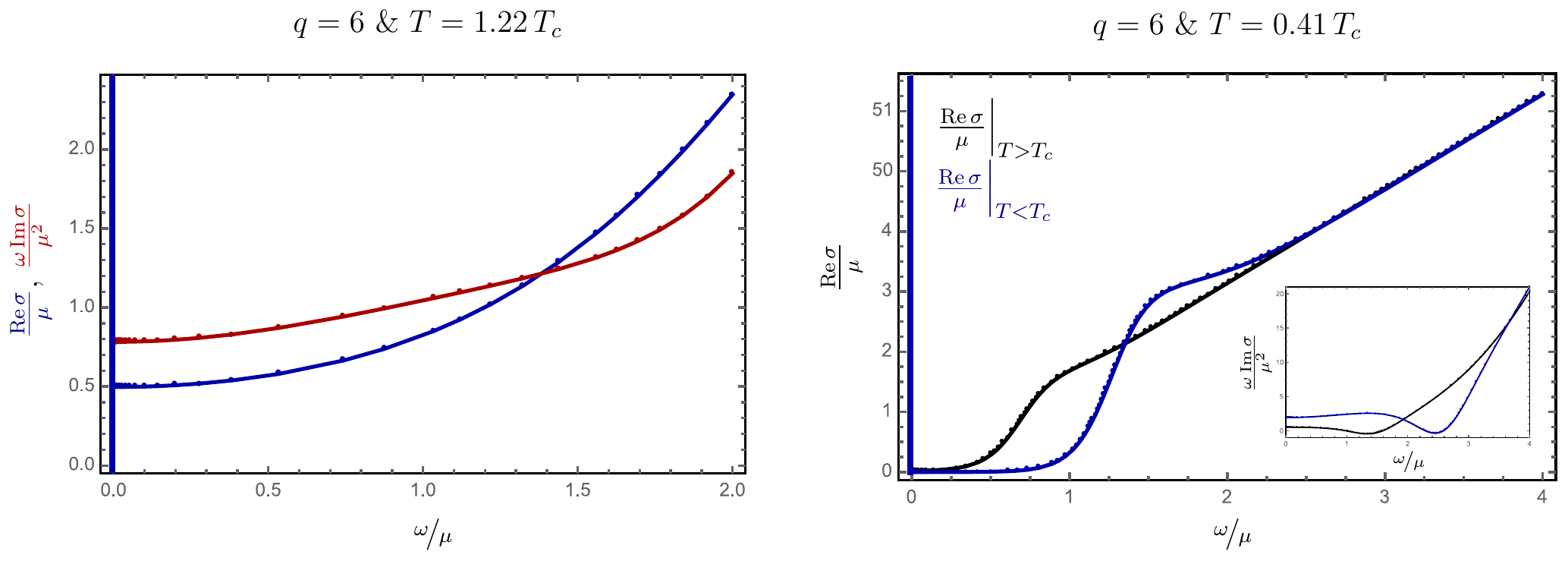}
  \caption{The left panel displays the optical conductivity in the translationally
    invariant system, \textit{i.e.}\@ $\nicefrac p\mu=0$, or $\nicefrac\lambda\mu=0$,
    respectively. The curves correspond to $q=6$ and $T=1.22\,T_c$. The $\nicefrac1\omega$
    pole indicating an infinite DC conductivity is also present in the normal phase due
    to momentum conservation. Accordingly, there is no momentum relaxation possible and
    hence there is no Drude peak at small frequencies. In the superconducting phase,
    shown in the right panel by the {\color{darkblue}blue} lines, the delta peak is
    enhanced by the superconducting degrees of freedom and the system is a mixture of an
    ideal metal and a superconductor for finite temperatures $T=0.41\,T_c$, as shown in
    the inset. Comparing the $\omega\to0$ value of the {\color{darkred}red} line in the
    left panel $\left.{\omega\im\sigma}/{\mu^2}\right|_{\omega=0}\approx0.8$ to the
    black in the inset in the right panel
    $\left.{\omega\im\sigma}/{\mu^2}\right|_{\omega=0}\approx0.5$, the ``strength'' of
    the ideal metal is reduced but it remains present, whereas the superfluid strength
    obtains a value of approximately $1.9$.\label{fig:helix-conductivity-p-zero}}
\end{figure}
%
In our helical setup, translational symmetry can be restored by setting $p=0$ and/or
$\lambda=0$.\footnote{Setting $p=0$ restores translational symmetry but the system is
  placed in an external magnetic field. In order to fully restore the plain
  holographic s-wave superconductor, the helix field needs to vanish, \ie
  $\lambda=0$. The magnetic field however points in the direction of the helix director
  (the x-direction), and hence does not lead to a gap in the conductivities considered
  in this paper.} In this case,  the helix field $w$ decouples from the system and we
obtain the classical holographic model of an s-wave superconductor as introduced in
\cite{Hartnoll2008,Hartnoll2008a}.
We can understand the translationally invariant case as a limit in which the relaxation
time $\tau$ tends to infinity: for $\omega\tau\gg1$, the Drude conductivity reduces to a
$\nicefrac1\omega$ pole in $\im(\sigma)$,
\begin{equation}
  \sigma=\ci\frac{\sigma_{DC}}{\omega\tau},
  \label{eq:helix-drude-tau-large}
\end{equation}
which indicates an infinite DC conductivity as explained below
\eqref{eq:kramers-kronig-relations}. Charge carriers which are accelerated by the
external field cannot dissipate their momentum and therefore the resulting response 
is infinite. Consequently, there are two physical mechanisms leading to an infinite DC
conductivity in the translationally invariant case. First, there is a contribution for $T<T_c$
which is caused by superconductivity. Additionally, there is a contribution due
to momentum conservation. It is, therefore, necessary to break translational symmetry in
order to determine the superconducting degrees of freedom separately. In
\figref{fig:helix-conductivity-p-zero}, we consider the translationally invariant case,
and indeed observe the absence of a Drude peak, and the presence of a delta peak
$\delta(\omega)$ both in the normal as well as superconducting phases. 

\begin{figure}[t]
  \includegraphics[width=\textwidth]{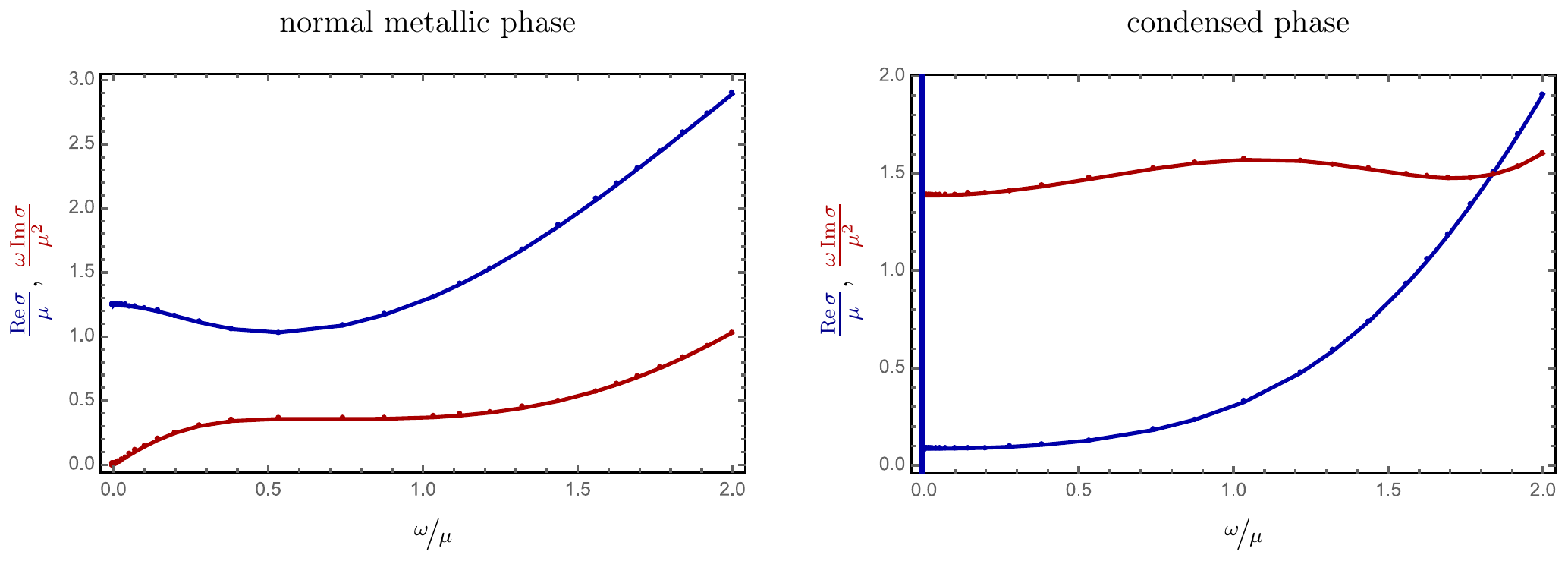}
  \caption{The optical conductivity for $q=0$, $\kappa=0$, $\nicefrac p\mu=2.4$,%
    \hyperref[fntext-fig:helix-conductivity-real-and-im-part]{$^{\thefootnote}$}
    $\nicefrac\lambda\mu=4$ and $T=0.57\,T_c$ in the normal phase
    (left panel) and in the thermodynamically preferred condensed phase (right
    panel). The real part of the optical conductivity exhibits a broad Drude peak for small
    frequencies in the normal phase related to the non-conserved linear momentum in the
    Bianchi type VII helical background. Comparing the normal phase real part of
    the optical conductivity ({\color{darkblue}blue} line) in the left panel with the
    condensed phase real part of the optical conductivity in the right panel, the
    developement of a soft gap becomes apparent. The gap scales
    algebraically with the temperature.
    \label{fig:helix-conductivity-real-and-im-part}}
\end{figure}

Turning on the helical structure $p\neq0$ and $\lambda\neq0$, linear momentum is no
longer conserved\footnote{The canonical momentum related to the Bianchi VII group
  translations is still conserved, albeit it is not accessible on the boundary field
  theory.} and we find, at last for a weak helix $\nicefrac\lambda\mu \ll 1$, a bona fide
Drude-model behavior.  In \figref{fig:helix-conductivity-real-and-im-part},
\addtocounter{footnote}{1}%
%
\footnotetext{This value of $\nicefrac p\mu$ allows for the minimal $T_c$ in the case
  $\kappa=0$ \cff\figref{fig:helix-finite-t-phase-trans-phase-diag}.
  \label{fntext-fig:helix-conductivity-real-and-im-part}}%
the optical conductivity is shown for $\nicefrac p\mu=2.4$, $\nicefrac\lambda\mu=1$, and
$q=6$ for a certain choice of temperature in the broken phase and for the transition
temperature $T_c$. For small frequencies, a Drude peak in the real part of the
conductivity is observed both in the normal phase at $T_c$ and in the superconducting
phase. Since we are not strictly at zero temperature, there is a remaining small
Drude-like peak after subtraction of the $\nicefrac\ci\omega$ pole.
%
\begin{figure}[t]
  \includegraphics[width=\textwidth]{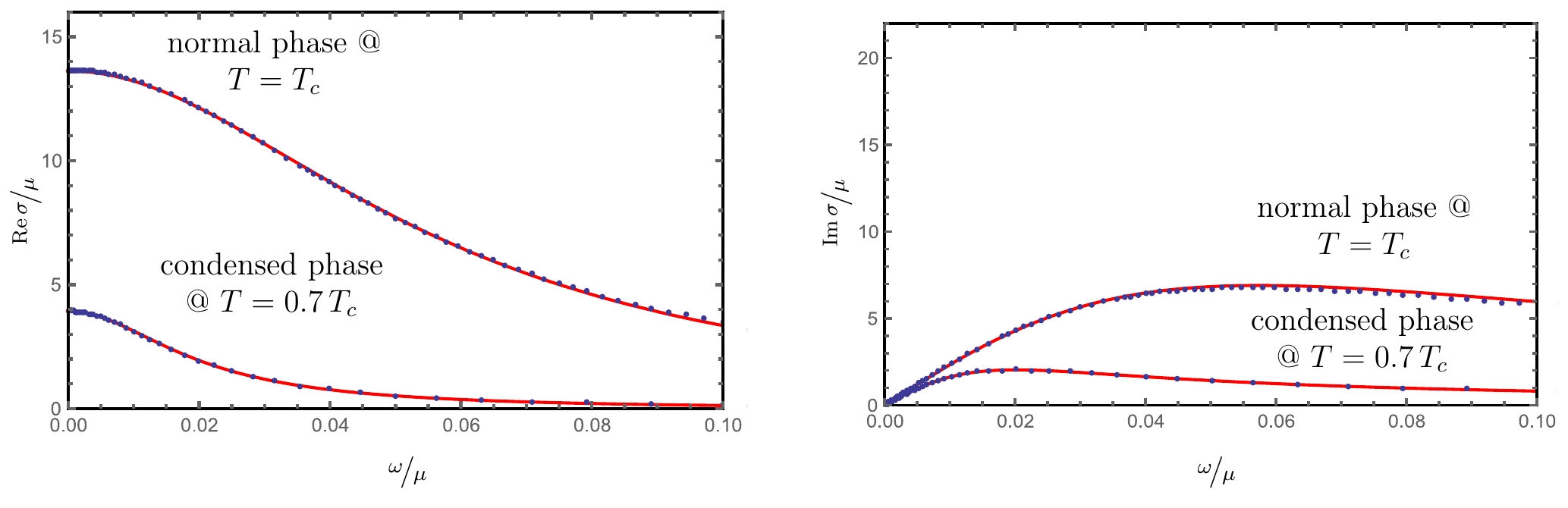}
  \caption{The small-frequency behavior of the optical conductivity in the metallic
    phase for $q=6$, $\kappa=0$, $\nicefrac p\mu=2.4$ and $\nicefrac\lambda\mu=1$. The
    solid lines are fits of the Drude model \eqref{eq:drude-model} to the numerically
    determined optical conductivity. The normal phase solution is given by
    $\sigma_{DC}=13.6873$ and $\tau=17.4352$. In the superconducting phase the
    $\nicefrac1\omega$ pole has been subtracted from $\im\sigma$ 
    and there is a remaining ``residual'' Drude-like peak $\sigma^{\text{reg}}$ as shown
    in the left panel. As explained in the main text and in \figref{fig:FGT-sum-rule},
    this residual contribution to the two-fluid model \eqref{eq:two-fluid-model} shows
    the coexistence of the superconducting phase and a ``normal'' holographic metal.
    \label{fig:helix-drude-fit}}
\end{figure}
\figref{fig:helix-drude-fit} shows the small-frequency regime of the optical
conductivity $\sigma$ and a corresponding fit to the Drude model. Furthermore, in the
condensed phase the imaginary part of the optical conductivity exhibits a
$\nicefrac1\omega$ pole for $T<T_c$ indicating a delta peak in the real part of the optical
conductivity related to an infinite DC conductivity, which is a characteristic of
superconductivity. This can be inferred from the Kramers-Kronig relation
\begin{equation}
  \im\sigma(\omega)=-\frac{2\omega}\pi\mathcal P\int_0^{\infty}\dd{\tilde\omega}
  \frac{\re\sigma(\tilde\omega)}{\tilde\omega^2-\omega^2}.
  \label{eq:kramers-kronig-relations}
\end{equation}
According to this relation, a $\nicefrac1\omega$ pole in the imaginary part of the
conductivity is related to a delta function at zero frequency in the real part by
\begin{alignat}{3}
  \im\sigma(\omega)&=\frac{2\rho_s}\pi\frac1\omega &\qquad &\longleftrightarrow &\qquad
  \re\sigma(\omega)&=\rho_s\delta(\omega).
  \label{eq:delta-func-pole}
\end{alignat}
This equation defines the {\it superfluid density} $\rho_s$  as the coefficient of
the zero frequency delta function in $\re(\sigma)$.\footnote{In the conventions of
  \cite{Homes2004,Homes2005,Erdmenger2012a}, $\rho_s$ is defined via
  $\re(\sigma)=\rho_s\delta(\omega)/8$, \ie it differs by a factor of $8$
  from the definition used here.\label{fn:correction-factor-superfluid}}
 
\begin{figure}[t]
  \includegraphics[width=\textwidth]{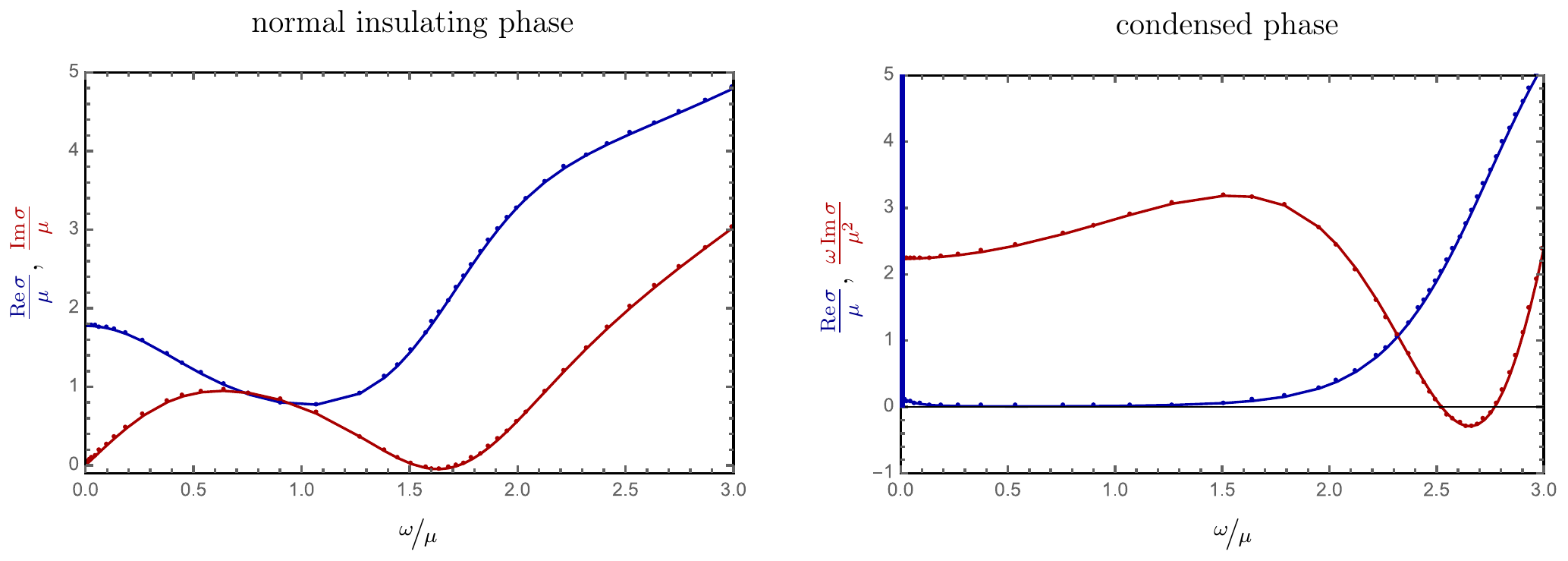}
  \caption{The left plot shows the ``insulating'' low-temperature phase optical
    conductivity with vanishing imaginary part as $\omega\rightarrow 0$, which indicates
    the absence of a delta peak at $\omega=0$ and allows for a proper Drude-like
    shape. In the right plot the superfluid phase shifts ``spectral weight'' from the
    finite frequency part which condensed in a $\omega=0$ delta peak, thus generating
    the gap in the superconducting state in the region
    $\nicefrac\omega\mu\in[0,1.5]$. Note that this is not a hard gap due to its
    algebraic temperature scaling on approaching the zero temperature limit and that
    there is a non-zero contribution coming from a residual Drude-like peak
    $\sigma^{\text{reg}}$ even at small $T$ in the superfluid phase. As expected from
    the Kramers-Kronig relations the imaginary part of the optical conductivity displays
    a $\nicefrac1\omega$ pole. Both plots are generated for $\kappa=\nicefrac1{\sqrt
      2}$, $q=6$, $\nicefrac p\mu=1.2$, $\nicefrac\lambda\mu=1.9$, and $T=0.46\,T_c$.
    \label{fig:superfluid-insulator-conductivity}}
\end{figure}

\begin{figure}[t]
  \includegraphics[width=\textwidth]{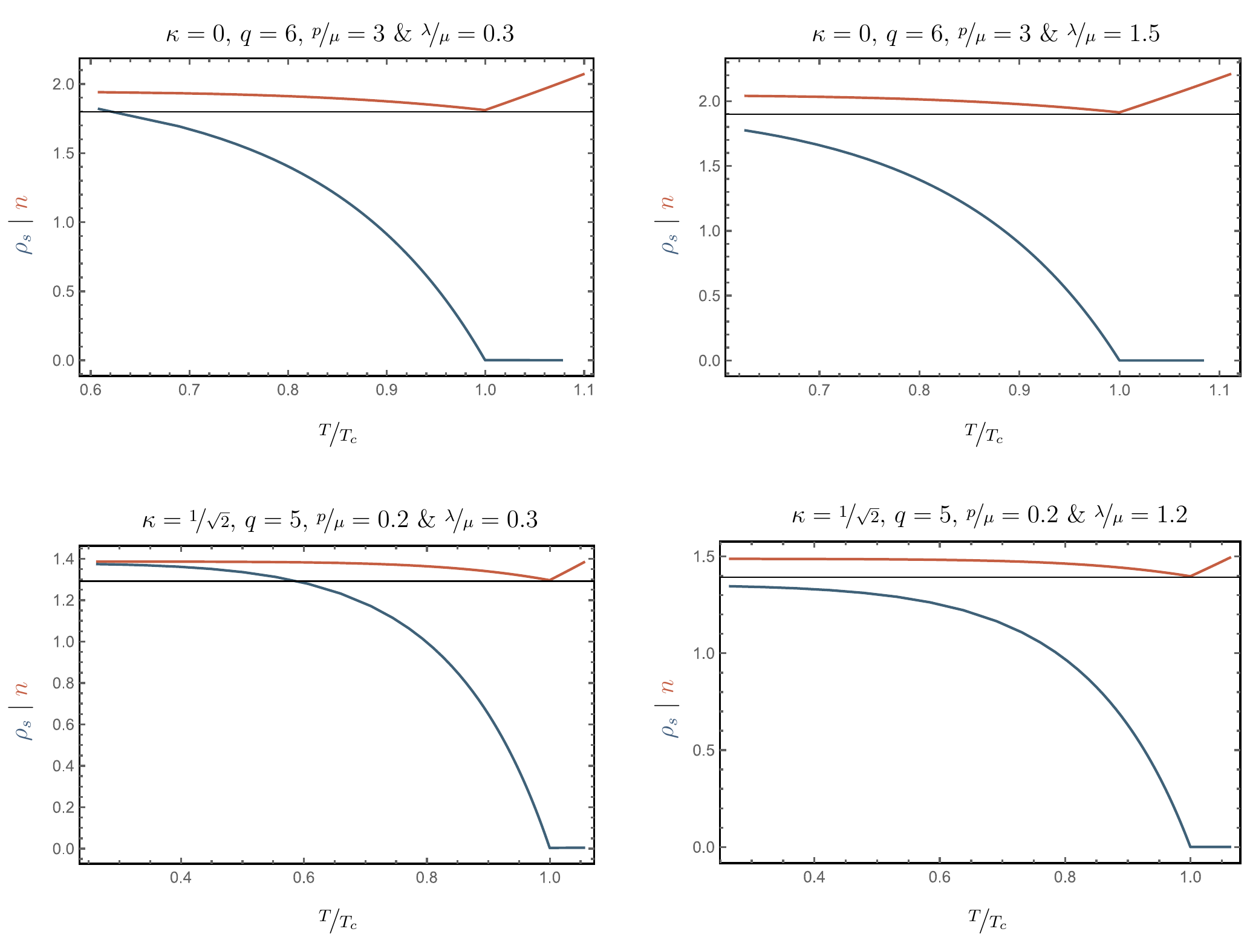}
  \caption{The superfluid density ${\color{DarkRainbow0p135}\rho_s}$ and the
    normalized\hyperref[fntext-fig:helix-rhosf-of-T]{$^{\thefootnote}$} charge density
    ${\color{DarkRainbow0p855}n}$ as a function of $\nicefrac T{T_c}$. For $T > T_c$,
    the superfluid density vanishes, \textit{i.e.}\@ the DC conductivity is finite. As
    $T$ is lowered, the superfluid density ${\color{DarkRainbow0p135}\rho_s}$
    increases, similar to the order parameter, \textit{c.f.}\@
    \figref{fig:helix-finite-t-phase-trans-thermodyn}, and curiously the normalized
    charge density ${\color{DarkRainbow0p855}n_s}$ as well. The gap between
    ${\color{DarkRainbow0p135}\rho_s}$ and ${\color{DarkRainbow0p855}n_s}$ is
    independent of the value of $\nicefrac p\mu$, but increases with increasing
    $\nicefrac\lambda\mu$. For small $\nicefrac\lambda\mu$, it is suggestive that at
    $T=0$ the superfluid density and the condensed phase charge density coincide. Thus,
    the longitudinal response \textit{i.e.}\@ the plasma frequency in the
    superconducting phase is sufficient to determine the superfluid strength.
    \label{fig:helix-rhosf-of-T}} 
\end{figure}

As shown in Figure \ref{fig:superfluid-insulator-conductivity}, a small Drude peak
remains present in the superconducting phase. To describe the system, it is thus
necessary to apply the {\it two-fluid model} \cite{Horowitz2013},  which supplements
\eqref{eq:delta-func-pole} with the metallic Drude model defined in
\eqref{eq:drude-model},
\begin{equation}
  \re\sigma(\omega)=\sigma^{\text{reg}}(\omega)+\rho_s\delta(\omega)
  =\left(\chi_n(T)\frac\tau{1+\omega^2\tau^2}+\frac\pi2\chi_s(T)\delta(\omega)\right),
  \label{eq:two-fluid-model}
\end{equation}
where $\chi_n(T)$ describes the Drude-like contribution resembling a normal fluid and
$\chi_s(T)$ the superconducting contribution. In the normal state, we have
$\chi_n(T>T_c)=n_n$ and $\chi_s(T>T_c)=0$, whereas a pure superconducting state would be
described by $\chi_n(T<T_c)=0$ and $\chi_s(T<T_c)=\rho_s$.\footnote{In order to restore the
  proper units of the two-fluid model, the charge density is given in units of
  $\nicefrac{e^2}{m^\ast}$, i.e. the number density and the charge density are related by
  $n_{\text{charge}}=e^2n_{\text{number}}/m^\ast$. Note that we work with charge densities and not
  number densities throughout the paper, as the quantities $e$ and $m^\ast$ are not
  directly accessible in holographic models. Furthermore, this choice of dimensions has
  the advantage that the superfluid strength $\rho_s$ and the charge density $n$ have
  the same units (in natural units).} Due to charge conservation
$\chi_n(T)+\chi_s(T)=n$.\footnote{Throughout the paper $n$ denotes a general charge
  density, while $n_n$ denotes the charge density in the normal phase, and $n_s$ the
  charge density in the superfluid phase.}
\\[\baselineskip]
Moreover, from Figure \ref{fig:superfluid-insulator-conductivity} we observe
that the conductivity in the superconducting state develops a gap at low frequencies,
\ie $\re(\sigma)$ is significantly suppressed. This gap is a characteristic of a
superconducting system; it indicates that low-energy charged degrees of freedom have
condensed into the delta function at $\omega=0$.
%
An important issue is whether $\sigma^{\text{reg}}(\omega)$ in \eqref{eq:two-fluid-model}
vanishes in the limit $T\rightarrow0$ for frequencies below $\omega_\mathrm{gap}$. This
would imply $n_s=\rho_s$ at $T=0$, with $n_s$ the thermodynamic density. In general,
the following scenarios are possible. One possibility is the presence of a \emph{hard
  gap}, in which at low frequencies $0<\omega<\omega_\mathrm{gap}$ we have an
exponential suppression $\sigma(\omega,T)\sim\exp((\omega-\omega_\mathrm{gap})/T)$. On
the other hand, for a \emph{soft gap} there is an algebraic (power law) scaling
$\sigma(\omega,T)\sim T^{c_1}\omega^{c_2} $. In this case, it is much harder to
determine numerically whether there exists an additional constant contribution
$\sigma_0$ to $\sigma(\omega,T)$. As we discuss below, for the model considered in this
paper, we find an algebraic scaling. Moreover, by calculating $n_s$ and $\rho_s$
independently, we find that $n_s=\rho_s$ for $T \rightarrow 0$ to good numerical
accuracy, at least for small $\nicefrac p\mu$ and $\nicefrac\lambda\mu$. This implies
that $\sigma_0=0$ in this case, \ie our system exhibits a soft gap.
\\[\baselineskip]
For translationally invariant holographic s-wave superconductors in $2+1$ dimensions, it is known
that, even though highly suppressed, $\re(\sigma)$ remains nonzero at $T=0$ but finite
frequency \cite{Horowitz2009} and one finds an algebraic scaling. On the other hand,
$p$-wave superconductors are reported to exhibit a hard gap \cite{Basu2010}.
Also for the helical lattice, by straightforwardly generalising the low frequency
analysis in the Appendix of \cite{Donos2013c}, we conclude that  the gap scales
algebraically in $\omega$.
\\[\baselineskip]
In addition, we compute $\rho_s$ and $n_s$ individually at very low temperatures and
find indications that they agree to good numerical accuracy. $\rho_s$ is read off from
the zero frequency pole of the conductivity, while the thermodynamic density $n_s$ is
obtained from \eqref{eq:ndef}. In \figref{fig:helix-rhosf-of-T},
\addtocounter{footnote}{1}
\footnotetext{Normalization of the charge density entails a rescaling by
  $\nicefrac{e^2}{m^\ast}$ compared to the usual number densities in the Drude model,
  thus the dimension of $n_s$ is $\text{(length)}^{-2}$ matching the dimension of
  $\rho_s=\nicefrac\pi2\lim_{\omega\to0}\omega\im\sigma(\omega)$, namely
  $\left[\rho_s\right]=\text{(length)}^{-2}$.\label{fntext-fig:helix-rhosf-of-T}}%
the superfluid density $\rho_s$ and the charge density $n_s$, \cff
Eq.\@ \eqref{eq:helix-omega-familiar}, are plotted as a function of temperature for two
sets of parameters. The superfluid density, being a measure for the superconducting
degrees of freedom, increases as the temperature is lowered beyond $T_c$ and it vanishes
for $T>T_c$. Of course, in order to finally conclude whether $\rho_s$ agrees with $n_s$
for $T =0$, as the extrapolation of our data suggests, we need to carefully analyze the zero
temperature transport properties, which we are planning to do in future work
\cite{WIP}. Nonetheless, for small $\nicefrac p\mu$ and $\nicefrac\lambda\mu$ the
difference becomes sufficiently small already at finite temperatures about
$T\approx0.6\,T_c$. The difference between $\rho_s$ and $n_s$ at this temperature seems
to be independent of the helix pitch\footnote{Technical problems arise for $\nicefrac
  p\mu\gg1$ due to numerical instabilities.}, parametrizing the helical lattice
constant, but grows with increasing $\nicefrac\lambda\mu$. This difference may be
accounted for by the residual contribution in the condensed phase for $\omega\to0$,
which is \emph{not} added to the zero mode delta peak. The optical conductivity of high
$T_c$ superconductors is known \cite{RevModPhys.77.721} to feature residual absorption at very
small frequencies, which gives rise to an additional contribution to the imaginary part
of $\sigma(\omega,T<T_c)$. For small helix strengths $\nicefrac\lambda\mu\ll 1$, this residual
part can be read off by a simple Drude-fit inside the superconducting gap as shown in
\figref{fig:helix-drude-fit}. The spectral weight inside the residual Drude peak
accounts exactly for the difference between $\rho_s$ and $n_s$. On the other hand, for
larger helix strengths, \ie stronger momentum dissipation, the gap cannot be accounted
for by the residual spectral weight inside the superconducting gap. We discuss two
possible reasons for this behavior in Section \ref{ssec:concl-transport}.


\subsubsection{Intermediate and High Frequency Regimes\label{sssec:frequency-regimes}}

\begin{figure}[t]
  \includegraphics[width=\textwidth]{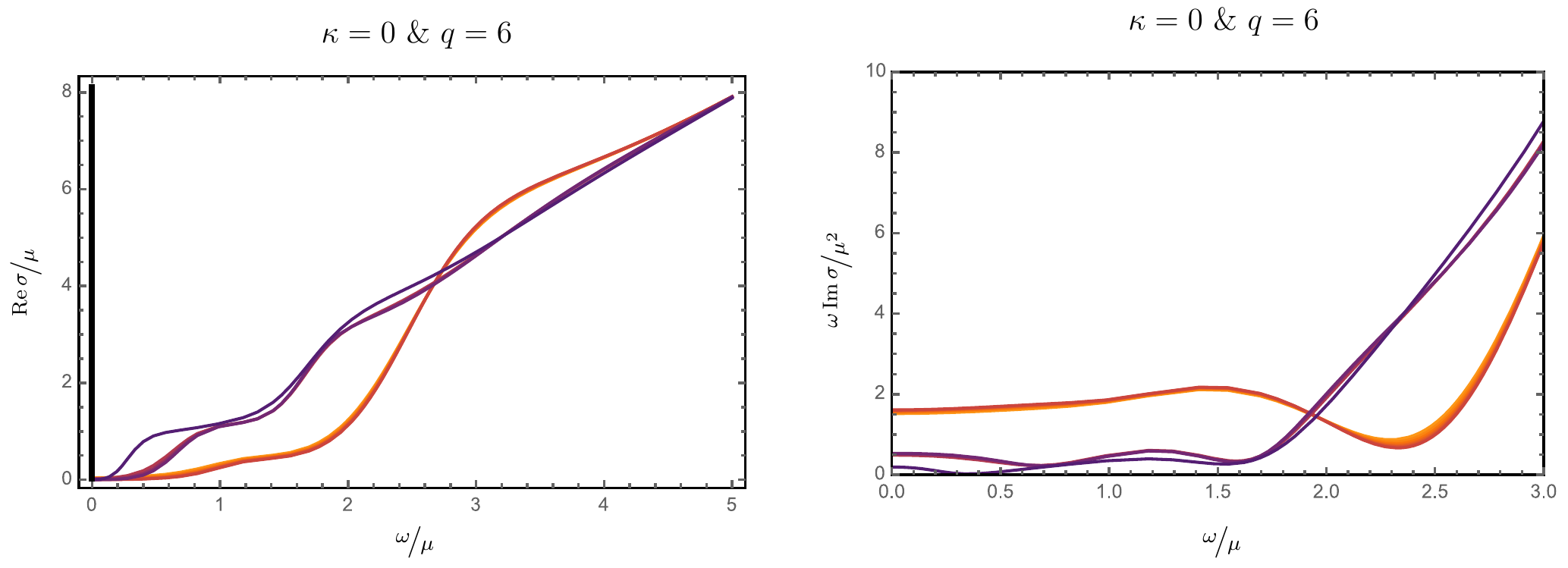}
  \caption{For $\kappa=0$, $\nicefrac p\mu=0.8$ and $\nicefrac\lambda\mu=3$ we find that
    the ``superconducting gap'' is replaced by a plateau-like decay in stark contrast to
    the right panel of \figref{fig:superfluid-insulator-conductivity}, which shows the
    typical form of the optical conductivity. It is suggestive to compare this result to
    Figure 16 in \cite{Andrade:2014xca} where the authors find a similar behavior for
    low temperatures. Again, the color coding for the temperature is as follows:
    $\nicefrac T{T_c}=\color{SunsetColors0p6}0.47$, $\color{SunsetColors0p55}0.42$,
    $\color{SunsetColors0p5}0.37$, $\color{SunsetColors0p45}0.33$,
    $\color{SunsetColors0p4}0.28$, $\color{SunsetColors0p35}0.23$,
    $\color{SunsetColors0p3}0.19$, $\color{SunsetColors0p25}0.14$,
    $\color{SunsetColors0p2}0.09$, $\color{SunsetColors0p15}0.05$. Counter-intuitively,
    the gap is drastically diminished with decreasing temperature, whereas one would
    expect the gap to become more pronounced for low temperatures.
    \label{fig:optical-conductivity-kappa0}
    }
\end{figure}

\begin{figure}[t]
  \includegraphics[width=\textwidth]{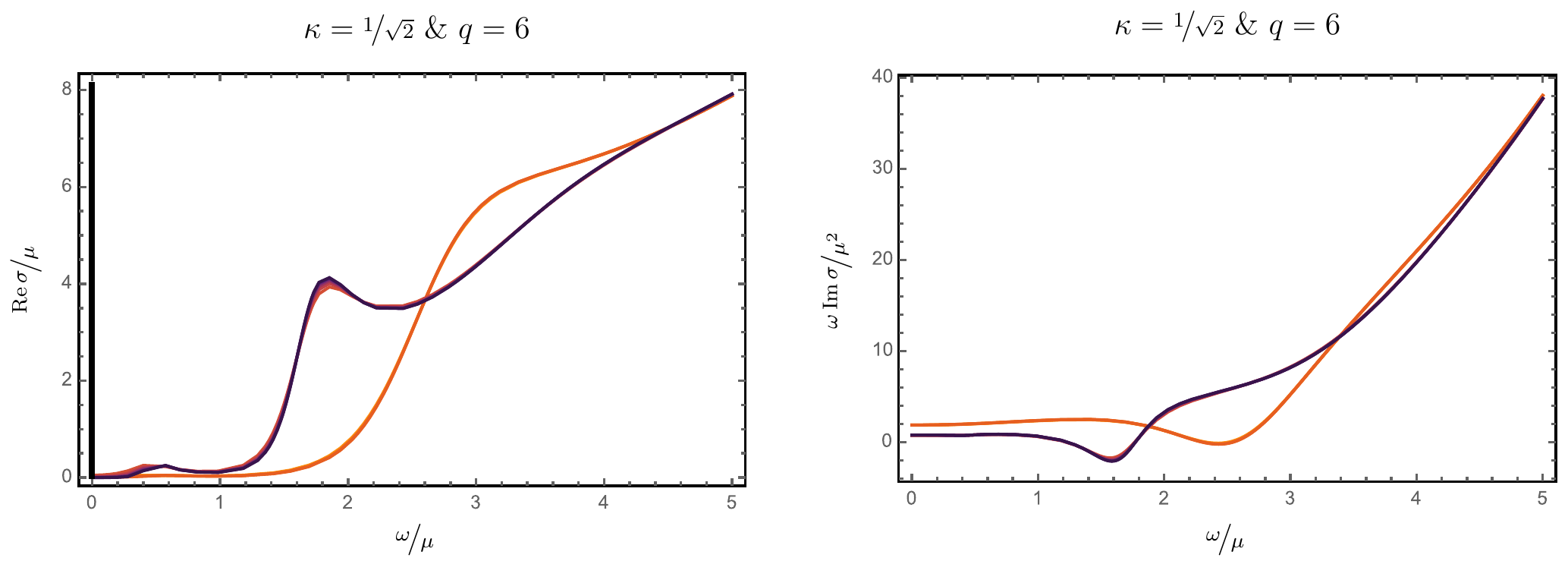}
   \caption{Different shapes of the optical conductivity with additional features
    compared to \figref{fig:superfluid-insulator-conductivity} for $\nicefrac p\mu=0.2$
    and $\nicefrac\lambda\mu=2$. In the case of $\kappa=\nicefrac1{\sqrt2}$ for
    decreasing temperature the gap seems to get smaller and a new peak arises,
    interestingly at lower temperatures. The color coding for the temperature is as
    follows: $\nicefrac T{T_c}=\color{SunsetColors0p6}0.41$,
    $\color{SunsetColors0p54}0.37$, $\color{SunsetColors0p49}0.33$,
    $\color{SunsetColors0p43}0.29$, $\color{SunsetColors0p38}0.25$,
    $\color{SunsetColors0p32}0.2$, $\color{SunsetColors0p27}0.16$,
    $\color{SunsetColors0p21}0.12$, $\color{SunsetColors0p16}0.08$,
    $\color{SunsetColors0p1}0.04$. Note that in all cases there is a delta peak in the
    real part indicating the superfluid phase.
    \label{fig:optical-conductivity-kappa1Sqrt2}}
\end{figure}

Let us also discuss the behaviour of the optical conductivity in further frequency
regimes. First we note that in the  intermediary frequency regime $T\ll\omega\ll\mu$,
we have not been able to find a scaling law of $|\sigma(\omega)|$. Such a scaling has been
observed in the strange metallic phase of the cuprates and interpreted as a consequence
of quantum criticality in \cite{Marel2003}. While some holographic models
\cite{Horowitz2012,Horowitz2012a,Vegh2013,Horowitz2013} seem to show such a scaling,
others \cite{Donos2013c} do not, and our model seems to be in the latter class. So far a
theoretical understanding of the origin of this scaling regime in holographic models is
still missing.\\[\baselineskip]
Concerning the nature of the superconducting gap, there are two more intriguing
features: in the case of $\kappa=0$, we find in the vincinity of particular parameter
values such as \eg $\nicefrac p\mu=0.8, \nicefrac\lambda\mu=3$, plateau-like solutions
where the energy scale of the gap as a well as the superfluid density $\rho_s$ is
drastically reduced, see \figref{fig:optical-conductivity-kappa0}. Curiously, these
solutions seem to arise for very low temperatures contrary to
the intuition that the gap should grow with decreasing temperature, \cff the
{\color{SunsetColors0p6}orange} line compared to the {\color{SunsetColors0p15}purple}
line in \figref{fig:optical-conductivity-kappa0}. 
A similar behavior has been found in Figure 16 of \cite{Andrade:2014xca}, although in our
case there is no Drude-like peak for higher temperatures close to the transition
temperature, which may be attributed to the fact that almost all degrees of freedom have
condensed.\footnote{See also the discussion below \figref{fig:FGT-sum-rule} and the
  FGT-Section \ref{ssec:helix-sum-rule}.} 
The same intriguing features are seen for $\kappa=\nicefrac1{\sqrt2}$ in the vicinity of
particular parameter values such as the ones shown in
\figref{fig:optical-conductivity-kappa1Sqrt2}.
%
Compared to the more generic result of the optical conductivity shown in
\figref{fig:helix-drude-fit} and \ref{fig:superfluid-insulator-conductivity}, the gap
seems to develop a new feature in the intermediary frequency regime that resembles a
peak at a non-zero frequency (as shown in
\figref{fig:optical-conductivity-kappa1Sqrt2} at $\nicefrac\omega\mu\approx1.78$) at 
lower temperatures in the superconducting phase. Again not only the gap is reduced but
also the superfluid density; this is evident since the Ferrell-Glover-Tinkham sum rule
holds, as proven in the following Section \ref{ssec:helix-sum-rule}. These curiosities
appear to be due to the contributions of additional resonances below the gap.
We plan to investigate these 
further in future work \cite{WIP}. Finally, in all cases considered, we find that for
large frequencies, $\omega\gg\mu$, the real part of the conductivity is proportional to
$\omega$. This behavior is a property of the ultraviolet fixed point of the field
theory: since $\sigma$ has energy dimension one, it scales linearly in $\omega$ once the
frequency is larger than any other scale of the system.


\subsection{Ferrell-Glover-Tinkham Sum Rule\label{ssec:helix-sum-rule}}

\begin{figure}[t]
  \figcapside{
    \caption{Visualization of the FGT sum rule as explained in the text. The blue area
      indicates the spectral weight which is transferred into the zero mode. Note the
      tiny regular contribution which resembles a key property of high-temperature
      superconductivity and might be responsible for the small possible offset in
      the computation of $\rho_s$, in particular it might account for the missing charge
      density in the superfluid phase, \textit{i.e.}\@ the offset between $n_s$ and
      $\rho_s$ displayed in \figref{fig:helix-rhosf-of-T}.\label{fig:FGT-sum-rule}}
  }{\includegraphics[width=.5\textwidth]{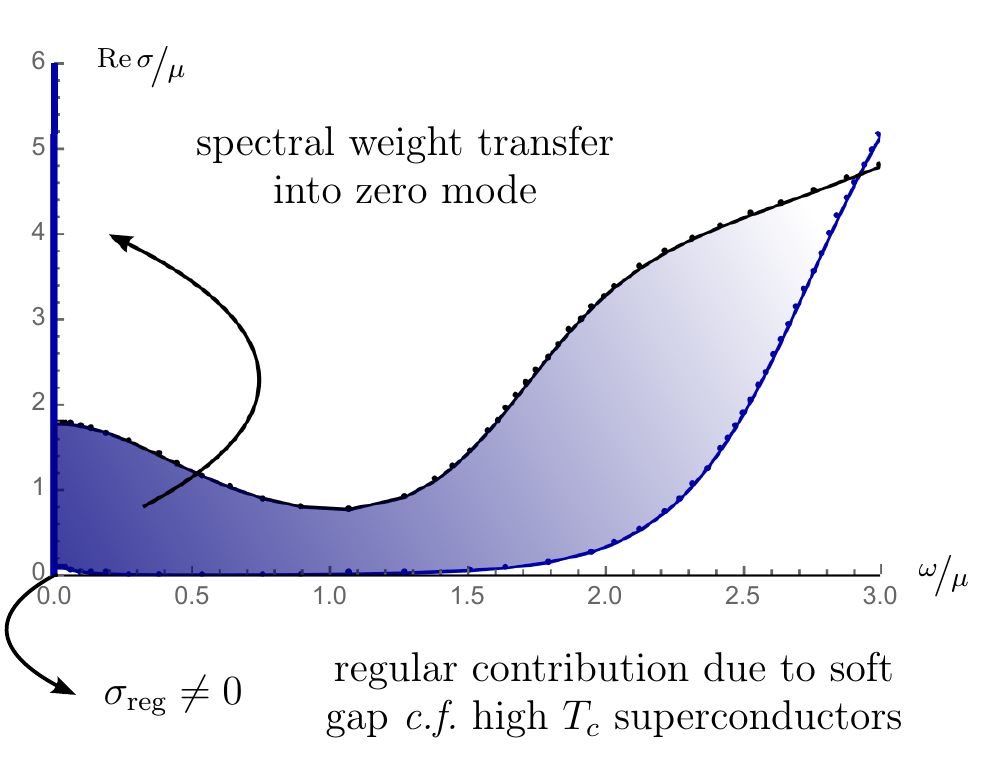}}
\end{figure}

\begin{figure}[t]
  \includegraphics[width=\textwidth]{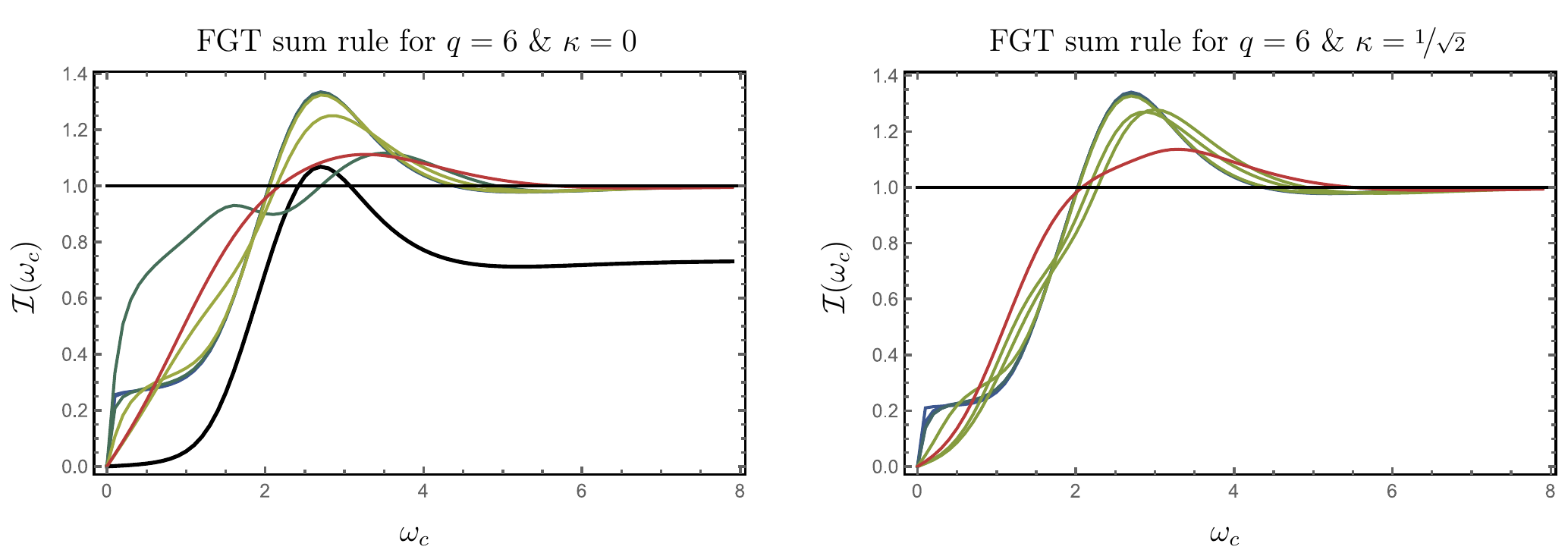}
  \caption{The Ferrell-Glover-Tinkham sum rule holds in the condensed phase for
    different values of $\nicefrac\lambda\mu$ color coded as in
    \figref{fig:helix-finite-t-phase-trans-phase-diag}. The left panel displays the FGT
    sum rule in the case $\kappa=0$ at the temperature $\nicefrac T\mu=0.1$ and 
    $\nicefrac\lambda\mu\,(\nicefrac p\mu)=\color{DarkRainbow0p045}0.3\,(0.4,1)$,
    $\color{DarkRainbow0p225}1.5\,(0.2,1.4)$,
    $\color{DarkRainbow0p495}3.3\,(0.2,1)$, $\color{DarkRainbow0p9}6\,(2.4)$,
    where the numbers in brackets denote the corresponding values of
    $\nicefrac p\mu$. The right panel shows the FGT sum rule for
    $\kappa=\nicefrac1{\sqrt2}$, $\nicefrac T\mu=0.05$ and
    $\nicefrac\lambda\mu\,(\nicefrac p\mu)=\color{DarkRainbow0p045}0.3\,(0.3,1)$,
    $\color{DarkRainbow0p18}1.2\,(0.2)$, $\color{DarkRainbow0p45}3\,(0.2,1,1.4)$,
    $\color{DarkRainbow0p9}6\,(1.4)$. The integral $\mathcal I(\omega_c)$ defined in
    \eqref{eq:sum-rule-in} measures the missing spectral weight up to the cutoff
    $\omega_c$ and is normalized so that, if the sum rule is satisfied, $\mathcal
    I(\infty)=1$. As can be seen \textit{e.g.}\@ from
    \figref{fig:helix-conductivity-p-zero} for $\nicefrac\omega\mu>8$ the optical
    conductivity enters the conformal regime, \textit{i.e.}\@ for $d=3+1$,
    $\sigma(\omega)\sim\omega$, irrespective of the existence of the ``superconducting
    gap''. In the conformal regime the normal phase and condensed phase optical
    conductivity becomes identical and thus will not contribute to $\mathcal
    I(\omega)$. Note that the thick black line in the left panel represents the
    translationally invariant case, $\nicefrac p\mu=\nicefrac\lambda\mu=0$, and as
    expected the FGT sum rule fails spectacularly, owing to the coexistence of a ideal
    metal and a superconductor. Thus, in this case the diamagnetic pole in the imaginary
    part of the optical conductivity includes not only the missing spectral weight, but
    also the ideal metal contribution.\label{fig:helix-sum-rule}}
\end{figure}

Sum rules are exact identities following from the analytic structure of Green
functions. The Ferrell-Glover-Tinkham sum rule
\cite{PhysRev.109.1398,PhysRevLett.2.331q} can be expressed as the integral over the
real part of the optical conductivity, being a constant regardless of the details of the
system,
\begin{equation}
  \int_{0}^{\omega_c
  }\dd\omega\re\sigma(\omega)=\text{const.},
  \label{eq:FGT-sum-rule}
\end{equation}
Note that the integral includes possible contributions from the lower bound in the form
of a delta function at $\omega=0$. Furthermore, while actual condensed matter systems typically become transparent at frequencies larger than the typical electronic energy scales and hence the integral in \eqref{eq:FGT-sum-rule} and \eqref{eq:PlasmaNormal} converges, in holographic setups it is generically UV divergent due to the UV conformal fixed point behavior of $\re\sigma(\omega)$. In order to regulate this divergence, we introduced a UV cutoff frequency $\omega_c$, to be taken larger than $\max(T,\mu)$. In appropriate formulations of the sum rule such as \eqref{eq:helix-sum-rule-second-form} the UV divergence cancels between the two integrals, and the regulator can be removed.\footnote{A more elegant way of regularizing \eqref{eq:FGT-sum-rule} and \eqref{eq:PlasmaNormal} is nicely described in \cite{Gulotta:2010cu}: Instead of working with the Green's functions obtained naively from a holographic calculation, which typically do not vanish in the upper half frequency plane and on the real axis for $\abs{\omega} \rightarrow \infty$, defining a subtracted Green's function with these problematic contributions removed ensures that the sum rules are valid. These local subtractions correspond to the addition of local finite counterterms to the holographically renormalized partition function. In this way a particular preferred renormalization scheme can be chosen without invoking additional requirements such as supersymmetry \cite{Karch:2005ms}. An example is described around eq.~12 of \cite{Horowitz2008}, where a local $i\omega/2$ term (also present in  \eqref{eq:helix-res-cond-in-modes}) was removed by a finite counterterm $\int d^4 x \sqrt{-\gamma} F_{ij}^2$. We thank Martin Ammon for pointing us to the latter reference. Note however that whatever renormalization scheme is chosen, UV counterterms can only affect the ultralocal terms in the Greens function and hence the UV asymptotics of the conductivity. The physical part of the conductivity which we are interested in, and which after renormalization should fulfill the Kramers-Kronig relations \eqref{eq:kramers-kronig-relations} (i.e. causality), comes from the current-current two point function at different points in space-time and hence cannot be affected by this choice, but must be  renormalization scheme invariant.} Physically, the sum rule expresses the conservation
of charged degrees of freedom, which are measured by the spectral weight, \ie the area
under $\mathrm{Re}(\sigma)$. For example, in the normal phase, Eq.\@
\eqref{eq:FGT-sum-rule} allows to identify 
the plasma frequency as a measure of the charge density in the system via
\begin{equation}
  \frac{\omega_{\text{Pn}}^2}8=\int_{0}^{\omega_c
  }\dd\omega\re\sigma_n(\omega).
  \label{eq:PlasmaNormal}
\end{equation}
In the superfluid phase, this definition excludes the delta function at $\omega=0$. 
In the case of the superconducting phase transition, where the
spectral weight is transferred into the delta function at $\omega=0$, the degrees
of freedom can rearrange themselves but they cannot be lost. The Ferrell-Glover-Tinkham
sum rule can also be expressed in the form 
\begin{equation}
  \rho_s=\frac\pi2\lim_{\omega\to0}\big[\omega\im\sigma_s(\omega)\big]
  =\int_{0^+}^{\infty}\dd\omega\big[\re\sigma_n(\omega)-\re\sigma_s(\omega)\big].
  \label{eq:helix-sum-rule-second-form}
\end{equation}
Here $\sigma_n$ denotes the optical conductivity in the normal phase, \ie for some
$T\ge T_c$, $\sigma_s$ the conductivity for some temperature below $T_c$, and $\rho_s$
is the superfluid density at that temperature. The contribution from
$\omega=0$ has been separated out explicitly giving rise to the term $\rho_s$ determined
by $\sigma_s$. According to \eqref{eq:helix-sum-rule-second-form}, the superfluid
density is equal to the missing spectral weight, \ie the difference in the area under
the conductivity curve in the normal and in the superconducting state,
\cff\figref{fig:FGT-sum-rule}.
%
Note that \eqref{eq:helix-sum-rule-second-form} assumes already that the translational
symmetry is broken, \ie a $\delta(\omega)$ contribution or diamagnetic pole in the
normal phase is absent. It is convenient to define\footnote{Additionally, the definition
  of $\rho_s$ in \eqref{eq:delta-func-pole} includes the factor of $\nicefrac\pi2$
  arising from the integration in the Kramers-Kronig relations, \ie
  $\rho_s=\nicefrac\pi2\lim_{\omega\to0}\omega\im\sigma(\omega)$.}
\begin{align}
  \mathcal I(\omega_c)=\frac1{\rho_s}\int_{0^+}^{\omega_c}\dd\omega
  \left[\re\sigma_n(\omega)-\re\sigma_s(\omega)\right],
  \label{eq:sum-rule-in}
\end{align}
in order to apply the sum rule to the numerically calculated conductivities. Here
$\omega_c$ is a cutoff frequency and the sum rule is satisfied if $\mathcal
I(\infty)=1$. In \figref{fig:helix-sum-rule},
%
$\mathcal I(\omega_c)$ is plotted in the condensed phase for two different temperatures
$\nicefrac T\mu=0.1,0.05$. As expected, $\mathcal I(\omega_c)$ approaches unity for
large enough cutoff frequencies. This confirms the sum rule for the system under
consideration and can be seen as a powerful consistency check of the holographic model
and of the calculation including the numerics. Physically, it shows that the charged
degrees of freedom of the system are conserved. In particular, it uncovers the main
obstacle in defining a proper superfluid density in the translational invariant system,
since the FGT sum-rule as defined in \eqref{eq:sum-rule-in} does not hold due to the
coexistence of the normal state ideal metal contributing to the diamagnetic pole, \cff
the black line in  \figref{fig:helix-sum-rule}. Once we turn on our helical structure
the ``spurious'' contribution due to momentum conservation are removed from the
diamagnetic pole and the FGT sum rule confirms the conservation of charged degrees of
freedom.


\subsection{Checking Homes' and Uemura's relations
  \label{ssec:homes-relation}}

\begin{figure}[t] 
  \includegraphics[width=.8\textwidth]{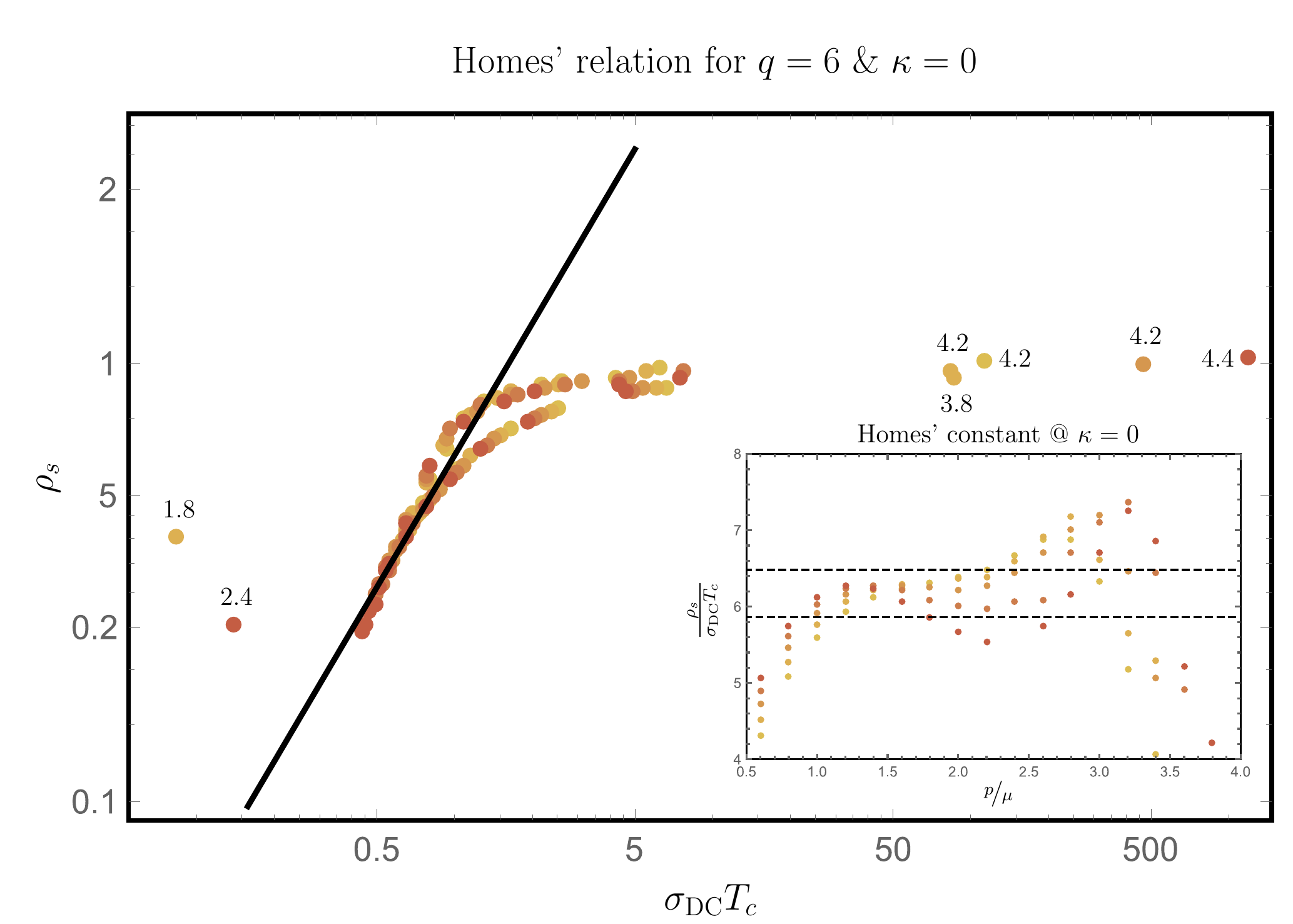}
  \caption{A log-log-plot of the superfluid density $\rho_s$ with respect to
    $\sigma_{\text{DC}}(T_c)T_c$.  The color coding for $\nicefrac\lambda\mu$ is
    identical to the phase-diagram plots presented in 
    \figref{fig:helix-finite-t-phase-trans-phase-diag},
    $\nicefrac\lambda\mu=\color{DarkRainbow0p675}4.5$, $\color{DarkRainbow0p72}4.8$,
    $\color{DarkRainbow0p765}5.1$, $\color{DarkRainbow0p81}5.4$,
    $\color{DarkRainbow0p855}5.7$ , whereas values of different $\nicefrac p\mu$ are not
    resolved, except for the outliers where the value of $\nicefrac p\mu$ is explicitly
    attached to the point. If Homes' relation holds, the points should roughly fall on a
    line with a slope of unity, according to $\log(\rho_s)=\log C
    +\log(\sigma_{\text{DC}}T_c)$ denoted by the black line.  
    The inset shows the value of Homes' constant $C$ for
    $\nicefrac\lambda\mu=\color{DarkRainbow0p675}4.5$, $\color{DarkRainbow0p72}4.8$,
    $\color{DarkRainbow0p765}5.1$, $\color{DarkRainbow0p81}5.4$,
    $\color{DarkRainbow0p855}5.7$. The relation is not expected to
    hold in the limits of $\nicefrac p\mu\to0$ and $\nicefrac p\mu\to\infty$. There the
    constant approaches zero due to the absence of momentum relaxation and the
    corresponding divergence of $\sigma_{\text{DC}}$. These data points may be
    faithfully discarded. Doing so, we see that, in the reasonably applicable range of
    $\nicefrac p\mu\in[1,2]$ Homes' relation seems to hold within the dashed lines given
    by $C\approx6.17\pm0.31$. This value for the constant is extracted from a
    least-squares fit represented by the thick black line in the main figure.
    \label{fig:homeslaw-const}}
\end{figure}

There are two very intriguing relations that were found
experimentally, namely Homes' relation \cite{Homes2004,Homes2005,Dordevic2013} and Uemura's
relation \cite{PhysRevLett.62.2317}.
The former is given by
\begin{align}
  \rho_s(T=0)&=C\sigma_{\text{DC}}(T_c)T_c.
  \label{eq:definition-homes-relation}
  \intertext{whereas the latter reads}
  \rho_s& = B T_c\,,
  \label{eq:definition-uemura-relation}
\end{align}
with $B$ being a proportionality constant of units $mass$ in natural units. Uemura's relation is found to hold for underdoped cuprates only, while, as demonstrated
in \cite{Homes2004,Homes2005}, Homes' relation holds for a much broader class of
materials. Concerning the units of Homes' constant, as defined in Section
\ref{ssec:drude-two-fluid}, $\rho_s$ is given in units of $\mu^2$ and
$\sigma_{\text{DC}}$ as well as $T$ in units of $\mu$. Thus, Homes' constant given by
$\nicefrac{\rho_s}{(\sigma_{\text{DC}}T_c)}$ is dimensionless in our unit system.
\\[\baselineskip]
Checking both Homes' and Uemura's relations by plotting $\rho_s$ against
$T_c$ and $\sigma_{\text{DC}}T_c$, we conclude that Uemura's relation does not hold in our helical
superconducting system. Homes' linear scaling relation, on the other hand, is
clearly visible in \figref{fig:homeslaw-const}, where
%
we show a log-log plot of $\rho_s$ vs $\sigma_{\text{DC}}T_c$ for various
$(\nicefrac\lambda\mu,\nicefrac{p}{\mu})$ with $\kappa=0$. In the range of
\begin{alignat}{3}
 \frac\lambda\mu&\approx4.5,\dotsc,6 &\quad &\text{and} &\quad 
 \frac p\mu&\approx1,\dotsc,2.
 \end{alignat}
the relation is linear and extracting Homes' constant, we find it to be  
\begin{equation}
  C\approx 6.2\pm0.3,
  \label{eq:HomesValue}
\end{equation}
Here the uncertainty is not statistical, but refers to the $\pm 5\%$
band bounded by the dashed lines in the inset in
\figref{fig:homeslaw-const}. Intriguingly, comparing our Homes' constant with the
experimentally found values \cite{Homes2004,Homes2005}, after correcting for the factor
$8$ in our definition of the superfluid density, \cff footnote
\ref{fn:correction-factor-superfluid}, the helical system seem to be interpolating
between the dirty limit BCS superconductors with $C=\nicefrac{65}8\approx8.1$ and the 
in-plane high $T_c$ cuprates result $C=\nicefrac{35}8\approx4.4$ \cite{Homes2005}. The
error bound may be retrieved from \cite{Homes2004} by converting from the dimensionful
constant in units of $\nicefrac{\text{cm}^{-1}}{\Omega^{-1}\text K}$ to our dimensionless unit
system\footnote{In the data analyzed in \cite{Homes2004} the unit of Homes' constant
  is given by
  \begin{equation*}
    [C]=\frac{\left[\rho_s\right]}{\left[\sigma_{\text{DC}}\right]\left[T_c\right]}
       =\frac{\text{cm}^{-2}}{\Omega^{-1}\text{cm}^{-1}\text K}
       =\frac{\text{cm}^{-1}}{\Omega^{-1}\text K}.
  \end{equation*}
  To convert to a dimensionless unit system used in our holographic system one needs to
  introduce the natural constants, \eg for the conversion of the temperature we have
  \begin{equation*}
    T[\text K]=\frac{c\cdot h}{K_{\text B}}\cdot 100\;T\left[\text{cm}^{-1}\right],
  \end{equation*}
  which amounts to $1\,\text K=0.695\,\text{cm}^{-1}$. Similarly,
  $1\,\Omega^{-1}\text{cm}^{-1}=4.935\,\text{cm}^{-1}$ and our final conversion factor reads
  $1\,\Omega^{-1}\text K=3.42983\,\text{cm}^{-1}$. Thus, the values given in
  \cite{Homes2004} are converted by
  \begin{equation*}
   (120\pm25)\frac{\text{cm}^{-1}}{\Omega^{-1}\text K}
   =\frac{120\pm25}{3.43}\approx35\pm7.3.
  \end{equation*}
  Taking into account the correction factor for our different definition of $\rho_s$ we
  arrive at $C=\nicefrac{35\pm7.3}8\approx4.4\pm0.9$.} and yields $C=4.4\pm0.9$. In
fact the value of $C\sim 6.2$ presented in \figref{fig:homeslaw-const}, seems to
be almost the arithmetic mean of the two experimentally determined values. Additionally,
one may compare to the most recent results found for organic superconductors in
\cite{Dordevic2013}, \ie $C=(110\pm60)\,\nicefrac{\text{cm}^{-1}}{\Omega^{-1}\text K}$,
again in dimensionful units. Converting to our dimensionless Homes' constant and
including the additional factor of $8$, we find $C=4\pm2.1$, which is very close to the
original result in \cite{Homes2004}.\\[\baselineskip]
Homes' relation appears to hold for high values of
$\nicefrac\lambda\mu\gtrsim4.5$. This is the regime where the Drude peak in the
optical conductivity becomes very broad such that the Drude regime, the intermediary
frequency regime and the conformal regime are shrinking together, indicating a mixing of
IR and UV degrees of freedom. One observation from the finite temperature thermodynamic
phase diagram, \cff\figref{fig:helix-finite-t-phase-trans-phase-diag}, is that for
such high values of $\nicefrac\lambda\mu$ we may hit a quantum critical point at a
critical value of $\nicefrac p\mu$. On the other hand, Homes' relation, as shown in the inset of
\figref{fig:homeslaw-const}, seems to work over a finite range of $\nicefrac p\mu$ beyond
the possible quantum critical point at $T=0$, which clearly calls for further investigation of the
zero temperature system. 
\\[\baselineskip]
Alternatively, according to the single scaling argument given in
\cite{2005PhRvL..95j7002P} Homes' relation seems to require two competing timescales. In
our system the helical lattice introduces an additional timescale for momentum
relaxation, controlled by $\nicefrac\lambda\mu$, which is very different from the
diffusive timescale in the original holographic s-wave superconductor
\cff\cite{Erdmenger2012a}, at least for small values of $\nicefrac\lambda\mu$. It is
compelling to speculate that in the large $\nicefrac\lambda\mu$ regime, where the
applicability of the Drude model may be problematic, these two timescales may become
almost identical. Let us stress that for a complete understanding of the aforementioned
scaling relations it is imperative to understand the zero temperature phases of the
helical system. Nonetheless, the optical conductivity with its broad Drude peak
resembles the dirty limit BCS superconductors, where Homes' relation follows naturally
from the missing spectral weight argument: due to the broad peak, we may think of the
missing spectral weight area roughly as a square spanned by $\sigma_{\text{DC}}$ and the
width of the gap, which is set by the universal gap equation at $T=0$ to be a number
times $T_c$, see also \cite{PhysRevB.73.180504}. We comment on this further in our
discussion in section~\ref{sec:conclusion}. 

\section{Zero Temperature Solutions and Holographic RG Flows 
  \label{sec:zero-temperature}}


In order to solve the system at zero temperature and to understand its zero temperature
phase diagram and quantum phase transition structure, it is necessary to identify the
correct infrared geometries. Classifying all possible IR geometries is in general a
complicated task which can only be done by restricting to certain Ans\"atze and symmetry
requirements, but within that class may yield interesting physical insights
\cite{Charmousis2010,Gouteraux:2012yr,Donos:2014oha}. As a possible candidate for such a
solution, we now generalize the insulating geometry of \cite{Donos2013c} to the case of
an additional massless charged scalar. We want to emphasize that this insulating
geometry is different from the usual gapped AdS-Soliton geometry: Instead of the
holographic direction ending at some particular point, the IR of this solution is an
anisotropic hyper-scaling violating Lifshitz throat. This anisotropy forces the system
to be a smectic material, \ie an insulator in the direction of the helix (the $x$
direction), and a metal in the other two orthogonal directions.
\\[\baselineskip]
For reasons explained in Section \ref{ssec:holographic-setup} we work with a
non-vanishing Chern-Simons coupling $\kappa=\nicefrac1{\sqrt 2}$ (but still
$m=m_\rho=0$).\footnote{One reason is that it seems harder to find IR scaling geometries
  for non vanishing masses. We plan to return to this question in the near future
  \cite{WIP}.}
The solution can be written as a power series in $r^{\nicefrac13}$ with the
leading terms being
\begin{equation}
  \begin{aligned}
    w&=w_0+w_1r^{\nicefrac43}+\dotsb, & \rho&=\rho_0+\rho_1r^{\nicefrac43}+\dotsb, & 
    a&=a_0r^{\nicefrac54}+\dotsb, \\
    \e[v_1]&=\e[v_{10}]r^{-\nicefrac13}+\dotsb, & \e[v_2]&=\e[v_{20}]r^{\nicefrac23}+\dotsb, & 
    U&=\frac{18}5r^2+\dotsb \\
    \e[v_3]&=\e[v_{30}]r^{\nicefrac13}+\dotsb. & &
  \end{aligned}
  \label{eq:app-t-zer-power}
\end{equation}
The coefficients of this expansion can be expressed in terms of the parameters
$\rho_0$, $v_{20}$ and $v_{30}$.\footnote{There is an additional free parameter, namely
  the expansion point $r_+$, which has been set to zero for simplicity. It can be
  reinstated by shifting $r \to r - r_+$.} In particular, it follows from the equations
of motion that
\begin{equation}
  \begin{aligned}
    a_0&=\frac{36\kappa\sgn(p)\e[2v_{20}-2v_{30}]}{5\left(6\kappa^2+q^2\rho_0^2-4\right)},
    & w_0&=\sqrt 3\e[2v_{30}-v_{20}], & \e[v_{10}]&=\frac12|p|\e[v_{30}-v_{20}], \\
    \rho_1&=-\frac{\kappa^2q^2\rho_0\e[4v_{20}-4v_{30}]}
    {\left(6\kappa^2+q^2\rho_0^2-4\right)^2}, & w_1&=\frac{\sqrt 3\left(
        q^2\rho_0^2-4\right)\e[3v_{20}-2v_{30}]}{2\left(6\kappa^2+q^2\rho_0^2-4\right)}.
  \end{aligned}
  \label{eq:zero-temperature-initial}
\end{equation}
This fixed point describes a cohesive IR geometry with a superconducting order parameter
turned on. Note that the charged scalar is not subleading or leading compared to the
original geometry without it, it rather has the same IR behavior as the helix field
$w$.\\[\baselineskip]
In order to understand whether this fixed point is stable under perturbations, we follow
\cite{Donos2013c} and calculate power law perturbations around
\eqref{eq:app-t-zer-power} by writing
\begin{equation}
  \begin{aligned}
    w&=w_0+w_1r^{\nicefrac43}(1+c_wr^{\delta}), & 
    \rho&=\rho_0+\rho_1r^{\nicefrac43}(1+c_\rho r^\delta), &
    a&=a_0r^{\nicefrac54}(1+c_ar^{\delta}), \\
    v_1&=v_{10}+\log(r^{-\nicefrac13})+c_1r^\delta, &
    v_2&=v_{20}+\log(r^{\nicefrac23})+c_2r^\delta, &
    U&=\frac{18}5r^2(1+c_Ur^\delta), \\
    v_3 &= v_{30}+\log(r^{\nicefrac13})+c_3 r^\delta. & &
  \end{aligned}
  \label{eq:zero-temperature-full-expansion}
\end{equation}
The equations of motion are linearized in the perturbations and solved to leading order
in $r$. All scaling exponents and the corresponding eigenvectors for the radial
perturbations around \eqref{eq:zero-temperature-full-expansion} are listed in Appendix
\ref{ssec:exponentsCondensed}. In summary, we find that the condensed insulating solution
\eqref{eq:zero-temperature-full-expansion} does not show any condensation instabilities
in which some of the IR operator dimensions $\delta$ violate the Breitenlohner-Freedman
bound by becoming complex. Instead, we find two IR irrelevant deformations, \ie
deformations with explicitly positive exponent, namely the $\delta_+$ mode of point 6
and 7 in Appendix \ref{ssec:exponentsCondensed},
\begin{align}
  \delta_1&=\frac16\left(\sqrt{145}-5\right), &
  \delta_2&=\frac16\left(\sqrt{185}-5\right).
  \label{eq:deformation-exponents}
\end{align}
These perturbations will be useful in generating the RG flows to the UV by shooting
numerically from the IR fixed point perturbed with these deformations (\cff Appendix
\ref{app:helix-numerical-method} for more details). These exponents are the same as the
ones  found in \cite{Donos2013c}. As will be explained in detail in Appendix
\ref{app:helix-numerical-method}, these two modes are sufficient to generate the
two-parameter family of zero temperature RG flows labeled by the chemical potential
$\nicefrac\mu p$ and the lattice strength $\nicefrac\lambda p$. 
\\[\baselineskip]
Besides the above superconducting IR geometry our model admits, at least for large
enough Chern-Simons couplings such as our choice $\kappa=\nicefrac1{\sqrt
  2}$,\footnote{For Chern-Simons couplings smaller than the critical value $\kappa_c
  \approx 0.57$ another unstable IR scaling fixed point appears \cite{Donos2013c}, which
  complicates the phase structure at zero temperature. Here we discuss only the simpler
  case of large $\kappa$.} two other IR fixed points: For a vanishing charged scalar,
there is a metallic $\text{AdS}_2\times\mathds R^3$ fixed point dominating for larger $p$
\cite{Donos2013c}, whose geometry including perturbations reads
\begin{equation}
  \begin{aligned}
    w(r)&=c_w r^\delta, & \rho(r)&=c_\rho r^\delta, & a(r)&=2 \sqrt 6r(1+c_ar^\delta), \\
    v_1(r)&=v_{10}(1+c_1r^{\delta}), &\quad v_2(r)&=v_{20}(1+c_2r^{\delta}), &\quad
    U(r)&=12{r^2}(1+c_Ur^{\delta}), \\
    v_3(r)&=v_{30}(1+c_3r^\delta). &&&&
  \end{aligned}
  \label{eq:MetallicAdS}
\end{equation}
The metallic $\text{AdS}_2\times\mathds R^3$ geometry has several deformation exponents,
which are spelled out together with the corresponding eigenvectors in Appendix
\ref{ssec:exponentsMetallic}. Here we focus solely on the condensation
instabilities. In particular, there are two modes corresponding to scalar condensation in
this $\text{AdS}_2$ near horizon geometry, with scaling exponents
\begin{equation}
  \delta_\pm=\frac16\left(-3\pm\sqrt{3(3-2q^2+{m_\rho}^2)}\right).
  \label{eq:AdS2scalarexp}
\end{equation}
If these exponents become complex, the charged scalar destabilizes the $\text{AdS}_2$,
and the system presumably flows to the superfluid IR geometry above. Furthermore, there
are two exponents connected to the condensation of the helix field,
\begin{equation}
  \delta_\pm=-\frac12\pm\frac{\sqrt{\left(\left(m^2+3\right)+p^2\e[-2v_{10}]
        -2\sqrt6\kappa p\e[v_{10}-2v_{20}]\right)}}{2\sqrt3}.
  \label{eq:AdS2wexp}
\end{equation}
Because the lattice is explicitly introduced, the crucial aspect for the physics is now
whether the exponent becomes relevant. This happens when $\delta_+<0$, see
\cite{Donos2013c}. In that case the system will flow to the insulating geometry. If
the exponents become complex, then the $\text{AdS}_2$ geometry can spontaneously destabilize
to the insulator, but we will not consider this particular case.
%
%
The insulating geometry of \cite{Donos2013c}, given by \eqref{eq:app-t-zer-power} with
the charged scalar $\rho$ switched off, is also unstable towards condensation of the
charged scalar within the system \eqref{eq:helix-action}, although in a slightly
different way. Analysing the radial perturbations for the case of vanishing scalar mass
$m_\rho=0$ one finds an additional mode for the charged scalar alone,
\begin{equation}
  \delta\rho=c_0+c_1r^{-\nicefrac53}.
  \label{eq:rhomodeDH}
\end{equation}
%
%
%
If the charged scalar had a non vanishing mass, its exponents would change from
\eqref{eq:rhomodeDH} to
\begin{equation}
  \delta\rho=c_0r^{-\frac56\left(1-\sqrt{1+\frac25m_\rho^2}\right)}
  +c_1r^{-\frac56\left(1+\sqrt{1+\frac25m_\rho^2}\right)}.
  \label{eq:rhomodeDHmass}
\end{equation}
Note that the IR dimension of the charged scalar in the insulating background of
\cite{Donos2013c} is independent of its charge $q$, due to the cohesive nature of the
extremal horizon. In the regime
\begin{equation}
  -4\leq m_\rho^2<-\frac52,
  \label{eq:rhomodeDHBFviolation}
\end{equation}
the charged scalar obviously violates the IR Breitenlohner-Freedman bound while
preserving the UV Breitenlohner-Freedman bound, and the condensation mechanism will be
analogous to the metallic $\text{AdS}_2\times\mathds R^3$ case. On the other hand, for
the massless case \eqref{eq:rhomodeDH}, no condensation instability is found. In this case,
condensation can still happen thermodynamically if the condensed zero temperature RG
flow obtained from the IR geometry \eqref{eq:app-t-zer-power} has a lower free energy
compared to the uncondensed one (Eq.\@ \eqref{eq:app-t-zer-power} with $\rho_0=0$). We
numerically constructed the holographic RG flow geometries up to the asymptotic AdS
boundary for both the insulating and superconducting fixed points for a certain range in
parameter space, and confirmed that they have lower free energy. For completeness, we
collect all the operator dimensions and the corresponding modes for each fixed point in
Appendix \ref{app:exponents}.

\section{Discussion and Outlook \label{sec:conclusion}}

In this work we analysed the transition to s-wave superconductivity in an anisotropic
five-dimensional holographic model with a helical Bianchi $\text{VII}_0$ symmmetry. This
corresponds to a 3+1 dimensional field theory in the presence of a helical lattice
\cite{Donos2013c}. The advantage of this model is that it allows us to cleanly separate
the IR dynamics in the system. This is hard to identify in the simplest holographic
superconductors for two reasons: Due to translation invariance there is already in the
normal phase a delta peak at zero frequency in the conductivity. In the superconducting
phase this mixes with the protected fluctuations of the order parameter. Secondly, most
well-known examples of holographic superconductors are accompanied by a remaining
gapless Lifshitz sector in the IR that mixes dynamically with the order parameter
physics. This is especially so at finite temperature. 
We improved on the former point by explicitly breaking translation invariance along one
of the field theory directions using the above-mentioned Bianchi $\text{VII}_0$ helix,
and on the latter by using the fact that this model \eqref{eq:helix-action} has an
anisotropic insulating ground state \cite{Donos2013c}. We established that this model
indeed undergoes a superconducting transition at low temperatures. Studying the optical
conductivity we can see that the IR dynamics is more cleanly controlled by order
parameter physics. This allowed us to extrapolate to a first holographic example where
Homes' relation holds. Let us discuss the physics of each of these points.

\subsection{Phases at Finite and Zero Temperature\label{ssec:concl-phases}}

\begin{figure}
  \centering
  \includegraphics[width=\textwidth]{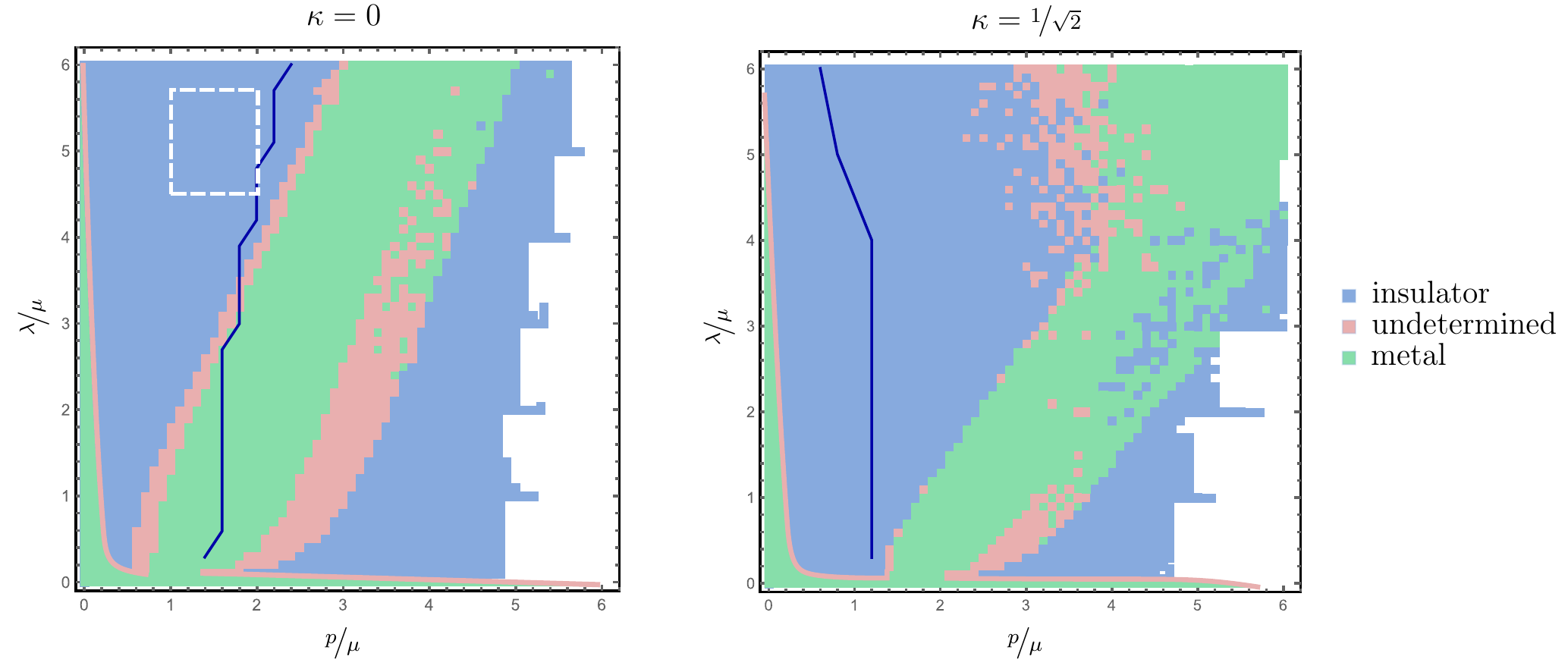}
  \caption{The nature of the zero-temperature ground state from
    the finite temperature conductivity. The surprise is that for fixed helix amplitude
    $\lambda$ the system transitions from an insulating to a metallic and then back to
    an insulating phase. For $\nicefrac p\mu\ll1$ and $\nicefrac\lambda\mu\ll1$, we
    expect a metallic phase designated by the shaded green area due to the fact that
    momentum relaxation is removed in the limit where either of these parameters
    vanishes. This part of the metallic phase could not be distinguished from the
    insulating phase since we are using very coarse measure to determine the nature of
    the ground state \textit{i.e.}\@ a qualitative measure of the conductivity. The
    thick {\color{darkblue}blue} line denotes the location of the minimal critical
    temperature $T_c^{\text{min}}$ extracted from
    \figref{fig:helix-finite-t-phase-trans-phase-diag}. In the case $\kappa=0$, shown in 
    the left panel, this minimum tracks qualitatively the metallic phase inferred from
    the conductivity, whereas for $\kappa=\nicefrac1{\sqrt2}$ the minimal critical
    temperature $T_c^{\text{min}}$ is invariant under changes in $\nicefrac p\mu$ and
    $\nicefrac\lambda\mu$. Note that at high values for $\nicefrac\lambda\mu$ the
    critical temperature is very low and thus our numerical code cannot reach
    $T_c^{\text{min}}$ anymore. Homes' relation holds in the region marked
    by the white dashed box.\label{fig:optical-phase-diagram}}
\end{figure}

The phase diagram of the holographic helical Bianchi $\text{VII}_0$ lattice model is
quite rich and this is reflected in the ways it approaches superconductivity. For large
enough charge $q$ of the scalar order parameter, both the insulating phase at small
helix pitch as well as the metallic phase at large helix pitch 
are unstable towards condensation of the charged scalar.
The second order mean field superfluid transition typically happens at a critical
temperature $T_c(\nicefrac\lambda\mu,\nicefrac p\mu)$, but the data in 
\figref{fig:helix-finite-t-phase-trans-phase-diag} suggests that a quantum phase
transition between condensed and uncondensed phases is possible for larger values of
$\nicefrac\lambda\mu$, similar to the situation in a recently investigated axion-based
system \cite{Kim:2015dna}.\\[\baselineskip]
The curious aspect is that the critical temperature does not have a monotonic behavior
as a function of the helix parameters. Naïvely the presence of a lattice should form an
obstacle for s-wave superconductivity. This is true at very small helix
parameters. There $T_c$ decreases compared to the translationally invariant
system. However, for a given value of the amplitude $\lambda$ there is a critical value
of the helix pitch $p$ beyond which $T_c$ starts to rise again. In the presence of the
Chern-Simons coupling, $\kappa=\nicefrac1{\sqrt 2}$, $T_c$ can even increase beyond its
isotropic value for very large $\nicefrac p\mu$. In the absence of the Chern-Simons
coupling, arguably, the tendency to return to its original homogeneous and isotropic
value for large $\nicefrac p\mu$ can be understood as the effect of the helix
diminishing if it rotates too fast around the x-axis: The valleys between the maxima
become so narrow that they do not influence the condensation dynamics any longer, and
homogeneity is approximately restored. It is an open question whether all observables
return to their homogeneous values at large $\nicefrac p\mu$, and at which rate.
\\[\baselineskip] 
In neither case, however, is the physics behind this behavior of $T_c$ very clear. There
is a strong indication, on the other hand, that it is correlated with the zero
temperature ground state of the system in the normal phase. We did not construct all of
these, but one can infer from the finite temperature optical conductivity qualitatively
whether the true ground state is insulating or conducting, see
\figref{fig:optical-phase-diagram}.\footnote{Note that the finite temperature solution is
  uniquely determined from the boundary conditions. It therefore already knows whether
  it originates from an insulating or a conducting zero-temperature geometry.} These do
indicate a second insulating phase occurs at large helix pitch $p$ or equivalently
small helix wavelengths. One now sees that there is a rough correlation between high
$T_c$ with an anisotropic insulating ground state in the normal phase and low $T_c$ and
a metallic ground state in the normal phase. The correlation is not exact,
however. Clearly, an independent analysis from thermodynamic quantities as well as a
complete calculation of the zero temperature phase diagram is required to establish this
concretely and unambiguously decide the fate of this new insulating phase.\footnote{We
  thank Aristomenis Donos for discussions on this point.} The correlation of the
behavior of $T_c$ with the zero-temperature normal phase ground states indicates that
the naïve insight that homogeneity is approximately restored is probably incorrect, as
then the system is expected to be in a conducting rather than an insulating phase.
%
A brief investigation into the possible zero temperature ground states, allowed us to
construct an IR geometry dual to the superconducting phase based on the original
insulating solution of \cite{Donos2013c}, \cff \figref{fig:rgflowsuco}. Interestingly,
the charged scalar shows the same approach to the IR as the helix field, indicating that
they might be able to compete in quantum phase transitions.
We analysed the static radial perturbations around these three IR fixed points, in order
to understand which  RG flows between them are allowed. The situation is summarised in
\figref{fig:rgflowsuco}: The metallic $\text{AdS}_2\times\mathds R^3$ IR geometry behaves
conventionally. It can be unstable towards either the insulating state and/or
superconductivity \cite{Hartnoll2008}. At the same time the condensed superconducting IR
geometry we constructed is nicely stable, indicating that it is the true ground state
\cite{Donos2013c}. The insulating IR geometry is indeed unstable towards
superconductivity, but curiously not for the mass of the scalar field considered here.
We suspect, however, that in this case the superconducting IR geometry, is still the
thermodynamically preferred ground state, \ie the state of lowest free energy. The
insulating but not superconducting geometry of \cite{Donos2013c} is hence dynamically
stable, but thermodynamically unstable. This would indicate that they are separated by a
first order transition. We will support this claim by an analysis of the thermodynamics
and transport at zero temperature in a forthcoming work \cite{WIP}.


\subsection{Transport\label{ssec:concl-transport}}

In our system, the linear momentum relaxation introduced by the Bianchi $\text{VII}_0$
structure of the geometry allows us to reliably analyse the physics behind the
low-frequency transport properties of our system. Our computation reveals that the
superconducting system is well described by a two-fluid Drude model at small frequencies
in the regime of weak momentum relaxation $\nicefrac\lambda\mu\ll 1$, a fact also
observed in the models of \cite{Andrade:2014xca,Kim:2015dna}. On the other hand, in the
regime of stronger momentum relaxation, $\nicefrac\lambda\mu\approx1$, the two-fluid
Drude model seems to work less and less well, again similar to
\cite{Andrade:2014xca,Kim:2015dna}.\\[\baselineskip]
It is the absence of a diamagnetic pole in the normal phase, that allows us to reliably
extract the superfluid density from the $\nicefrac1\omega$ pole in the imaginary part of
the optical conductivity in the superconducting phase. Naïvely the cohesiveness of the
superconducting phases in our system at zero temperature literally forces all charge
carriers in the system to condense into the matter fields outside the extremal black
hole horizon.\footnote{Note that a translationally invariant cohesive phase still can
  have a zero frequency delta function due to the presence of the charge density, as
  explicitly shown in \cite{Lippert:2014jma}.}
Since the $U(1)$ gauge field sourcing the helix itself is not charged under the
`charge' $U(1)$ and since the Chern-Simons term in \eqref{eq:helix-action} only induces
currents but not a charge density for the `helix' $U(1)$, one would expect all the
charge density (as carried by the `charge' $U(1)$) in the system at zero temperature to
be carried completely by the charged scalar dual to the superconducting order
parameter. The charge simply has no other place to go within this system. By comparing
the charge density in the superfluid phase at very low temperatures to the superfluid
density as extracted from the optical conductivity (\cff \figref{fig:helix-rhosf-of-T}),
we showed numerically that these indeed become identical for small lattice strengths
$\nicefrac\lambda\mu\ll1$. In more detail, we found that when the Drude model is a good
approximation to the optical conductivity at small frequencies, \ie when
$\nicefrac\lambda\mu\ll1$, the difference between the charge density at small
temperatures and the superfluid density is accounted for by the additional spectral
weight residing in a small residual Drude peak at low frequencies. The fact that we are
analyzing the system correctly is confirmed by the fact that the
Ferrell-Glover-Tinkham sum rule holds (\cff \figref{fig:helix-sum-rule}) when crossing
the phase transition from the normal to the superconducting phase, \ie that there is no
missing spectral weight in our system. This is to be contrasted with the translationally
symmetric case (the black line in the left panel of \figref{fig:helix-sum-rule}), in
which the sum rule fails spectacularly due to the non-accounted spectral weight
residing in the $\delta(\omega)$ poles related to momentum conservation in both the
normal and superconducting phases.\\[\baselineskip]
For weak lattice strengths $\nicefrac\lambda\mu\ll 1$, the model considered here is
therefore in several respects under better control compared to the simple holographic
superconductor analyzed \eg in \cite{Erdmenger2012a}. A puzzle appears for stronger
momentum relaxation $\nicefrac\lambda\mu\approx1$. Now the charge density and superfluid
density do not approach each other up to the small temperatures probed in our numerics,
and the difference cannot be accounted for any more by a normal fluid Drude
component. Even so, the FGT sum rule continues to hold. A possibility is that some of
the low frequency spectral weight gets transferred to intermediate frequencies rather
than the superfluid pole. Another explanation for the failure could be that our
identification of $\rho_s$ from the optical conductivity does not match with the
superfluid density as calculated from the transverse response via the magnetic/London
penetration depth, which is another important cross-check \cite{WIP}.
This deserves further study. Based on the new insulating phase found at high $\nicefrac
p\mu$, a distinct possibility is that the phase structure of the system is more
complicated and interesting for strong lattice potentials at high
$\nicefrac\lambda\mu$. It is precisely in this novel regime where
we can find a region in parameter space in which Homes' relation is valid to a good accuracy.


\subsection{About Homes' Relation\label{ssec:concl-homes}}

\begin{figure}[t]
  \includegraphics[width=\textwidth]{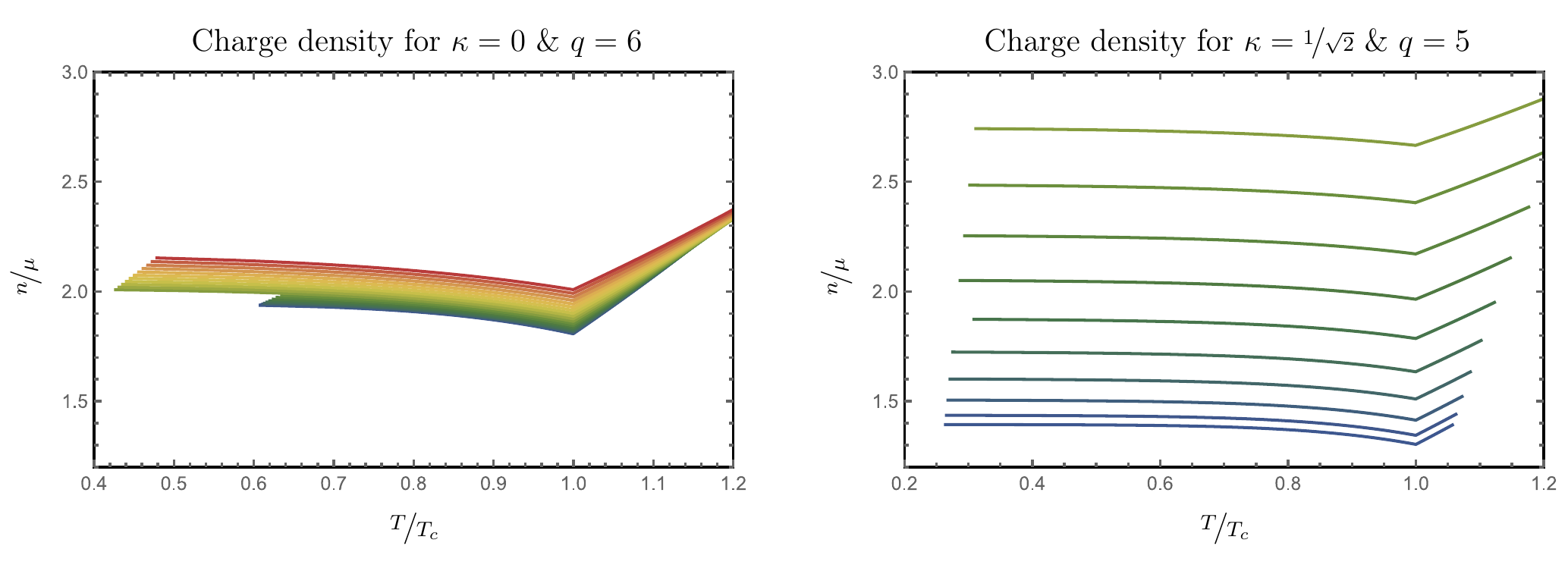}
\caption{The charge density as a function of $\nicefrac T{T_c}$ for $\nicefrac
    p\mu=0.4$ and for various values of $\nicefrac\lambda\mu$, using the same color
    coding as in \figref{fig:helix-finite-t-phase-trans-phase-diag}:
    $\color{DarkRainbow0p045}0.3$, $\color{DarkRainbow0p09}0.6$,
    $\color{DarkRainbow0p135}0.9$, $\color{DarkRainbow0p18}1.2$,
    $\color{DarkRainbow0p225}1.5$, $\color{DarkRainbow0p27}1.8$,
    $\color{DarkRainbow0p315}2.1$, $\color{DarkRainbow0p36}2.4$,
    $\color{DarkRainbow0p405}2.7$, $\color{DarkRainbow0p45}3$,
    $\color{DarkRainbow0p495}3.3$, $\color{DarkRainbow0p54}3.6$,
    $\color{DarkRainbow0p585}3.9$, $\color{DarkRainbow0p63}4.2$,
    $\color{DarkRainbow0p675}4.5$, $\color{DarkRainbow0p72}4.8$,
    $\color{DarkRainbow0p765}5.1$, $\color{DarkRainbow0p81}5.4$,
    $\color{DarkRainbow0p855}5.7$, $\color{DarkRainbow0p9}6$ in the case of $\kappa=0$
    (left panel) and $\color{DarkRainbow0p045}0.3$, $\color{DarkRainbow0p09}0.6$,
    $\color{DarkRainbow0p135}0.9$, $\color{DarkRainbow0p18}1.2$,
    $\color{DarkRainbow0p225}1.5$, $\color{DarkRainbow0p27}1.8$,
    $\color{DarkRainbow0p315}2.1$, $\color{DarkRainbow0p36}2.4$,
    $\color{DarkRainbow0p405}2.7$, $\color{DarkRainbow0p45}3$ for
    $\kappa=\nicefrac1{\sqrt2}$ (right panel). Curiously, the charge density at the
    phase transition seems to be minimal and hence $n_s(T<T_c)>n_n(T_c)$. Note that this
    is more prominent at vanishing Chern-Simons coupling $\kappa$. The decrease of the
    charge density in the normal phase can be attributed to the effects of the
    insulating phase, \textit{i.e.}\@ there are less charged degrees of freedom
    available for lower temperature. However, in the superconducting phase the system
    turns completely into the condensed state which is growing with lower
    temperatures. Therefore, the superconducting degrees of freedom are increasing for
    lower temperatures and may be drawn from the metallic directions, due to 
    partial restoration of the isotropy.\label{fig:charge-density-normal-super}}  
\end{figure}

{Homes' relation} \cite{Homes2004,Homes2005} -- the experimental result that for
high $T_c$ superconductors as well as conventional BCS superconductors, there is a
universal relation of the form
\begin{equation}
  \rho_s(T=0)=C\sigma_{\text{DC}}(T_c)\cdot T_c,
  \label{eq:Homes}
\end{equation}
with a nearly universal, material independent constant $C$ --- would follow naturally
from an argument based on the shortest possible time scale in a strongly coupled quantum
critical state \cite{Marel2003,Zaanen2004}, a so-called {\it Planckian dissipator}.
The basic idea is that in a quantum critical system there is no other scale than the
temperature, and hence the relaxation time responsible for the finite electric
conductivity in the system must depend on the inverse of the temperature alone, up to
numerical factors of $O(1)$. Holography is unique in its ability to model interacting
quantum critical systems and the notion of Planckian dissipation is clearly visible in
the universal holographic result for the ratio of shear viscosity over entropy density
\cite{Kovtun:2004de}. A holographic foundation for Homes' relation therefore has the
potential to confirm that a similar universal mechanism is at work in
superconductivity.\footnote{Of course, though actual transport in high $T_c$ cuprates
  does have Planckian dissipative features, it cannot be a pure quantum critical state,
  see \eg \cite{2005PhRvL..95j7002P,Hartnoll:2014lpa}.}
\\[\baselineskip]
In our system, however, the timescale is not the intrinsic timescale associated with the
strongly coupled dynamics of the system. Instead it is the external momentum relaxing
timescale introduced by the lattice. This is evident from the validity of the Drude
response, where $\sigma_{\text{DC}} \sim \tau_{\text{momentum}}$. The relaxation time
scale that controls the low-frequency conductivity is thus a function of the lattice
strength and helix pitch,
%
\begin{equation}
  \tau_{\text{relax}}=\frac{f(T/\mu,\nicefrac\lambda\mu,\nicefrac p\mu, \dots)}{T}.
  \label{eq:PlanckianDissipator}
\end{equation}
We have extracted an explicit factor of $T$ such that the function $f$ is dimensionless.
A Planckian dissipator would have a mostly constant function $f$ of order $O(1)$ in a
CFT. Clearly, for weak momentum relaxation, the relaxation time is typically very large,
$\tau T \gg 1$, non-universal and far from a Placnkian dissipator.
\\[\baselineskip]
The breaking of translation invariance and the introduction of an external scale is,
however, important in studying Homes' relation with gauge/gravity duality. As we
emphasized, in a translationally invariant system there is already an infinite
$\delta$-function contribution to the DC-conductivity. To extract the superfluid density
reliably, one needs to resolve this either into a Drude peak behaviour for weak breaking
or to something beyond
\cite{Hartnoll:2007ih,Hartnoll:2008hs,Lucas:2015vna}. Additionally the system should be
in a cohesive phase at low temperatures in order to prevent the existence of additional
IR charged degrees of freedom, hidden behind the extremal horizon. It was found in
\cite{Horowitz2012,Horowitz2012a,Horowitz2013} that a modulated chemical potential is
not sufficient to realise Homes' relation in the simple model considered there. In view
of the above, these works probably did not access the regime of strong momentum
relaxation that could potentially make Homes' relation work. For example, in
\cite{Horowitz2012,Horowitz2012a,Horowitz2013} the two-fluid Drude model works well for
all lattices considered, pointing to a regime in which momentum relaxation is 
weak. Instead, in this work a clearer picture emerges: Due to the cohesive nature of the
superconducting ground state, as well as the broken translation invariance in the helix
direction, we were able to eliminate, respectively, the additional charged degrees of
freedoms in the IR and the zero frequency delta function, hence obtaining a clearer
account of the reshuffling of charge density as well as spectral weight during the
superfluid transition. As noted already in \cite{Erdmenger2012a}, to successfully
analyse Homes' relation in a holographic model it is essential to be able to keep track of
all charged degrees of freedom at low/zero temperatures. Since our helical lattice model
admits an insulator/superconductor transition with a cohesive phase at low temperatures,
this setup allows for a fresh look at Homes' relation, addressing both of these points. 
\\[\baselineskip]
\figref{fig:helix-rhosf-of-T}, which shows the agreement between the superfluid density
and the total charge density in the limit $T\rightarrow 0$, illustrates this cleanly,
together with our verification that the Ferrell-Glover-Tinkham sum rule holds. 
The validity of the sum rule is of particular importance, as it can be used to obtain
Homes' relation if the underlying system is a Planckian dissipator \cite{Erdmenger2012a}.
With these results as well as \eqref{eq:PlanckianDissipator}, \eqref{eq:Homes} can be
reformulated 
as\footnote{\label{fn:EffectiveMasses}Note that compared to the standard condensed
  matter notation we absorbed the effective masses and numerical factors in the
  respective phases (which are not directly accessible in holographic models) into $n$
  by $n_ne^2/m^\ast\mapsto n_n$ and $ n_se^2/m^\ast\mapsto n_s$, \ie we used
  $\sigma_{\text{DC}}=n\tau$ and $\rho_s=n_s$. This redefinition gives $n_n$ and $n_s$
  units of $\text{(length)}^{-2}$, which coincides with charge and not number densities
  in five dimensions.}
\begin{equation}
  \frac{n_s(T=0)}{n_n(T=T_c)}=C(\nicefrac\lambda\mu,\nicefrac p\mu)
  f(T_c/\mu,\nicefrac\lambda\mu,\nicefrac p\mu,\dotsc).
  \label{eq:HomesReformulated}
\end{equation}
with $C$ the coefficient in Homes' relation.
%
In simple systems, with a single species of charge carriers and a reasonable gap in the
superconducting phase, the charge density at zero temperature in the superfluid phase is
approximately equal to the charge density at the critical temperature in the normal
phase. From \eqref{eq:HomesReformulated}, Homes' constant must hence vary inversely with
the function $f$ parametrising the momentum relaxation time scale,
\begin{equation}
  C \sim 1/f.
\end{equation}
In a \emph{bona fide} Planckian dissipator, $f$ would be a universal constant of $O(1)$,
and is in this way seen to explain Homes' relation. 
%
We also investigated the LHS of
\eqref{eq:HomesReformulated}. \figref{fig:charge-density-normal-super} shows that 
numerically the charge density in the system does not vary much between $T\ll T_c$ and
$T_c$.\footnote{\figref{fig:charge-density-normal-super} does indicates that for most
  choices of $\nicefrac\lambda\mu$ and $\nicefrac p\mu$ our system has slightly ($\sim
  5\dotso10\%$) more charged degrees of freedom in the superfluid than in the normal
  phase. While in the normal phase the system is anisotropic, the system becomes more
  isotropic in the superfluid phase due to the isotropic s-wave order parameter. An
  explanation for the rise could be that charged degrees of freedom from the metallic
  transverse directions also contribute to the charge density in the helix direction.}
Hence in combination with our results from Section \ref{ssec:drude-two-fluid} that the
charge density and superfluid density are approximately equal at low $T$, also the ratio
of charge densities in \eqref{eq:HomesReformulated} is seen to be close to unity. This is
so in the weak momentum relaxation regime where $\nicefrac\lambda\mu\ll 1$, but also for
rather large $\nicefrac\lambda\mu$, \ie strong momentum relaxation in which the Drude
model is not readily applicable. in this case Homes' relation may hold. As we found in
Section \ref{ssec:homes-relation}, explicitly computing the functional relation between
$\rho_s$ and $\sigma_{\text{DC}}(T_c)T_c$ we indeed find that in our holographic
superconductor in a helical lattice Homes' relation holds with coefficient 
\begin{align} 
  C\approx6.2\pm0.3
\end{align}
in the range of parameters 
\begin{alignat}{5}
  \frac\lambda\mu&\approx4.5,\dotsc,6, &\quad &\text{and} &\quad 
  \frac p\mu&\approx1,\dotsc,2, & \quad &\text{for} &\quad \kappa&=0.
  \label{eq:HomesRegime} 
\end{alignat}
It can, however, not be explained as a consequence of the simple ratio of charge
densities at $T_c$ and $T=0$. This is also the regime where a difference between
$\rho_s$ and $n_s$ opens (\cff \figref{fig:helix-rhosf-of-T}), and the simple relation
between \eqref{eq:HomesReformulated} and \eqref{eq:Homes} breaks down.
\\[\baselineskip]
As we now argue, the fact that Homes' relation does hold in the parameter
range given above may on the contrary be due to a strong deviation
from the simple Planckian dissipation behavior.\footnote{We are grateful to
  E.\@ Kiritsis for pointing out this possibility.}
We may parametrize the difference between $\rho_s$ and $n_s$ as
\begin{equation}
  \rho_s(T=0)=n_s(T=0)+\delta\rho_s(T=0).
  \label{eq:rhosnsgap}
\end{equation}
Furthermore, let us assume\footnote{We make this assumption in hindsight of the
  parameter regime in which Homes' relation is valid (see below) to be not too far into the
  regime of strong scattering, such that this Drude-like approximation still should
  yield reasonable results. Also, it is reasonable to assume that a system conducts
  better if more charged degrees of freedom are present.}
the DC conductivity is still roughly proportional to the charge density times a
relaxation time,
\begin{equation}
  \sigma_{DC}=n_n\tau.
  \label{eq:DCcond}
\end{equation}
Substituting Homes' relation on the LHS of \eqref{eq:rhosnsgap} together with
\eqref{eq:DCcond}, and using that the ratio \eqref{eq:HomesReformulated} is close to
unity, we arrive at
\begin{equation}
  Cf(T_c,\nicefrac\lambda\mu,\nicefrac p\mu,\dotsc)
  =1+\frac{\delta\rho_s(T=0,\nicefrac\lambda\mu,\nicefrac p\mu,\dotsc)}
  {n_n(T_c,\nicefrac\lambda\mu,\nicefrac p\mu,\dotsc)}.
  \label{eq:HomesReformulated2}
\end{equation}
Then, for Homes' relation \eqref{eq:Homes} to be valid and $C$ to be a univeral constant of
$O(1)$, we see that the gap $\delta \rho_s$ and the function $f$ cannot vary
independently from each other, but must conspire. As $\delta \rho_s$ varies with the
parameters, $f$ cannot be constant, and hence the system is not a Planckian
dissipator. A more detailed investagation of the function $f$ directly as well as other
relaxational scales is hence of great interest, 
as well as how the gap $\delta\rho_s$ behaves in the regime of strong(er)
scattering. Since this is no longer Drude physics, it requires a more detailed response
analysis to extract these. We do note that the regime of $\nicefrac p\mu$ and
$\nicefrac\lambda\mu$ where we find Homes' relation to hold
is near the apparent insulator-metallic quantum phase transition in the normal phase,
see \figref{fig:optical-phase-diagram}. It will hence be very interesting to investigate
in more detail the behavior of the above quantities in the zero temperature ground
states \cite{WIP}.\\[\baselineskip]
To conclude, we would like to emphasize that in our model we have found an example of
the validity of Homes' relation in a holographic model with strong momentum
relaxation. This is similar to the experimental result \cite{Homes2004,Homes2005} in
which the c-axis high $T_c$ cuprates as well as the dirty limit BCS superconductors,
both materials with strong scattering and momentum relaxation, follow the same Homes'
relation. 
As explained above, in the strong momentum relaxation regime there is a nontrivial
difference between $\rho_s$ and $n_s$ at low temperatures, whose origin needs to be
investigated in more detail \cite{WIP}.
An important consistency check will be to compute the superfluid strength not from the
longitudinal response, \ie the plasma frequency and the Drude weight, but from the
transverse response by determining the magnetic/London penetration depth. This requires
to solve for the transverse propagator by finding a solution at small but non-zero
momentum. Indeed, an analysis of the dynamical conductivity
$\sigma_{xx}(\omega,\vec{k})$ and possible finite momentum instabilities \cite{WIP2}
will allow us to determine the spectrum of quasi-particle excitations, and also whether
the superconducting phase \eqref{eq:app-t-zer-power} is really the thermodynamically
preferred ground state in our system. There is another interesting aspect to consider,
that is the case of spontaneously generated helical ground states. Similar to the case
of spontaneous generated charge density waves
\cite{Nakamura:2009tf,Donos:2011bh,Withers:2013loa,Ling:2014saa} this might even
dynamically fix the preferred helix pitch $p$, leaving fewer free UV
parameters. If Homes' relation hold for all such spontaneous helix
models, then one would have a more satisfying reason to explain its universality.
At this stage, all these results are still preliminary and for $\kappa=0$ only, and we
plan to analyze the question under which exact conditions Homes' relation holds  and
related questions, in a future work \cite{WIP}.
%


\acknowledgments

We would like to thank Aristomenis Donos, Blaise Gout\'eraux, Sean Hartnoll, Elias
Kiritsis and Jan Zaanen for helpful discussions during the course of this work. SK and
KS are very grateful to Department of Physics at Harvard University for extensive
hospitality. KS is supported in part by a VICI grant of the Netherlands Organization for
Scientific Research (NWO), by the Netherlands Organization for Scientific
Research/Ministry of Science and Education (NWO/OCW) and by the Foundation for Research
into Fundamental Matter (FOM). SK is supported by a grant from the John Templeton
foundation. The opinions expressed in this publication are those of the authors and do
not necessarily reflect the views of the John Templeton foundation. The work of RM was
supported by World Premier International Research Center Initiative (WPI), MEXT, Japan.

\appendix

\section{Equations of Motion for S-Wave  Superconductors on a
  Helical Lattice\label{app:eq-helix-background}}

The equations of motion for the metric fields $U, v_1, v_2, v_3$, the Maxwell
fields $A$ and $B$, and the scalar field $\rho$ following from the
action \eqref{eq:helix-action} are
\begin{equation}
  \begin{aligned}
    0={}&a''+a'\left(v_1'+v_2'+v_3'\right)-\frac{2aq^2\rho^2}U+\kappa p\e[-v_1-v_2-v_3]ww', \\
    0={}&w''+w'\left(\frac{U'}U+v_1'-v_2'+v_3'\right)+\frac wU
    \left(\kappa p\e[-v_1+v_2-v_3]a'-m^2-p^2\e[-2\left(v_1-v_2+v_3\right)]\right), \\ 
    0={}&2\rho^2\left(m_{\rho }^2-\frac{a^2q^2}U\right)+a'^2+w^2\left(m^2\e[-2 v_2]
      +p^2\e[-2 \left(v_1+v_3\right)]\right)+4p^2\e[-2 v_1]\sinh^2\left(v_2-v_3\right) \\ 
    &+2U\left(v_1'+v_2'+v_3'\right)-U\left(2\rho'^2+\e[-2 v_2]w'^2
      -4v_1'v_2'-4v_1'v_3'-4v_2'v_3'\right)-24, \\ 
    0={}&2\rho^2\left(\frac{m_{\rho }^2}{U}-\frac{a^2q^2}{U^2}\right)
    -\frac{a'^2}U+\frac{w^2}U\left(m^2\e[-2 v_2]-p^2\e[-2 \left(v_1+v_3\right)]\right) \\  
    &+\frac{p^2}U\left(-2\e[-2 v_1]+3\e[-2\left(v_1+v_2-v_3\right)]
      -\e[-2\left(v_1-v_2+v_3\right)]\right)+2\rho'^2+\frac{2U''}U \\  
    &+4\left(\frac{U'}U\left(v_1'+v_2'\right)+v_1'^2+v_2'v_1'+v_2'^2\right)
    -\frac{24}U+\e[-2 v_2]w'^2+4\left(v_1''+v_2''\right), \\ 
    0={}&\frac{2w^2}U\left(p^2\e[-2\left(v_1+v_3\right)]-m^2\e[-2v_2]\right)
    +\frac{4p^2}U\left(\e[-2\left(v_1-v_2+v_3\right)]
      -\e[-2\left(v_1+v_2-v_3\right)]\right) \\ 
    &+\frac{4U'}U\left(v_3'-v_2'\right)-2\e[-2 v_2]w'^2
    +4\left(-v_2'^2-v_1'v_2'+v_3'^2+v_1'v_3'\right)+4\left(v_3''-v_2''\right), \\ 
    0={}&\frac{p^2}U\left(\e[-2 v_1]-\e[-2\left(v_1+v_2-v_3\right)]\right)
    +\frac{U'}U\left(v_3'-v_1'\right)-v_1'^2+v_3'^2-v_1'v_2'+v_2'v_3'-v_1''+v_3'', \\ 
    0={}&\rho''+\rho'\left(\frac{U'}U+v_1'+v_2'+v_3'\right)
    +\rho\left(\frac{a^2q^2}{U^2}-\frac{m_{\rho }^2}U\right).  
  \end{aligned}
  \label{eq:appendix-helix-background-eoms}
\end{equation}
The third equation is first order and originates from the $rr$-component of the
Einstein equations. There is another second order equation
\begin{equation}
  \frac{a^2q^2\rho^2}{U^2}+\rho'^2+\frac12\e[-2v_2]w'^2+v_1'^2+v_2'^2+v_3'^2
  +v_1''+v_2''+v_3''=0,
  \label{eq:appendix-helix-background-eom-further-second-order-eq}
\end{equation}
which follows from the above equations. Therefore, the first order equation is a
constraint. In the above equations, we have made use of the $U(1)$ symmetry
associated with $A_\mu$ to choose $\rho$ real.


\subsection{Linear Response for S-Wave Superconductors on a
  Helical Lattice\label{ssec:appendix-helix-fluct}}

We consider linearized fluctuations around the background solution. Therefore, we write
\begin{equation}
  \begin{alignedat}{2}
    g_{\mu\nu}&=g^{b}_{\mu\nu}+h_{\mu\nu}, &\qquad\qquad A_{\mu}&=A^{b}_{\mu}+A^{f}_{\mu}, \\
    B_{\mu}&=B^{b}_{\mu}+B^{f}_{\mu}, & \rho&=\rho^b+\rho^f,
  \end{alignedat}
  \label{eq:helix-appendix-fluct-all}
\end{equation}
where the fields with a superscript `$b$' denote the background solutions, $h$ denotes
the metric fluctuation, and fields with a superscript `$f$' denote the matter
fluctuations. The metric and the vector fields are expanded in the basis 
$(\dd t,\dd r,\omega_1,\omega_2,\omega_3)$. The background fields are $r$-dependent only
and are written in the Ansatz of Eqs.\@ \eqref{eq:helix-ansatz-A},
\eqref{eq:helix-ansatz-B} and \eqref{eq:helix-metric-ansatz}. The fluctuation fields are
chosen to depend on $r$ and $t$ because the retarded Green function leading to the
conductivity is evaluated at zero spatial momentum, \cf
Eq.\@ \eqref{eq:helix-kubo-form}. Expanding the action of Eq.\@ \eqref{eq:helix-action} to
second order in the fluctuations, we obtain an action $S_{q}$ that determines the
linearized equations of motion. The term linear in the fluctuations vanishes upon use of
the equations of motions for the background fields. Analyzing the action $S_{q}$, we can
determine which fields couple to each other. The result of this analysis is summarized
in \tabref{tab:appendix-helix-matrix-of-couplings} on
page \pageref{tab:appendix-helix-matrix-of-couplings}. The block of fluctuations which
contains $A^f_1$ and decouples from all other fluctuations is 
\begin{equation}
  \left(\mathcal A\equiv A^f_1,\mathcal B\equiv B^f_3,h_{t1},h_{23},h_{r1}\right).
  \label{eq:helix-appendix-relevant-block}
\end{equation}
 It is more convenient to work with the fields
\begin{align}
  \mathcal E&=(g^b)^{11}\;h_{t1}=\e[-2v_1]h_{t1}, \\
  \mathcal F&=(g^b)^{22}\;h_{23}=\e[-2v_2]h_{23},
  \label{eq:helix-metric-fields-redef}
\end{align}
instead of $h_{t1}$ and $h_{23}$. The reason is that $\mathcal E$ and $\mathcal F$ have
a finite limit for $r\to\infty$, whereas $h_{t1}$ and $h_{23}$ are proportional to $r^2$
for large values of $r$. The equations of motion for the fluctuations are obtained by
varying the action $S_q$. After variation, we can set $h_{r1} = 0$ choosing \emph{radial
  gauge} in which all radial field components vanish. Carrying out a Fourier transform
of the time coordinate, \ie choosing a harmonic time dependence $\e[-\ci\omega t]$, we
obtain the following linearly coupled ordinary differential equations in $r$ for the
fluctuation fields $\mathcal A$, $\mathcal B$, $\mathcal E$, and $\mathcal F$:
\begin{equation}
  \begin{aligned}
    0={}&\mathcal A''+\mathcal A'\left(\frac{U'}U-v_1'+v_2'+v_3'\right)
    +\mathcal A\left(\frac{\omega^2}{U^2}-\frac{2q^2\rho^2}U\right) \\  
    &+\frac{\ci\mathcal B\kappa\e[v_1-v_2-v_3]\omega w'}U
    -\frac{\kappa p\mathcal E\e[v_1-v_2-v_3]ww'}U+\frac{\e[2v_1]a'\mathcal E'}U, \\
    ={}&\mathcal B''+\mathcal B'\left(\frac{U'}U+v_1'+v_2'-v_3'\right)
    +\mathcal B\left(\frac{\kappa pa'}U\e[v_3-v_1-v_2]-\frac{m^2}U
      -\frac{p^2}U\e[2\left(v_3-v_1-v_2\right)]+\frac{\omega^2}{U^2}\right) \\
    &+\frac{\kappa}U\left(\mathcal Fp\e[-v_1+v_2-v_3]wa'
      -\ci\mathcal A\e[-v_1-v_2+v_3]\omega w'\right)
    -\frac{\mathcal Fp^2w}U\left(\e[-2 v_1]+\e[-2\left(v_1-v_2+v_3\right)]\right) \\
    &+2\mathcal Fw'\left(v_3'-v_2'\right)+\frac{\ci p\mathcal Ew\omega}{U^2}
    -\mathcal F'w' \\
    0={}&\mathcal F''+\mathcal F'\left(\frac{U'}U+v_1'+3v_2'-v_3'\right)
    -\frac{\ci p\mathcal E\e[2 v_3-2 v_2]\omega}{U^2}
    +\frac{\ci p\mathcal E\omega}{U^2}+\e[-2 v_2]\mathcal B'w' \\
    &+\mathcal F\bigg[\frac{\omega^2}{U^2}-\e[-2v_2]w'^2-\frac{m^2\e[-2 v_2]w^2}U
    -\frac{2p^2}U\left(\e[-2v_1]+\e[-2\left(v_1+v_2-v_3\right)]\right)\bigg] \\
    &+\frac{\mathcal Bm^2\e[-2v_2]w}U-\frac{\mathcal Bp^2\e[-2\left(v_1+v_2\right)]w}U \\ 
    0={}&\mathcal E'+\mathcal A\e[-2v_1]a'+\frac{\ci\mathcal FpU}\omega
    \left(\e[-2\left(v_1+v_3\right)]ww'+2\e[-2v_1]v_2'-2\e[-2 v_1]v_3'\right) \\
    &+\frac{\ci pU}\omega\left(\mathcal B\e[-2\left(v_1+v_2\right)]w'
      -\e[-2\left(v_1+v_3\right)]w\mathcal B'\right)+\frac{\ci pU\mathcal F'}\omega
    \left(\e[-2v_1]-\e[-2\left(v_1-v_2+v_3\right)]\right).
  \end{aligned}
  \label{eq:appendix-helix-fluct-eom} 
\end{equation}
The first three equations are second order. The last one -- the constraint -- is
first order and originates from the metric mode $h_{r1}$ after choosing radial
gauge. There is a forth second order equation,
\begin{align}
  0={}&\mathcal E''+\mathcal E'\left(3v_1'+v_2'+v_3'\right)+\e[-2v_1]a'\mathcal A'
  +\frac{2a\mathcal Aq^2\rho^2\e[-2v_1]}U \notag\\ 
  &+\mathcal E\frac{p^2}U\left(2\e[-2v_1]-\e[-2\left(v_1+v_3\right)]w^2
    -\e[-2\left(v_1+v_2-v_3\right)]-\e[-2\left(v_1-v_2+v_3\right)]\right) \notag\\
  &+\ci\frac{p\omega}U\left(\mathcal B\e[-2\left(v_1+v_3\right)]w
    -\mathcal F\e[-2v_1]+\mathcal F\e[-2\left(v_1-v_2+v_3\right)]\right),
\label{eq:appendix-helix-fluc-add-eom}  
\end{align}
which follows from the above equations and the equations of motion for the
background fields. The background equations of motion have been used to
eliminate second order derivatives of background fields in the above equations. 

\begin{sidewaystable}[hp]
  \centering
  \small
  \begin{tabular}{@{}S{2em}*{28}{C@{\hspace{1em}}}@{}}
    \toprule
    & h_{tt} & h_{rr} & h_{11} & h_{22} & h_{33} & h_{tr} & {\color{darkred}h_{t1}}
    & h_{t2} & h_{t3} & {\color{darkred}h_{r1}} & h_{r2} & h_{r3} & h_{12} & h_{13}
    & {\color{darkred}h_{23}} & A^f_t & A^f_r & {\color{darkred}A^f_1} & A^f_2
    & A^f_3 & B^f_t & B^f_r & B^f_1 & B^f_2 & {\color{darkred}B^f_3} & \rho^f 
    & \left(\rho^f\right)^\ast \\
    \midrule
    h_{tt} & \bullet & \bullet & \bullet & \bullet & \bullet & \bullet 
    &  &  &  &  &  &  &  &  &  & \bullet & \bullet &  &  &  &  
    &  &  & \bullet &  & \bullet & \bullet \\
    h_{rr} &  & \bullet & \bullet & \bullet & \bullet & \bullet &  &  
    &  &  &  &  &  &  &  & \bullet & \bullet &  &  &  &  &  &  
    & \bullet &  & \bullet & \bullet \\
    h_{11} &  &  & \bullet & \bullet & \bullet & \bullet &  &  &  &  
    &  &  &  &  &  & \bullet & \bullet &  &  &  &  &  &  & \bullet
    &  & \bullet & \bullet \\
    h_{22} &  &  &  & \bullet & \bullet & \bullet &  &  &  &  &  &  
    &  &  &  & \bullet & \bullet &  &  &  &  &  &  & \bullet &  
    & \bullet & \bullet \\
    h_{33} &  &  &  &  & \bullet & \bullet &  &  &  &  &  &  &  &  
    &  & \bullet & \bullet &  &  &  &  &  &  & \bullet &  & \bullet 
    & \bullet \\
    h_{tr} &  &  &  &  &  & \bullet &  &  &  &  &  &  &  &  &  &  
    & \bullet &  &  &  &  &  &  & \bullet &  & \bullet & \bullet \\
    {\color{darkred}h_{t1}} &  &  &  &  &  &  & {\color{darkred}\bullet} &  &  
    & {\color{darkred}\bullet} &  &  &  &  & {\color{darkred}\bullet} &  &  
    & {\color{darkred}\bullet} &  &  &  &  &  &  & {\color{darkred}\bullet} &  
    &  \\
    h_{t2} &  &  &  &  &  &  &  & \bullet &  &  & \bullet &  &  & \bullet
    &  &  &  &  & \bullet &  & \bullet & \bullet &  &  &  &  &  \\
    h_{t3} &  &  &  &  &  &  &  &  & \bullet &  &  & \bullet & \bullet &  
    &  &  &  &  &  & \bullet &  &  & \bullet &  &  &  &  \\
    {\color{darkred}h_{r1}} &  &  &  &  &  &  &  &  &  & {\color{darkred}\bullet} 
    &  &  &  &  & {\color{darkred}\bullet} &  &  & {\color{darkred}\bullet} &  &  
    &  &  &  &  & {\color{darkred}\bullet} &  &  \\
    h_{r2} &  &  &  &  &  &  &  &  &  &  & \bullet &  &  & \bullet &  &  
    &  &  & \bullet   &  &  & \bullet &  &  &  &  &  \\
    h_{r3} &  &  &  &  &  &  &  &  &  &  &  & \bullet & \bullet &  &  &  
    &  &  &  & \bullet &  &  & \bullet &  &  &  &  \\
    h_{12} &  &  &  &  &  &  &  &  &  &  &  &  & \bullet &  &  &  &  &  
    &  &  &  &  & \bullet &  &  &  &  \\
    h_{13} &  &  &  &  &  &  &  &  &  &  &  &  &  & \bullet &  &  &  &  
    &  &  &  &  &  &  &  &  &  \\
    {\color{darkred} h_{23}} &  &  &  &  &  &  &  &  &  &  &  &  &  &  
    & {\color{darkred}\bullet} &  &  &  &  &  &  &  &  &  
    & {\color{darkred}\bullet} &  &  \\
    A^f_t &  &  &  &  &  &  &  &  &  &  &  &  &  &  &  & \bullet & \bullet 
    &  &  &  &  &  &  & \bullet &  & \bullet & \bullet \\
    A^f_r &  &  &  &  &  &  &  &  &  &  &  &  &  &  &  &  & \bullet &  &  
    &  &  &  &  & \bullet &  & \bullet & \bullet \\
    {\color{darkred} A^f_1} &  &  &  &  &  &  &  &  &  &  &  &  &  &  &  &  &
    & {\color{darkred}\bullet} &  &  &  &  &  &  & {\color{darkred}\bullet} &  
    &  \\
    A^f_2 &  &  &  &  &  &  &  &  &  &  &  &  &  &  &  &  &  &  & \bullet &  
    & \bullet & \bullet &  &  &  &  &  \\
    A^f_3 &  &  &  &  &  &  &  & \text{*} &  &  &  &  &  &  &  &  &  &  &  
    & \bullet &  &  & \bullet &  &  &  &  \\
    B^f_t &  &  &  &  &  &  &  &  &  &  &  &  &  &  &  &  &  &  &  &  
    & \bullet & \bullet &  &  &  &  &  \\
    B^f_r &  &  &  &  &  &  &  &  &  &  &  &  &  &  &  &  &  &  &  &  &  
    & \bullet &  &  &  &  &  \\
    B^f_1 &  &  &  &  &  &  &  &  &  &  &  &  &  &  &  &  &  &  &  &  &  &  
    & \bullet &  &  &  &  \\
    B^f_2 &  &  &  &  &  &  &  &  &  &  &  &  &  &  &  &  &  &  &  &  &  &  &  
    & \bullet &  &  &  \\
    {\color{darkred}B^f_3} &  &  &  &  &  &  &  &  &  &  &  &  &  &  &  &  &  &  &  
    &  &  &  &  &  & {\color{darkred}\bullet} &  &  \\
    \rho^f  &  &  &  &  &  &  &  &  &  &  &  &  &  &  &  &  &  &  &  &  &  &  &
    &  &  &  & \bullet \\
    \left(\rho^f\right)^\ast &  &  &  &  &  &  &  &  &  &  &  &  &  &  &  &  &  &
    &  &  &  &  &  &  &  &  & \\
    \bottomrule
  \end{tabular}
  \caption{Couplings between the fluctuation fields. Couplings between two fields in the
    quadratic action $S_q$, which determines the linearized equations of motion, are
    indicated by $\bullet$. The block of coupled fluctuations containing $A^f_1$ is
    marked in {\color{darkred}red}.\label{tab:appendix-helix-matrix-of-couplings}}
\end{sidewaystable}

\section{Asymptotic Expansions\label{app:appendix-helix-fluct-asymp-exp}}

Asymptotic expansions of the fluctuation fields are computed near the thermal horizon
$r_h$ and near the boundary. The expansion around $r_h$ has the leading terms 
\begin{equation}
  \begin{aligned}
    \mathcal A&=(r-r_h)^{-\nicefrac{\ci\omega}{(4\pi T)}}\left(A^h_0+A^h_1(r-r_h)+\dotsb\right), \\ 
    \mathcal B&=(r-r_h)^{-\nicefrac{\ci\omega}{(4\pi T)}}\left(B^h_0+B^h_1(r-r_h)+\dotsb\right), \\ 
    \mathcal E&=(r-r_h)^{-\nicefrac{\ci\omega}{(4\pi T)}}\left(E^h_1(r-r_h)+E^h_2(r-r_h)^2
      \dotsb\right), \\
    \mathcal F&=(r-r_h)^{-\nicefrac{\ci\omega}{(4\pi T)}}\left(F^h_0+F^h_1(r-r_h)+\dotsb\right).
  \end{aligned}
  \label{eq:appendix-horizon-fluct-leading}
\end{equation}
Infalling wave boundary conditions have been chosen, which lead to the retarded Green
function. The expansion has three free parameters which are chosen to be $A^h_0$,
$B^h_0$, and $F^h_0$. All remaining expansion coefficient can be expressed in terms of
these three parameters by means of the equations of motion. We can choose three linearly
independent points in the ($A^h_0$, $B^h_0$, $F^h_0$)-space to define initial conditions
for a numerical integration starting from a point $r_h+\delta$ with $\delta$ numerically
small. In this way, we obtain three linearly independent sets of solutions. These can be
linearly combined to satisfy three conditions at the boundary. Two of these conditions
are given by requiring that gauge invariant fields are built from $\mathcal B$ and
$\mathcal F$, and that these have no source at the boundary. The third condition
corresponds to the normalization of the solution. At the boundary, a double expansion in
$\nicefrac 1r$ and $\nicefrac{\log(r)}r$ is carried out. The leading terms of the
expansion\footnote{For the sake of clarity, the shift parameter $\alpha$ is set to zero
  here. It can be reinstated using the transformation of
  Eq.\@ \eqref{eq:helix-reinstate-alpha} and additionally $F_4^b\to F_4^b-3\alpha^2
  \nicefrac{(4p^2-\omega^2)}{16}$.} are
\begin{equation}
  \begin{aligned}
    \mathcal A&=A_0^b+\frac{A_2^b+\omega^2\log(r)\nicefrac{A_0^b}2}{r^2}+\dotsb, \\
    \mathcal B&=B_0^b+\frac{2B_2^b+\log(r)\left[\left(\omega^2-p^2\right)B_0^b
        +p\lambda\left(\ci\omega E_0^b-2pF_0^b\right)\right]}{2r^2}+\dotsb, \\
    \mathcal E&=E_0^b+\frac{E_4^b-p\lambda\log(r)
      \nicefrac{\left(p\lambda E_0^b-\ci\omega B_0^b\right)}4}{r^4}+\dotsb, \\
    \mathcal F&=F_0^b+\frac{\left(\omega^2-4p^2\right)F_0^b}{4r^2}+\frac{F_4^b
      +\log(r)\left[\left(\omega^2-4p^2\right)^2F_0^b-4p^2\lambda B_0^b\right]/16}{r^4}
    +\dotsb.
  \end{aligned}
\label{eq:appendix_boundary_fluct-leading}
\end{equation}
This expansion has seven free parameters, namely $A_0^b$, $A_2^b$, $B_0^b$, $B_2^b$,
$E_0^b$, $F_0^b$, and $F_4^b$. The coefficient $E_4^b$ and higher order coefficients can
be expressed in terms of these parameters.


\subsection{Residual Gauge Transformations and Physical Degrees of
  Freedom\label{ssec:appendix-helix-residual-gauge-tr}}

In order to determine the physical degrees of freedom corresponding to the fluctuation
fields, the residual gauge transformations of the radial gauge and their action on the
fluctuation fields are worked out. The physical fields are those being invariant with
respect to a residual gauge transformation. To determine them, we follow a similar
calculation carried out in \cite{Erdmenger2012} in the context of the holographic
p-wave model. The gauge transformations of the action of Eq.\@ \eqref{eq:helix-action}
with a massless helix field and vanishing Chern-Simons coupling are diffeomorphisms
$x^{\mu}\to x'^{\mu}=x^{\mu}-\Sigma^{\mu}(x)$, $U(1)$ transformations of $A_\mu$, and
$U(1)$ transformations of $B_{\mu}$. Their infinitesimal action on the fluctuation
fields is given by
\begin{equation}
  \begin{aligned}
    \delta h_{\mu\nu}&=\nabla_{\mu}\Sigma_{\nu}+\nabla_{\nu}\Sigma_{\mu}, \\
    \delta A^f_{\mu}&=(\partial_\mu\Sigma^\nu)A^b_\nu+(\partial_\nu A^b_{\mu})\Sigma^{\nu}
    +\partial_\mu\Lambda, \\
    \delta B^f_{\mu}&=(\partial_\mu\Sigma^\nu)B^b_\nu+(\partial_\nu B^b_{\mu})\Sigma^{\nu}
    +\partial_\mu \Gamma, \\
    \delta\rho^f&=(\partial_\nu\rho^b)\Sigma^\nu+\ci q\Lambda\rho^b.
  \end{aligned} 
   \label{eq:appendix-gauge-trafos}
\end{equation}
The vector $\Sigma_{\mu}$ parameterizes the diffeomorphisms, and the scalars $\Lambda$
and $\Gamma$ the $U(1)$ transformations. The transformations depend implicitly on the
background metric. Since we work at zero spatial momentum, we can focus on $r$- and
$t$-dependent diffeomorphisms and $U(1)$ transformations. Furthermore, we work in
frequency space assuming a harmonic time dependence $\e[-\ci\omega t]$ of
$\Sigma_\mu$, $\Lambda$ and $\Gamma$. The residual gauge transformations of radial gauge
are those satisfying
\begin{alignat}{3}
  \delta h_{r\mu}&=0, &\qquad \delta A^f_r&=0, &\qquad \delta B^f_r&=0.
  \label{eq:appendix-helix-residual-gauge-trafos-requirement}
\end{alignat}
Using the transformation rules of Eq.\@ \eqref{eq:appendix-gauge-trafos}, a system
of differential equations for $\Sigma$, $\Lambda$ and $\Gamma$ is obtained. It
has the solution
\begin{equation}
  \begin{alignedat}{3}
    \Sigma_t&=K_t-\ci K_r\omega\int_1^r \frac{\dd{\bar r}}{U(\bar r)^{3/2}}, &\qquad 
    \Sigma_r&=K_r\sqrt U,  &\qquad \Sigma_x&=K_x, \\ 
    \Sigma_y&=K_y, &\qquad \Sigma_z&=K_z, &\qquad \Gamma&=K_\Gamma r, \\
    \Lambda&=K_\Lambda+\ci K_r\omega\int_1^r\frac{\dd{\bar r}a(\bar r)}{U(\bar r)^{3/2}},
  \end{alignedat}
  \label{eq:appendix-helix-residual-gauge-trafos-sol}
\end{equation}
with $K_t$, $K_r$, $K_x$, $K_y$, $K_z$, $K_\Lambda$, and $K_\Gamma$ being
constants. Having determined the residual gauge transformations, we can write
down their action on the fluctuation fields:
\begin{equation}
  \begin{alignedat}{2}
    \delta\mathcal A&=0, &\qquad\qquad \delta\mathcal B&=-K_xpw, \\
    \delta h_{t1}&=-\ci\e[2v_1]K_x\omega,& \delta h_{23}&=K_xp(\e[2v_3]-\e[-2v_2]).
  \end{alignedat}
  \label{eq:appendix-helix-resid-gauge-trafos-on-fields}
\end{equation}
The gauge transformation of $\mathcal A$ vanishes and, since $\e[2v_2]=r^2$ and
$\e[2v_3]=r^2$ for large values of $r$, the metric fluctuation $h_{23}$
is also gauge invariant at the boundary, where the Green function is read
out. $h_{t1}$ and $\mathcal B$ can be combined into the gauge invariant field
\begin{align}
  \mathcal G=-\ci\omega\mathcal B+wp\e[-2v_1]h_{t1}.
  \label{eq:appendix-helix-gauge-inv-g}
\end{align}
We therefore have three physical fluctuation fields, namely $\mathcal A$, $\mathcal G$,
and $\mathcal F=\e[-2v_2]h_{23}$. The field $\mathcal E=\e[-2v_1]h_{t1}$ is not gauge
invariant and does, therefore, not carry dynamical degrees of freedom.


\section{Radial Perturbations of Zero-Temperature Fixed Points\label{app:exponents}}

We list below the radial perturbations (IR operator dimensions) and their corresponding
eigenvectors in the three IR geometries found in our system \eqref{eq:helix-action}. The
below notation for the eigenvectors $\vec{v}$ is connected to the notation in
\eqref{eq:app-t-zer-power} respectively by
\begin{equation}
  (c_U,c_1,c_2,c_3,c_a,c_w,c_\rho)=c\vvec v=c(v_1,v_2,v_3,v_4,v_5,v_6,v_7).
\label{eq:EVnotation}
\end{equation}
Note that the definitions of the $(c_U,c_1,c_2,c_3,c_a,c_w,c_\rho)$ slightly change in
\eqref{eq:app-t-zer-power}, which has to be taken into account when using the
here-quoted results. In the below tables, M denotes an IR marginal, R an IR
relevant, and I and IR irrelevant mode.


\subsection{Metallic $AdS_2\times \mathds{R}^3$ Fixed Point\label{ssec:exponentsMetallic}}

Within the system \eqref{eq:helix-action} including the charged scalar, the insulating
geometry of \cite{Donos2013c} has the following radial perturbations:
\begin{enumerate}
\item x-rescalings (M): $\delta=0$ and $\vvec v=(0,1,0,0,0,0,0)$,
\item Combined y/z rescalings (M): $\delta=0$ and $\vvec v=(0,0,1,1,0,0,0)$,
\item Constant shift of the chemical potential (R): $\delta=-1$ and $\vvec v=(0,0,0,0,1,0,0)$,
\item Combined $v_i$ mode (R): $\delta=-1$ and
  $\vvec v=\left(0,-2\frac{v_{30}}{v_{10}},1,1,0,0,0\right)$,
\item Mode in the blackening factor (R): $\delta=-1$ and $\vvec v=(1,0,0,0,0,0,0)$,
\item Constant shift of the blackening factor (R): $\delta=-2$ and $\vvec v=(1,0,0,0,0,0,0)$,
\item Combined gauge field and geometry mode (I):\\  $\delta=1$ and
  $\vvec v=\left(\frac{14}9,-\frac2{3v_{10}},-\frac2{3v_{30}},-\frac2{3v_{30}},1,0,0\right)$,
\item Scalar condensation mode (if real: $\delta_-$ (R), $\delta_+$ (I) for
  $m_\rho^2>2q^2$ and (R) for $m_\rho^2<2q^2$): $\vvec v=(0,0,0,0,0,0,1)$ and
  $\delta_\pm=\frac16\left(-3\pm\sqrt{3(3-2 q^2+m_\rho^2)}\right)$. Note that
  for the massless charged scalar used in this work, the $\delta_+$ mode is always
  relevant if the exponents are real, \ie for $q^2<\nicefrac32$.
\item Helix condensation mode  (if real: $\delta_-$ (R), $\delta_+$ (I) for
  $m^2+p^2\e[-2v_{10}]-2\sqrt6\e[-v_{10}]p\kappa>0$ and (R) for
  $m^2+p^2\e[-2v_{10}]-2\sqrt6\e[-v_{10}]p\kappa<0$):\\ 
  $\delta_\pm=-\frac12\left(1\mp\sqrt{\frac13\left(3+m^2+p^2\e[-2v_{10}]
        -2\sqrt6\e[-v_{10}]p\kappa\right)}\right)$ and $\vvec v=(0,0,0,0,0,1,0)$. Note
  that for the massless helix field chosen in this work, $\delta_+$ in tendency will be
  irrelevant if $p\e[-v_{10}]$ is large, but the actual value of $v_{10}$ is of course
  given by UV data. 
\item Additional geometry mode ($\delta_-$ (R), $\delta_+$ (I)):
  $\delta_\pm=-\frac12\left(1\mp\sqrt{1+\frac43p^2\e[-2v_{10}]}\right)$ and
  $\vvec v=(0,0,-1,1,0,0,0)$. Note that while for the other modes considered above the
  $U''(r)$ equation enforces $c_2=c_3$, this is not the case for this mode, since here
  the $U''(r)$ equation is proportional to the quadratic polynomial in $\delta$ which
  vanishes for the solutions $\delta_\pm$ considered here, and hence is automatically
  fulfilled to first order in the perturbations.  
\end{enumerate}
Note that the background has $v_{20}=v_{30}$ in the deep IR. Interestingly, only the
last mode breaks this as one flows up to the UV. Also, all modes are either marginal,
have $\delta=-1$, or come in pairs which sum up to $-1$. This is obvious for the modes
8,9,10, but in fact modes 6 and 7 are also a pair arising from the polynomial
$\delta^2+\delta-2$. Finally, note that several modes (5,6,7) contribute to the
perturbation of $U(r)$ and hence can contribute to the temperature perturbation. This
means that one must continue these IR perturbations to the UV by constructing the
perturbed RG flow in order to understand the different contributions of these modes. We
will come back to this in future work \cite{WIP}.


\subsection{Insulating Fixed Point\label{ssec:exponentsInsulating}}

By switching off the scalar by setting $\rho_0=0$ in the zero temperature solution
\eqref{eq:app-t-zer-power}, \eqref{eq:zero-temperature-initial} implies $\rho_1=0$ and
hence the Ansatz \eqref{eq:zero-temperature-full-expansion} would have no radial
perturbation for the charged scalar $\rho$. In calculating the above modes we hence used
a slightly different Ansatz for $\rho$ compared to
\eqref{eq:zero-temperature-full-expansion}, namely 
\begin{equation}
  \rho=\underbrace{\rho_0+\rho_1r^{\nicefrac43}}_{=0}+c_\rho r^\delta.
  \label{eq:newrhoansatz}
\end{equation}
Within the system \eqref{eq:helix-action} including the charged scalar with mass
$m_\rho$, the insulating geometry of \cite{Donos2013c} (\ie \eqref{eq:app-t-zer-power}
with $\rho_0=0$) has the following radial perturbations:
\begin{enumerate}
\item Blackening factor mode (R): $\delta=-\nicefrac53$ and $\vvec v=(1,0,0,0,0,0,0)$,
\item Gauge field mode (R): $\delta=-\nicefrac53$ and $\vvec v=(0,0,0,0,1,0,0)$,
\item Constant shift of the leading helix parameter $w_0$ (M): $\delta=-\nicefrac43$ and \\
  $\vvec v=(0,0,0,0,0,1,0)$,
\item Charged scalar mode (if real: $\delta_-$ (R), $\delta_+$ (I) if $m_\rho^2>0$
  or (R) if $m_\rho^2<0$, (M) if $m_\rho^2=0$):
  $\delta_\pm=-\frac56\left(1\mp\sqrt{1+\frac25m_\rho^2}\right)$ and $\vvec v=(0,0,0,0,0,0,1)$.
\item Combined gauge and helix field mode:
  $\delta_\pm=\frac16\left(-9\pm\sqrt{1+120\kappa^2}\right)$ and \\
  $\vvec v_\pm=\left(0,0,0,0,\frac{-1\pm \sqrt{1+120\kappa^2}}{12\kappa^2},1,0\right)$,
with $\delta_-<0$ always (R), and $\delta_+<0$ (R) ($\delta_+>0$ (I)) for
$|\kappa|<\sqrt{\nicefrac23}\approx0.817$ ($|\kappa|>\sqrt{\nicefrac23}\approx0.817$), 
\item Combined matter and geometry mode:
  $\delta_\pm=\frac16\left(-5\pm\sqrt{145}\right)$
  ($\delta_+$ (I), $\delta_-$ (R)) and
  \begin{align*}
    \vvec v_+&=\left(\frac2{17}\left(14-\sqrt{145}\right),-1,
      -\frac1{17}\left(3+\sqrt{145}\right),1,\right.\\
    &\left.-\frac{10\left(3\left(116105+9643\sqrt{145}\right)
          \kappa^2+1081\sqrt{145}+13103\right)}{17\left(15\left(26309+2185\sqrt{145}\right)
          \kappa^2-70009 \sqrt{145}-843005\right)},\right.\\
    &\left.\frac{4\left(-12\left(7720+641\sqrt{145}\right)\kappa^2
          +8071\sqrt{145}+97211\right)}{17\left(3\left(4105+341\sqrt{145}\right)\kappa^2
          -2185\sqrt{145}-26309\right)},0\right)\\
    \vvec v_- &=\left(\frac2{17}\left(14+\sqrt{145}\right),-1,
      -\frac1{17}\left(3-\sqrt{145}\right),1,\right.\\
    &\left.\frac{10\left(\left(348315-28929\sqrt{145}\right)\kappa^2
          -1081\sqrt{145}+13103\right)}{17\left(15\left(2185\sqrt{145}-26309\right)\kappa^2
          -70009\sqrt{145}+843005\right)},\right.\\
    &\left.\frac{4\left(\left(92640-7692\sqrt{145}\right)\kappa^2
          +8071\sqrt{145}-97211\right)}{17\left(3\left(341\sqrt{145}-4105\right)\kappa^2
          -2185\sqrt{145}+26309\right)},0\right),
  \end{align*} 
\item Combined geometry mode: $\delta_\pm =  \frac{1}{6}\left( -5\pm \sqrt{185} \right)$
  ($\delta_+$ (I), $\delta_-$ (R)) and
  \begin{align*}
    \vvec v_\pm=&\left(-\frac2{13}\left(71\mp5\sqrt{185}\right),
      \frac1{13}\left(31\mp2\sqrt{185}\right),-\frac2{13}\left(31\mp2\sqrt{185}\right),
      1,\right.\\
    &\left.\frac1{26}\left(205\mp17\sqrt{185}\right),
      -\frac4{13}\left(31\mp2\sqrt{185}\right),0\right),
  \end{align*}
\item Combined matter and geometry mode (M): $\delta = 0$ and $\vvec v=(0,-1,2,1,2,4,0)$,
\item Combined matter and geometry mode (R): $\delta = -1$ and $\vvec v=(6,-1,2,1,5,4,0)$.
\end{enumerate}
Here modes 1,6,7,9 can contribute to the temperature mode. Besides the obvious pairs
(4,5,6,7) which sum up to $-\nicefrac53$ (4,6,7) and $-3$ (5), there seem to be single
modes (3,8,9) as well.


\subsection{Condensed Fixed Point\label{ssec:exponentsCondensed}}

For convenience, we switch back to the Ansatz for the perturbations
\eqref{eq:zero-temperature-full-expansion}. Within the system
\eqref{eq:helix-action} including the charged scalar, our insulating geometry with
charged scalar hair \eqref{eq:app-t-zer-power} has the following radial perturbations:
\begin{enumerate}
\item Combined blackening factor, gauge field and charged scalar mode (R):
  $\delta=-\nicefrac53$ and $\vvec v=(-1,0,0,0,-1,0,5)$ (corresponds together with mode
  2 below to mode 1 and 2 in Appendix \ref{ssec:exponentsInsulating}),
\item Combined blackening factor, gauge field and helix mode (R):
  $\delta=-\nicefrac53$ and $\vvec v=\left(\frac{9(q^2\rho_0^2-4)}{10q^2\rho_0^2},
    0,0,0,\frac{13(q^2\rho_0^2-4)}{20q^2\rho_0^2},1,0\right)$
  (corresponds together with mode 1 above to mode 1 and 2 in Appendix
  \ref{ssec:exponentsInsulating}, $\rho_0\rightarrow 0$ not obvious), 
\item Constant shift of the leading helix parameter $w_0$ (M): $\delta=-\nicefrac43$ and \\
  $\vvec v=(0,0,0,0,0,1,0)$,
\item Charged scalar modes:\\
  Constant shift in $\rho_0$ (M): $\delta=-\nicefrac43$ and $\vvec v=(0,0,0,0,0,0,1)$
  (corresponds to $\delta_+$ of mode 4 in Appendix \ref{ssec:exponentsInsulating}),\\ 
  IR relevant scalar mode: $\delta=-3$ and $\vvec v=(0,0,0,0,0,0,1)$ (corresponds to
  $\delta_-$ of mode 4 in Appendix \ref{ssec:exponentsInsulating} for $m_\rho^2=0$, as for
  this mode $\delta\rho\sim r^{-\nicefrac53}$), 
\item Combined gauge, helix and charged scalar mode:\footnote{Compared to
    \cite{Donos2013c}, there is a shift of $\nicefrac23$ in this exponent because the
    perturbations of the matter fields are written differently with respect to the
    background.}\\
  $\delta_\pm=\frac16\left(-9\pm\sqrt{1+120\kappa^2+20q^2\rho_0^2}\right)$ and \\
  $\vvec v_\pm=\left(0,0,0,0,\frac{(q^2\rho_0^2-4)(-1\pm\sqrt{1+120\kappa^2+20q^2
        \rho_0^2})}{48\kappa^2},1,\frac{3(q^2\rho_0^2-4)(-9\pm\sqrt{1+120\kappa^2+20q^2
        \rho_0^2})}{10\kappa^2(-4+6\kappa^2+q^2\rho_0^2)}\right)$, 
  with $\delta_-< 0$ always (R), and $\delta_+<0$ (R) ($\delta_+>0$ (I)) for
  $1+120\kappa^2+20q^2\rho_0^2<81$ ($1+120\kappa^2+20q^2\rho_0^2>81$),  
\item Combined matter and geometry mode: $\delta_\pm=\frac16\left(-5\pm\sqrt{145}\right)$
  ($\delta_+$ (I), $\delta_-$ (R)) and
  {\small
  \begin{align*}
    \vvec v_+={}&\left(1,-\frac73-\frac{\sqrt{145}}6,\frac16\left(-11-\sqrt{145}\right),
      \frac16\left(14+\sqrt{145}\right),\right.\\
    &\frac{30\left(71+\sqrt{145}\right)\kappa^2
      +5\left(71+\sqrt{145}\right)q^2\rho_0^2+178\sqrt{145}
      -2050}{6\left(6\left(7\sqrt{145}-115\right)\kappa^2
        +\left(7\sqrt{145}-115\right)q^2\rho_0^2-10\sqrt{145}+514\right)},\\
    &-\frac{2\left(6\kappa^2\left(\left(1295+\sqrt{145}\right)q^2\rho_0^2
          -4\left(1565+91\sqrt{145}\right)\right)\right)}
    {3\left(q^2\rho_0^2-4\right)\left(6\left(47\sqrt{145}-335\right)\kappa^2
        +\left(47\sqrt{145}-335\right)q^2\rho_0^2-242\sqrt{145}-46\right)}\\
    &-\frac{2\left(q^2\rho_0^2-4\right)\left(\left(1295+\sqrt{145}\right)q^2\rho_0^2
        -2\left(5209+263\sqrt{145}\right)\right)}
    {3\left(q^2\rho_0^2-4\right)\left(6\left(47\sqrt{145}-335\right)\kappa^2
        +\left(47\sqrt{145}-335\right)q^2\rho_0^2-242\sqrt{145}-46\right)},\\
    &\left.\frac{2\left(6\left(41\sqrt{145}-1985\right)\kappa^2
          +\left(41\sqrt{145}-1985\right)q^2\rho_0^2-614\sqrt{145}+10262\right)}
      {18\left(205+23\sqrt{145}\right)\kappa^2
        +\left(615+69\sqrt{145}\right)q^2\rho_0^2-798\sqrt{145}-8922}\right)\\
    \vvec v_-={}&\left(1,\frac16\left(\sqrt{145}-14\right),
      \frac16\left(\sqrt{145}-11\right),
      \frac73-\frac{\sqrt{145}}6,\right.\\
    &\left.\frac{30\left(\sqrt{145}-71\right)\kappa^2
        +5\left(\sqrt{145}-71\right)q^2\rho_0^2
        +178\sqrt{145}+2050}{6\left(6\left(115+7\sqrt{145}\right)\kappa^2
          +\left(115+7\sqrt{145}\right)q^2\rho_0^2
          -2\left(257+5\sqrt{145}\right)\right)},\right.\\ 
    &\left.-\frac{2\left(6\kappa^2\left(\left(\sqrt{145}-1295\right)q^2\rho_0^2
            -364\sqrt{145}+6260\right)\right)}
      {3\left(q^2\rho_0^2-4\right)\left(6\left(335+47\sqrt{145}\right)\kappa^2
          +\left(335+47\sqrt{145}\right)q^2\rho_0^2-242\sqrt{145}+46\right)}\right.\\
    &\left.-\frac{2\left(\left(q^2\rho_0^2-4\right)
          \left(\left(\sqrt{145}-1295\right)q^2\rho_0^2
            -526\sqrt{145}+10418\right)\right)}{3\left(q^2\rho_0^2-4\right)
        \left(6\left(335+47\sqrt{145}\right)\kappa^2
          +\left(335+47\sqrt{145}\right)q^2\rho_0^2-242\sqrt{145}+46\right)},\right.\\
    &\left.\frac{2\left(6\left(1985+41\sqrt{145}\right)\kappa^2
          +\left(1985+41\sqrt{145}\right)q^2\rho_0^2
          -2\left(5131+307\sqrt{145}\right)\right)}
      {18\left(23\sqrt{145}-205\right)\kappa^2
        +\left(69\sqrt{145}-615\right)q^2\rho_0^2-798\sqrt{145}+8922}\right),
  \end{align*}
  }
\item Combined  geometry mode: $\delta_\pm=\frac16\left(-5\pm\sqrt{185}\right)$
  ($\delta_+$ (I), $\delta_-$ (R)) and
  {\small
    \begin{align*}
      \vvec v_+={}&\left(1,\frac1{64}\left(-27-\sqrt{185}\right),
        \frac1{32}\left(27+\sqrt{185}\right),
        \frac1{64}\left(-71-5 \sqrt{185}\right),\right.\\
      &\left.\frac{10\left(3\left(35+3\sqrt{185}\right)\kappa^2
            +\left(67+5\sqrt{185}\right)q^2\rho_0^2
            -22\left(13+\sqrt{185}\right)\right)}{30\left(21+\sqrt{185}\right)\kappa^2
          +5\left(21+\sqrt{185}\right)q^2\rho_0^2-4\left(355+23 \sqrt{185}\right)},\right.\\
      &\left.\frac{\left(q^2\rho_0^2-4\right)\left(5\left(87+7\sqrt{185}\right)q^2\rho_0^2
            -524\sqrt{185}-6940\right)-120\kappa^2\left(3q^2\rho_0^2+7\sqrt{185}+87\right)}
        {\left(q^2\rho_0^2-4\right)\left(30\left(31+3\sqrt{185}\right)\kappa^2
            +5\left(31+3\sqrt{185}\right)q^2\rho_0^2
            -4\left(665+53\sqrt{185}\right)\right)},\right.\\
      &\left.\frac{2\left(60\left(83+9\sqrt{185}\right)\kappa^2
            +5\left(1057+81\sqrt{185}\right)q^2\rho_0^2
            -176\left(85+7\sqrt{185}\right)\right)}{30\left(479+35\sqrt{185}\right)\kappa^2
          +5\left(479+35\sqrt{185}\right)q^2\rho_0^2-4\left(9225+677\sqrt{185}\right)}\right)\\
      \vvec v_-={}&\left(1,\frac1{64}\left(\sqrt{185}-27\right),
        \frac1{32}\left(27-\sqrt{185}\right),\frac1{64}\left(5\sqrt{185}-71\right),\right.\\
      &\left.\frac{10\left(3\left(3\sqrt{185}-35\right)\kappa^2
            +\left(5\sqrt{185}-67\right)q^2\rho_0^2-22\left(\sqrt{185}
              -13\right)\right)}{30\left(\sqrt{185}-21\right)\kappa^2
          +5\left(\sqrt{185}-21\right)q^2\rho_0^2-92\sqrt{185}+1420},\right.\\
      &\left.\frac{\left(q^2\rho_0^2-4\right)\left(5\left(7\sqrt{185}-87\right)q^2\rho_0^2
            -524\sqrt{185}+6940\right)-120\kappa^2\left(-3q^2\rho_0^2
            +7\sqrt{185}-87\right)}{\left(q^2\rho_0^2-4\right)
          \left(30\left(3\sqrt{185}-31\right)\kappa^2
            +5\left(3\sqrt{185}-31\right)q^2\rho_0^2-212\sqrt{185}+2660\right)},\right. \\
      &\left.\frac{2\left(60\left(9\sqrt{185}-83\right)\kappa^2
            +5 \left(81\sqrt{185}-1057\right)q^2\rho_0^2
            -176\left(7\sqrt{185}-85\right)\right)}{30 \left(35\sqrt{185}-479\right)\kappa^2
          +5\left(35\sqrt{185}-479\right)q^2\rho_0^2-2708\sqrt{185}+36900}\right),
    \end{align*}
  }
\item Combined matter and geometry mode (M): $\delta = 0$ and $\vvec v=(0,-1,2,1,2,4,4)$,
\item Combined matter and geometry mode (R): $\delta = -1$ and $\vvec v=(6,-1,2,1,5,4,4)$.
\end{enumerate}
Here modes 1,2,6,7,9 can contribute to the temperature mode. Besides the obvious pairs
(4,5,6,7) which sum up to $-\nicefrac53$ (4 after taking into account the different
Ansatz for the fluctuations,6,7), and $-3$ (5), there seem to be single modes (3,8,9) as
well.


\section{Numerical Method for Background and Fluctuations
  \label{app:helix-numerical-method}}

From a numerical perspective, we have to solve a boundary value problem for a
set of coupled, non-linear, ordinary differential equations. This is done using
a shooting method consisting of the following steps:
\begin{enumerate}
\item Choosing an initial guess of horizon parameters, the asymptotic horizon
  expansion is used to set up initial conditions at $r_h+\delta$ with $\delta$
  numerically small. The horizon radius $r_h$ can be set to one by a radial
  rescaling (\cff the scaling symmetries discussed below).
\item The equations of motion are integrated numerically between $r_h+\delta$
  and $r_b\gg r_h$ using Mathematica's numerical integrator \texttt{NDSolve}.
\item The difference between the numerical solution and the desired boundary
  values at $r_b$ is read out.
\item The integration between the horizon and the boundary is iterated. Using
  Broyden's method \cite{press1992numerical} as a root finding algorithm, the
  horizon parameters for which the numerical solutions satisfies the boundary
  conditions are determined.
\item The boundary parameters are determined by matching the asymptotic boundary
  expansion to the numerical solution.
\end{enumerate}
%
Alternatively, some of the boundary conditions can be imposed making use of the
following scaling symmetries of the equations of motion:
\begin{equation}
  \begin{alignedat}{5}
    &(\text{I}) & r&\to\gamma r, & t&\to\frac t\gamma, & U&\to\gamma^2U, & 
    a&\to\gamma a, \\
    &(\text{II}) & x&\to\gamma x, & p&\to\frac p\gamma, &
    \e[2v_1]&\to\frac{\e[2v_1]}{\gamma^2}, \\    
    &(\text{III}) &\qquad (y,z)&\to\gamma(y,z), &\qquad w&\to\frac w\gamma, 
    &\qquad \e[2v_{2,3}]&\to\frac{\e[2v_{2,3}]}{\gamma^2}. &\qquad &
  \end{alignedat}
  \label{eq:helix-scaling-symmetries}
\end{equation}
Using scaling (II) and (III), we can set $\e[2v_1]=r^2$ for $r\gg r_h$ and either
$\e[2v_2]=r^2$ or $\e[2v_3]=r^2$. In this way, the scaling symmetries reduce the number
of boundary conditions which have to be imposed on the numerical solution by means of
the shooting method. They are in particular useful for finding a first solution to the
equations of motion.  However, the scalings (II) and (III) change the value of $p$ and
the asymptotic value of the helix field $w$. Once a branch of solutions is found, it is
therefore more convenient to use the shooting method as explained above to satisfy all
boundary conditions. Once the backgrounds are generated, we use the following numerical
procedure to calculate the conductivity for a given solution:
\begin{enumerate}
\item Using Mathematica's \texttt{NDSolve}, three linearly independent sets of solutions
  are constructed by numerical integration between $r_h+\delta$ and $r_b$ (with
  $\delta\ll1$ and $r_b\gg r_h$). 
\item The three sets of solutions are linearly combined to a solution for which
  $\mathcal G$ and $h_{23}$ have a vanishing source and which satisfies the
  (arbitrary) normalization condition $\mathcal A(r_b)=1$.
\item The boundary expansion modes of the fluctuations are determined by
  matching the asymptotic expansion valid for $r\gg r_h$ to the numerical
  solution.
\end{enumerate}

Given that the work of \cite{Donos2013c} was not very explicit about the number of free
UV and IR parameters in the zero temperature RG flows of our model, let us conclude this
section by a more detailed explanation of the situation: As noted above, due to the
broken conformal symmetry \eqref{eq:helix-conformal-anomaly} and the five asymptotic
conditions
\begin{alignat}{3}
  U(r)&\sim r^2, &\qquad\quad v_i(r)&\sim\ln r, \qquad i=1,2,3, &\qquad\quad  J_\rho&=0,
  \label{eq:UVconditions}
\end{alignat}
we expect that the zero temperature RG flows are labeled by the two parameters
$(\lambda,\mu)$. In the IR, the conditions \eqref{eq:UVconditions} are fixed by the five
IR parameters ($\rho_0$, $v_{20}$ and $v_{30}$, and the two coefficients of the two IR
irrelevant modes \eqref{eq:deformation-exponents}). We are hence left with a
two-parameter family of solutions, labeled by the two parameters $(\lambda,\mu)$. 
The vacuum expectation values in the UV are then fixed by the requirement of regularity
in the IR, \ie by the vanishing of the coefficients of the IR relevant perturbations
around the condensed geometry (the modes under point 1,2,4,5,6,7,9). Dropping the
charged scalar both in the UV and in the IR, the same argument applies to the
uncondensed insulating solutions. The metallic $\text{AdS}_2\times\mathds R^2$
solutions however are unique (\ie a zero-parameter family) due to the vanishing of the
helix, $\lambda=0$, as well as the charged scalar, and hence the restored conformal
symmetry \eqref{eq:helix-conformal-anomaly}. This is matched in the IR by two free
parameters $v_{10}$, $v_{20}$ ($v_{20}=v_{30}$) and the coefficients of four IR
irrelevant perturbations (modes 7,8,9,10 in Appendix \ref{ssec:exponentsMetallic}), and
hence there are no free parameters left (except of $p$) in this holographic RG flow. In
particular the vanishing of mode 8 and 9 in Appendix \ref{ssec:exponentsMetallic} sets
the charged scalar and the helix field to zero, respectively. An important outcome of
this whole discussion is that the helix pitch $p$ should not be counted as an independent
UV parameter, as it is not a source in the usual quantum field theoretic sense, but
rather a parameter of the boundary geometry. 


\section{Holographic Renormalization and Operator Mixing
  \label{app:helix-renorm-and-op-mixing}}

In order to calculate the retarded Green function, the on-shell action of the
fluctuations is evaluated. As for the background, the on-shell action can be
reduced to a boundary term by partial integration since the bulk term vanishes
upon use of the equations of motion. Fourier transformation of the fluctuation
fields,
\begin{equation}
  \mathcal A(t,r)=\int\frac{\dd\omega}{2\pi}\e[-\ci\omega t]\mathcal A(\omega,r)
  \label{eq:helix-appendix-ft}
\end{equation}
and similarly for the remaining fluctuation fields, results in an action
bilinear in the Fourier modes. It has the form
\begin{align}
  S_q=\int\frac{\dd\omega\dd r\dd[3]{\vvec x}}{2\pi}&\bigg[
  \bf\Phi_{-\omega}''M_1\bf\Phi_{\omega}+\bf\Phi_{-\omega}M_2\bf\Phi_{\omega}''
  +\bf\Phi_{-\omega}'M_3\bf\Phi_{\omega}'+\bf\Phi_{-\omega}'M_4\bf\Phi_{\omega}
  \notag\\
  &+\bf\Phi_{-\omega}M_5\bf\Phi_{\omega}'+\bf\Phi_{-\omega}M_6\bf\Phi_{\omega}
  \bigg]
  \label{eq:appendix-helix-sq-schematic}
\end{align}
where $\bf\Phi$ is the vector of fluctuation fields and $M_i$ are matrices of
$r$-dependent functions containing the background fields. The derivatives of modes that
are evaluated at $(-\omega)$ can be eliminated by partial integration. Upon use of the
equations of motion the bulk term vanishes and $S_q$ reduces to a boundary term. The
horizon contributions to this boundary term are discarded following the prescription of
\cite{Son2002}. To regularize divergences, we introduce an ultraviolet cutoff $r_b$,
which will be removed eventually. The full on-shell action of the fluctuations
$S^f_{\text{os}}$ is given by the sum of $S_q$, the Gibbons-Hawking term
\begin{equation}
  S_{\text{GH}}=2\int\dd t\dd[3]{\vvec x}\sqrt{-\gamma}\nabla_{\mu}n^{\mu},
 \label{eq:appendix-helix-gb-term}
\end{equation}
expanded to second order in  the fluctuations, and appropriate counterterms
$S^f_{\text{ct}}$, ensuring that $S^f_{\text{os}}=S_q+S_{\text{GH}}+S^f_{\text{ct}}$ is
finite in the limit $r_b\to\infty$. The counterterms needed to make $S^{f}_{\text{os}}$
finite are
\begin{align}
  S^f_{\text{ct}}=\int\dd t\dd[3]{\vvec x}\sqrt{-\gamma}\left[-6-\frac12R_\gamma
    +\log(r_b)\left(\frac14F_{ab}F^{ab}+\frac14W^{ab} W_{ab}
      -\frac14R^{ab}_\gamma R_{\gamma,ab}\right)\right].
  \label{eq:appendix-helix-fluct-ct}
\end{align}
Here $\gamma_{ab}$ denotes the induced metric at $r=r_b$, $R_{\gamma}$ is the Ricci
scalar of the induced metric, and $R_{\gamma,ab}$ is the Ricci tensor. The
studies \cite{Erdmenger2012, Erdmenger2013} were taken as references in finding possible
counterterms. In order to extract the retarded Green function, we can either switch off
the sources of the physical fields $\mathcal G$ and $\mathcal F$ by hand, or use the
holographic operator mixing method \cite{Kaminski2010,Erdmenger2012} as explained in the
following. Arranging the physical fields in a vector,
$\bf\Phi^{\text{(phys)}}=(\mathcal A,\mathcal G,\mathcal F)$, the terms of the on-shell
action containing these fields can be written as
\begin{equation}
  S^f_{\text{os}}\supset V\int\frac{\dd\omega}{2\pi}\left[
    \bf\Phi^{\text{(phys)}}_{-\omega}M_A\bf\Phi^{\text{(phys)}'}_{\omega}
    +\bf\Phi^{\text{(phys)}}_{-\omega}M_B\bf\Phi^{\text{(phys)}}_{\omega}\right]_{r = r_b}.
 \label{eq:appendix-helix-on-shell-for-phys-fields}
\end{equation}
Here $M_A$ and $M_B$ are the two matrices
\begin{align*}
  -\frac U2
  \begin{pmatrix}
    \e[v_2+v_3-v_1] & 0 & 0 \\
    0 & \frac{\e[v_1+v_2-v_3]}{\omega^2} & 0 \\
    0 & 0 & \e[v_1+3v_2-v_3] \\
  \end{pmatrix}
  \text{ and }
  \begin{pmatrix}
    -\frac{\e[-v_1+v_2+v_3]\omega^2\log(r)}{2\sqrt U} & \frac{\kappa w}4 & 0 \\ 
    \frac{\kappa w}4 & * & * \\
    0 & * & * \\
  \end{pmatrix},
  \label{eq:appendix-helix-operator-mixing-matrices}
\end{align*}
respectively. The entries marked with an asterisk correspond to rather longish functions
of $r$ containing the background fields. Their precise form is not needed since we are
only interested in the retarded Green function corresponding to $\mathcal A$.
Numerically, we can construct three sets of linearly independent solutions
$\bf\Phi^{\text{phys}}_1$,$\bf\Phi^{\text{phys}}_2$, and $\bf\Phi^{\text{phys}}_3$ by
integration of the equations of motion with three linearly independent initial
conditions. Arranging the solutions in a matrix,
\begin{align}
  H&=
  \begin{pmatrix}
    | & | & | \\
    \bf\Phi^{\text{phys}}_1 & \bf\Phi^{\text{phys}}_2 &
    \bf\Phi^{\text{phys}}_3 \\
    | & | & | 
  \end{pmatrix},
\end{align}
the matrix of the retarded Green function corresponding to ($\mathcal A,\mathcal G,
\mathcal F$) can be calculated as
\begin{equation}
  G_R(\omega)=2\left(-M_A(\omega,r_b)H'(\omega,r_b)H^{-1}(\omega,r_b)-M_B(\omega,r_b)\right)
  \bigg|_{r_b \to \infty}.
 \label{eq:appendix-helix-gf-master-formula}
\end{equation}
A derivation of this result can be found in \cite{Erdmenger2013,Kaminski2010}. The basic
idea consists in constructing sets of solutions to the equations of motion with each set
sourcing only one particular fluctuation on the boundary. Finally, a generalization of the
prescription of \cite{Son2002} to the case of multiple fluctuation fields results in the
above formula for the retarded Green function.

\bibliography{s-wave-helical-lattice}

\end{document}